\AddToHook{package/hyperref/before}{\RequirePackage{float}}
\documentclass[pdflatex,sn-mathphys-num,12pt]{sn-jnl}
\usepackage{graphicx}%
\usepackage{multirow}%
\usepackage{array}%
\usepackage{tabularx}%
\usepackage{amsmath,amssymb,amsfonts}%
\usepackage{bm}%
\usepackage{amsthm}%
\usepackage{mathrsfs}%
\usepackage[title]{appendix}%
\usepackage{xcolor}%
\usepackage{microtype}%
\usepackage{textcomp}%
\usepackage{manyfoot}%
\usepackage{booktabs}%
\usepackage{algorithm}%
\usepackage{algorithmicx}%
\usepackage{algpseudocode}%
\usepackage{listings}%
\usepackage[capitalise,nameinlink,noabbrev]{cleveref}
\usepackage{placeins}
\usepackage{docmute}
\usepackage{setspace}

\crefname{figure}{Figure}{Figures}
\crefname{table}{Table}{Tables}
\crefname{section}{Section}{Sections}
\crefname{equation}{Equation}{Equations}
\newcounter{methods}

\crefname{methods}{Methods}{Methods}
\Crefname{methods}{Methods}{Methods}
\crefformat{methods}{#2Methods#3}
\Crefformat{methods}{#2Methods#3}

\let\table\tableorg
\let\endtable\endtableorg

\begin{document}

\title{{Training-free hydrostatic adjustment and limiters allow stable multi-year rollouts of AI weather models}}

\author[1]{\fnm{Xin} \sur{Wang}}\email{xin.w24@nus.edu.sg}
\equalcont{These authors contributed equally to this work.}

\author[2]{\fnm{Zinuo} \sur{Zhang}}\email{zzn24@mails.tsinghua.edu.cn}
\equalcont{These authors contributed equally to this work.}

\author[4]{\fnm{Juntao} \sur{Yang}}\email{yjuntao@nvidia.com}

\author[3]{\fnm{Zhuoyuan} \sur{Li}}\email{zy.li@nus.edu.sg}

\author[1]{\fnm{Yuxuan} \sur{Yang}}\email{y.yang.65@u.nus.edu}

\author[1]{\fnm{Chenyu} \sur{Dong}}\email{chenyu.dong@u.nus.edu}

\author[4]{\fnm{Jeff} \sur{Adie}}\email{jadie@nvidia.com}

\author[4]{\fnm{Simon} \sur{See}}\email{ssee@nvidia.com}

\author*[1]{\fnm{Gianmarco} \sur{Mengaldo}}\email{mpegim@nus.edu.sg}

\affil*[1]{\orgdiv{Department of Mechanical Engineering, College of Design and Engineering}, \orgname{National University of Singapore}, \orgaddress{\street{9 Engineering Drive 1}, \postcode{117575}, \country{Singapore}}}

\affil[2]{\orgdiv{Xingjian College},
  \orgname{Tsinghua University},
  \orgaddress{\street{30 Shuangqing Road, Haidian District},
    \city{Beijing}, \postcode{100084}, \country{China}}}

\affil[3]{\orgdiv{Institute for Functional Intelligent Materials}, \orgname{National University of Singapore}, \orgaddress{\street{4 Science Drive 2}, \postcode{117544}, \country{Singapore}}}

\affil[4]{
\orgdiv{NVIDIA AI Technology Centre}, 
\orgname{NVIDIA Corporation}, 
\country{Singapore}}

%ABSTRACT 
\abstract{
Artificial intelligence (AI) weather models now match or exceed state-of-the-art numerical weather prediction at medium range. 
Rolling these models out over months or years would extend their use to subseasonal-to-seasonal and climate prediction, but several become unstable during long rollouts.
Here we show that in FourCastNet~3 (FCN3) hydrostatic imbalance among the model's own predicted variables, largest in the tropical lower stratosphere, precedes rollout failure. 
We use this information to activate HADRIAN, a training-free physical corrector that leaves the model weights and its medium-range forecasts unchanged. 
HADRIAN passes predictions through untouched until the hydrostatic residual exceeds a fixed threshold, then restores hydrostatic balance and applies targeted corrections to the tropical stratospheric wind and near-surface temperature before each state is fed back.
All 40 corrected FCN3 members initialized in 2021 and 2023 remain stable for two years, whereas all 40 of their uncorrected controls fail before the end of the second year. 
Over the two years, the corrected rollouts are on average 0.14~K (2021 initializations) and 0.44~K (2023 initializations) colder than ERA5 in global-mean 2-m temperature and retain 93--97\% of its storm-track variance.
In preliminary adaptations to five other models, each tested in a single rollout with targeted corrections matched to it, HADRIAN restores polar temperature cycles in GraphCast and, for about 20 months, in GenCast, reduces the cold biases of Aurora and suppresses the late Antarctic warming of AIFS, but does not stabilize FuXi.
In released AI weather models, long-rollout failures that originate in hydrostatic imbalance can thus be anticipated from the models' own predictions and contained at inference time.
}

\keywords{AI weather models, autoregressive rollout stability, hydrostatic balance, physical consistency, inference-time correction, subseasonal-to-seasonal prediction}

\maketitle
\doublespacing

%%%%%%%%%%%%%%%%%%%%%%%%%%%%%%%%%%%%%%%%%
% INTRODUCTION
%%%%%%%%%%%%%%%%%%%%%%%%%%%%%%%%%%%%%%%%%
\clearpage
\section{Introduction}\label{intro}
Artificial intelligence (AI) weather models now rival the skill of numerical weather prediction (NWP) systems at lead times of up to two weeks~\citep{pathak2022fourcastnet, bi2023accurate, lam2023learning, chen2023fuxi, chen2025fengwu, price2025probabilisti, bonev2025fourcastnet, rasp2023weatherbench}.
These AI models forecast autoregressively.
A neural network trained to advance the atmosphere by a short step ahead is applied repeatedly, with each prediction serving as the input for the next step, a procedure known as rollout.
Because this step can be repeated indefinitely, the same AI model could in principle be rolled out far beyond two weeks, extending its use from weather to subseasonal, seasonal and climate timescales.
Such an extension would allow AI models to operate seamlessly across timescales~\citep{palmer2008seamless, hoskins2013potential}, a goal that operational centres have made a strategic priority, as exemplified by the 2025 to 2034 strategy of the European Centre for Medium-Range Weather Forecasts~\citep{ecmwf2025ecmwf}.

Recent AI studies have begun to bridge the divide between weather and climate timescales~\citep[e.g.,][]{ling2024fengwu, kochkov2024neural, campsvalls2026bridging, dong2026s2s}, but whether AI models tuned for medium-range forecasting can themselves be rolled out for months or years and still produce meaningful forecasts remains an open question.
Long rollouts are challenging because each step starts from the previous step's output, allowing errors to compound over time, potentially undermining forecasting accuracy.
Indeed, several publicly available $0.25^\circ$ AI weather models tuned for medium-range forecasting drift, lose their seasonal cycle or blow up during long rollouts~\citep{lehmann2026can}.

Efforts to prevent such issues either focus on the model design and training strategy, or leave the model untouched and correct its predicted states during rollouts.

In the first category, pure AI climate emulators have been built for multi-decadal simulations, either without explicit conservation constraints~\citep{wattmeyer2023ace} or with hard global conservation constraints~\citep{wattmeyer2024ace, chapman2025camulator}, while hybrid physics-AI models couple an equation-based dynamical core with machine-learned subgrid parameterizations~\citep{wang2022stable, kochkov2024neural, wang2026condensnet}.
Other approaches either impose physical constraints, such as conservation laws, output bounds, and hydrostatic balance, through the model architecture or training strategy, mainly to improve physical consistency and forecast skill~\citep{beucler2021enforcing, harder2023hard, hess2022physically, sha2025improving, moldovan2026aifs, bonev2025fourcastnet, subramaniam2025imposing}, or design dedicated architectures and training losses for stable rollouts~\citep{li2022dissipative, mccabe2023stability, bonev2023spherical, guan2025lucie}.
All of these require custom model design or retraining, which is costly for users of publicly released $0.25^\circ$ AI weather models, since these are distributed as fixed weights trained for medium-range forecasting~\citep{rasp2023weatherbench, lehmann2026can}.

Inference-time corrections avoid this cost, but existing methods either act regardless of need or require additional training.
Fixed rules, such as tracer mass fixers, output clipping, and dry-air mass guidance, are applied at every step~\citep{diamantakis2014global, moldovan2026aifs, harder2023hard, aich2026wind}.
Other corrections rely on trained components, such as a denoiser activated by a learned noise estimate in a turbulence emulator~\citep{pedersen2025thermalizer} or a wrapper that nudges a $1.5^\circ$ weather model towards climatology~\citep{cheon2026rescene}.
Intervening only when needed has been explored in hybrid models, which restrict their machine-learned components when a novelty detector~\citep{sanford2023improving} or a learned confidence estimate~\citep{heuer2025beyond} flags them as unreliable.
None of these methods offers a training-free way to decide when a released AI weather model needs correction.
\begin{figure}[h!]
  \centering
  \includegraphics[width=\linewidth]{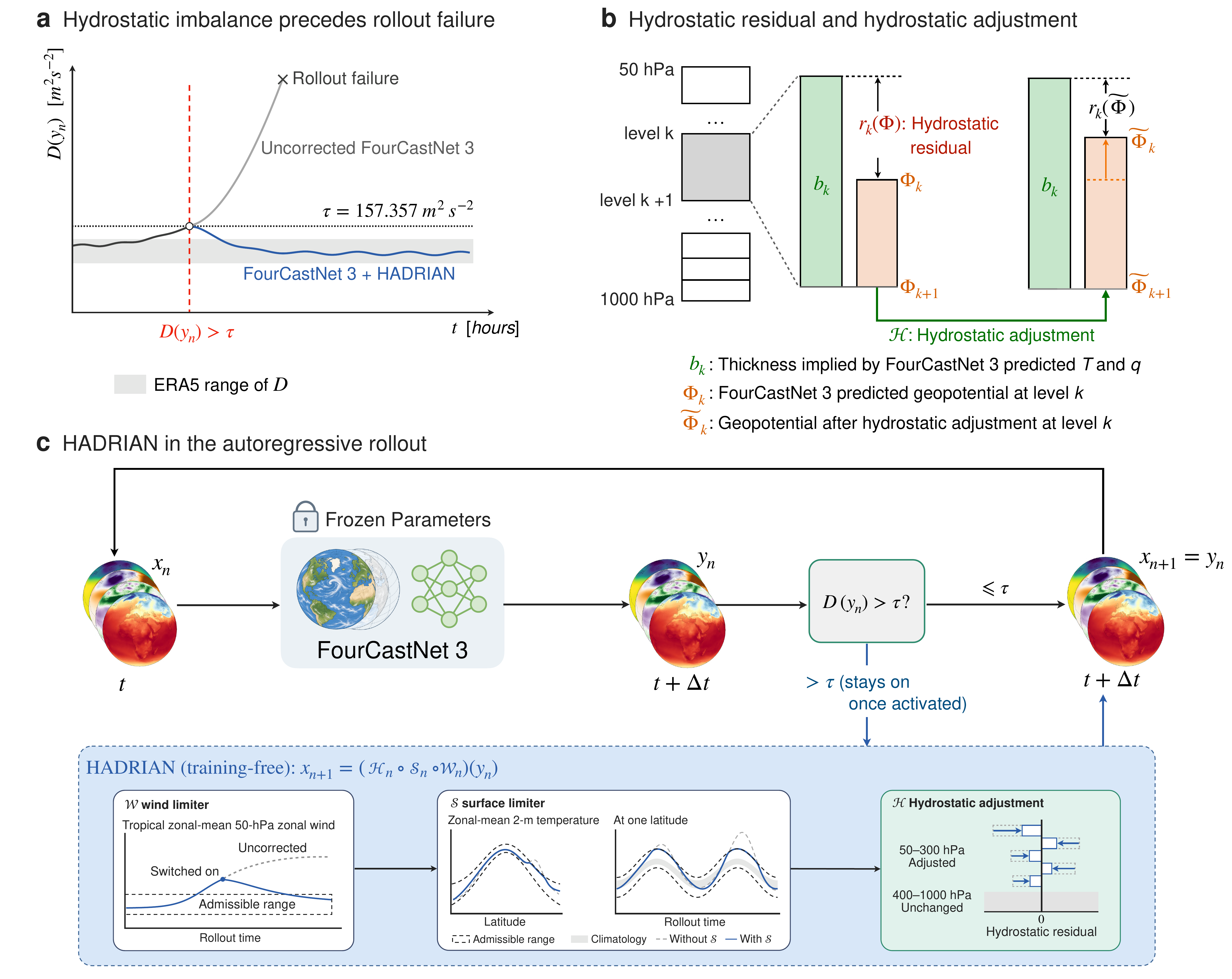}
  \caption{\textbf{FCN3 + HADRIAN: physical correction of the predicted state before it is fed back as input.} (a)~The global root-mean-square hydrostatic residual $D$ of each prediction exceeds the threshold $\tau$ before an uncorrected rollout fails; with HADRIAN, $D$ returns below $\tau$ after the first threshold crossing. (b)~The hydrostatic residual $r_k$ of the layer between levels $k$ and $k+1$ is the difference between the predicted geopotential thickness and the thickness $b_k$ implied by the predicted temperature and specific humidity. Hydrostatic adjustment holds temperature and specific humidity fixed and changes only geopotential, which moves the residual towards its target value. (c)~Predictions pass through unchanged until the first threshold crossing. Thereafter, the wind limiter ($\mathcal W$), surface limiter ($\mathcal S$) and hydrostatic adjustment ($\mathcal H$) correct every prediction at its valid time before it is fed back as the next input. The wind limiter and the surface limiter act only when the tropical zonal-mean 50-hPa zonal wind or the zonal-mean 2-m temperature of a latitude row lies outside its admissible range; hydrostatic adjustment changes geopotential at 50--300~hPa and leaves 400--1000~hPa unchanged. The correction is training-free and leaves FCN3's weights unchanged. All curves, columns and profiles are schematic.}
  \label{fig:architecture}
\end{figure}
Such a decision requires a criterion that can be evaluated on each predicted state as it is produced, so that the need for correction can be flagged while the rollout is still running.
However, most of what is known about how AI weather models fail in long rollouts comes from statistics accumulated over months or years, such as biases, the seasonal cycle, and variability~\citep{chattopadhyay2023challenges, karlbauer2024advancing, henn2026aimip, singh2026evaluating}, none of which can be evaluated state by state as the rollout proceeds.
These statistics show when and how models fail~\citep{lehmann2026can} and point to spectral bias and missing forcing as possible causes~\citep{chattopadhyay2023challenges, karlbauer2024advancing}, but only in retrospect.
Physical consistency, by contrast, can be checked on each state as it is produced. 
For instance, hydrostatic balance, one of the relations that govern the atmosphere, holds closely at every instant on the scales these models resolve.

Balance diagnostics~\citep{bonavita2024some, kasteleyn2026physmetrics, pasquini2026assessing, subramaniam2025imposing} and controlled dynamical tests~\citep{hakim2024dynamical} have assessed physical consistency only at lead times from hours to two months.
Over the first ten days, deterministic AI models are at least as close to hydrostatic balance between 850 and 500~hPa as the ERA5 reanalysis~\citep{hersbach2020era5, kasteleyn2026physmetrics}.
Whether imbalance grows before a long rollout fails has not been examined.
If it does, hydrostatic balance could serve both as a diagnostic of departures from physical consistency and as a constraint for correcting them.
Any such correction must stabilize the rollout without damping atmospheric variability, which was weakened in an early free-running AI weather model that otherwise reproduced a year-long seasonal cycle~\citep{weyn2020cubed}.
Here we show that, in long rollouts of a released AI weather model, failure is preceded by a measurable departure from physical consistency, specifically from hydrostatic balance.
We use this departure to decide when the predicted state needs correction.
The correction requires no retraining, leaves the model weights unchanged, and does not alter the forecast before it is applied.

We use FourCastNet~3 (FCN3)~\citep{bonev2025fourcastnet}, a $0.25^\circ$ probabilistic model that typically becomes unstable and fails within one year of rollout (\cref{sec:failure}).
To identify a precursor of failure, we first examine 20 uncorrected members initialized in 2021, all of which fail.
In every one of them, the hydrostatic imbalance between the geopotential thickness of atmospheric layers and their virtual temperature exceeds a fixed threshold before failure, with a median lead time of 66.25 days.
On this basis, we introduce HADRIAN (Hydrostatic ADjustment and Range limits at Inference, As Needed; \cref{fig:architecture}), a training-free method that leaves the FCN3 predictions unchanged until the threshold is exceeded, and thereafter restores hydrostatic balance and applies targeted corrections to the tropical stratospheric wind and near-surface temperature.
We then evaluate HADRIAN on 40 members run for 2 years.
All remain stable when corrected, whereas all fail within two years when uncorrected.

The corrected 2-year FCN3 rollouts follow the ERA5 seasonal cycle of global-mean 2-m temperature and mean sea-level pressure.
They also keep storm-track variance and the zonal power spectra of temperature, wind and geopotential close to those of ERA5 up to wavenumber 250 (a wavelength of about 160~km at the equator).
By contrast, two AI weather models that remain stable without correction, namely SFNO~\citep{bonev2023spherical} and Aurora~\citep{bodnar2025foundation}, lose spectral power in temperature and wind towards higher wavenumbers (\cref{fig:series}, \cref{tab:survival}).
The stability gained with HADRIAN is therefore not obtained by damping atmospheric variability.
Five further corrected members, initialized on 1 January 2017, run for 50 years without blowing up (\cref{sec:longruns}).
In preliminary tests on five other AI weather models, hydrostatic imbalance develops in all five, and HADRIAN, with targeted corrections matched to the model, improves stability or long-term temperature evolution in four (\cref{sec:crossmodel}).
These results indicate that the failure of a released AI weather model in long rollouts can be anticipated from the physical consistency of its own predictions, and that such failures can be prevented at inference time.

%%%%%%%%%%%%%%%%%%%%%%%%%%%%%%%%%%%%%%%%%
% RESULTS
%%%%%%%%%%%%%%%%%%%%%%%%%%%%%%%%%%%%%%%%%
\FloatBarrier
\section{Results}\label{sec:exp}

\subsection{Hydrostatic imbalance in FCN3}\label{sec:failure}
Most AI weather models, including FCN3, predict geopotential, temperature and specific humidity on pressure levels as separate output channels, with no constraint linking them.
In the atmosphere, by contrast, hydrostatic balance determines the geopotential thickness of a layer from its mean virtual temperature, which combines temperature and humidity.
Each predicted state therefore gives two thicknesses for every layer between adjacent pressure levels, one from its geopotential and one from its temperature and humidity (\cref{fig:architecture}b).
Their difference, the hydrostatic residual, measures how far a prediction departs from hydrostatic balance.

In AI weather models, such balance relations have so far been examined only at lead times much shorter than those at which rollouts fail.
Balance diagnostics and idealized dynamical tests cover lead times from hours to 20 days~\citep{bonavita2024some, hakim2024dynamical, kasteleyn2026physmetrics, pasquini2026assessing}.
Between 850 and 500~hPa, deterministic models stay at least as close to hydrostatic balance as ERA5 over the first ten days~\citep{kasteleyn2026physmetrics}.
The only study, to our knowledge, that imposes hydrostatic balance on an AI weather model adds it as a loss term during fine-tuning, relates the imbalance to forecast skill and follows it over 60 days~\citep{subramaniam2025imposing}.

\begin{figure}[!htb]
  \centering
  \includegraphics[width=\textwidth]{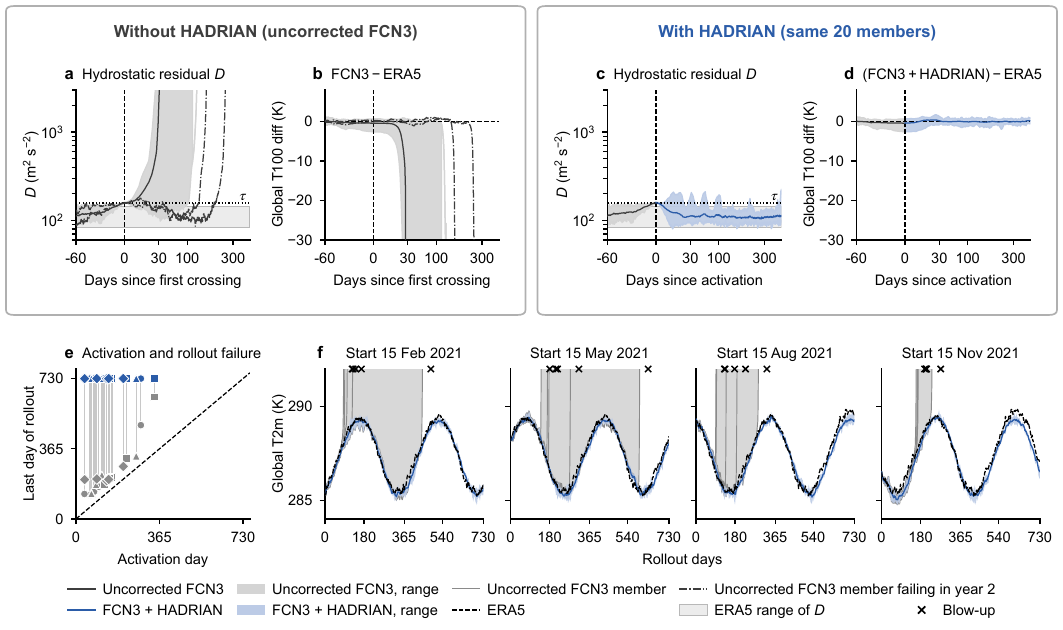}
  \caption{\textbf{Hydrostatic imbalance precedes rollout failure, and correcting it keeps FCN3 stable.} Two-year rollouts of 20 FCN3 members from four 2021 initialization dates, each run without correction and with HADRIAN. (a,b)~Hydrostatic residual $D$ (a) and global-mean 100-hPa temperature minus ERA5 (b) of uncorrected FCN3, aligned to the first crossing of $\tau$ (dotted). Dash-dotted lines show the two members that fail in the second year, which are excluded from the shading. (c,d)~As in a,b for FCN3 + HADRIAN, aligned to activation; $D$ is evaluated on the raw prediction. Lines show the mean over the members still running (geometric mean for $D$), shading the member range and the grey band the 2000--2015 ERA5 range of $D$; time axes are square-root scaled. (e)~Activation day versus last day of each rollout, with lines joining each corrected member (blue) to its control (grey); symbols mark initialization dates (circle, 15 February; square, 15 May; triangle, 15 August; diamond, 15 November). (f)~Global-mean T2m: FCN3 + HADRIAN five-member mean and range (blue), uncorrected members and their range (grey) and ERA5 (dashed). Crosses mark blow-up.}
  \label{fig:imbalance_hadrian}
\end{figure}

To examine hydrostatic balance at longer lead times, we first establish how free-running FCN3 fails.
We initialize FCN3 from ERA5~\citep{hersbach2020era5} at 00 UTC on 15 February, 15 May, 15 August and 15 November 2021, with five random seeds per date, and roll out the resulting 20 members for up to two years without correction or data assimilation.
Following~\citet{lehmann2026can}, we distinguish three forms of \emph{rollout failure}: blow-up, in which predicted values grow without bound; drift, in which the rollout stays finite but its mean state departs from that of ERA5 beyond physically plausible limits; and loss of seasonality, in which the rollout continues but loses its seasonal cycle and variability.
All 20 members blow up, 18 of them within the first year (\cref{fig:imbalance_hadrian}e,f).
We date each blow-up to the first step at which the predicted state contains a non-finite value (NaN or Inf); our framework checks every step and terminates the rollout there.

To determine whether imbalance precedes these failures, we test each FCN3 prediction for hydrostatic balance at every step of the rollout, using only its own predicted variables.
The test quantity is the hydrostatic residual, the difference between the two thicknesses that a prediction gives for each layer (\cref{fig:architecture}b).
With the 13 pressure levels of FCN3 numbered from $p_1 = 50$~hPa to $p_{13} = 1000$~hPa, the residual of the layer between levels $k$ and $k+1$ at grid point $i$ of a state $x$ is
\begin{equation}
\begin{aligned}
r_{k,i}(x) &= b_{k,i} - \bigl(\Phi_{k,i} - \Phi_{k+1,i}\bigr), \\[0.8em]
b_{k,i} &= \frac{R_d}{2}\bigl(T_{v,k,i} + T_{v,k+1,i}\bigr) \ln\frac{p_{k+1}}{p_k}, \\[0.8em]
T_v &= T\left[1 + \left(\frac{R_v}{R_d} - 1\right)q\right],
\end{aligned}
\label{eq:hydrostatic_residual}
\end{equation}
where $\Phi$, $T$ and $q$ are the predicted geopotential, temperature and specific humidity, $T_v$ is the virtual temperature, and $R_d$ and $R_v$ are the gas constants of dry air and water vapour.
Here $b_{k,i}$ is the thickness implied by the predicted temperature and humidity, with the layer-mean virtual temperature approximated by the average of its two bounding levels, and $\Phi_{k,i} - \Phi_{k+1,i}$ is the predicted geopotential thickness.
The residual is therefore positive where the predicted geopotential thickness is smaller than the implied one.
The global magnitude of the residual is its unweighted root mean square (RMS) over all $N$ grid points and the twelve layers,
\begin{equation}
D(x) = \left[\frac{1}{12N}\sum_{i=1}^{N}\sum_{k=1}^{12}r_{k,i}(x)^2\right]^{1/2}.
\label{eq:residual_statistic}
\end{equation}
The two-level approximation of the layer mean leaves a residual even in a hydrostatic atmosphere. 
In daily ERA5 states over 2000--2015, $D$ ranges from 83.2 to 144.7~$\mathrm{m^2\,s^{-2}}$ (\cref{app:precursor}).

We compare $D$ of each prediction with a fixed threshold $\tau \approx 157\,\mathrm{m^2\,s^{-2}}$, equivalent to about 16.0~m of geopotential height.
The threshold is the 99th percentile of an area-weighted form of the residual in ERA5 states from 2000--2015 (\cref{app:parameters}).
Because $D$ is unweighted, $\tau$ lies above the whole ERA5 range of $D$, at 1.09 times its largest value.

In every uncorrected member, $D$ grows during the rollout and exceeds $\tau$ at least 58 days before the rollout fails (\cref{fig:imbalance_hadrian}a,e; shown schematically in \cref{fig:architecture}a).
The first crossing occurs between 37 and 342 days into the rollout (\cref{fig:imbalance_hadrian}e).
At that point, the median departures of global-mean temperature from ERA5 are still relatively small, namely $-0.11$~K for 100-hPa temperature (T100), which ranges from $-2.6$ to $+0.7$~K across members, and $+0.06$~K for 2-m temperature (T2m; \cref{app:precursor}, \cref{tab:hydrostatic_evolution}).
Over the next ten days, $D$ grows by about 40\%, while the median departures of T100 and T2m change by less than 0.1~K.
The median T100 then falls sharply, reaching 2.70~K below ERA5 twenty days after the crossing (\cref{fig:imbalance_hadrian}b).

Global-mean T2m first departs from ERA5 by more than 10~K about 29 days after the crossing (median value with range 21--255 days), and the rollout stops with a numerical error about 66 days after it (median value with range 58--293 days).
The rollout has therefore left physically plausible states well before it blows up.

The longest lead, 293 days, reflects a temporary first crossing.
In that member $D$ falls back below $\tau$ and exceeds it continuously only from 84 days before failure, and $D$ also returns below $\tau$ for a time in nine other members (\cref{app:precursor}).
That the first crossing precedes failure does not depend on the choice of threshold: with the 95th or 99.9th percentile of the calibration distribution (144.9 or 164.3~$\mathrm{m^2\,s^{-2}}$) in place of $\tau$, it still precedes failure in all 20 members.
The median lead, however, changes by about three days with the 99.9th percentile and increases to about 94 days with the 95th percentile, which lies only $0.2\,\mathrm{m^2\,s^{-2}}$ above the largest ERA5 value of $D$ (\cref{app:precursor}).

This imbalance is concentrated in the tropical lower stratosphere, just above the 100-hPa level, which later cools.
Over the ten days before the first crossing, the tropical ($20^\circ$S--$20^\circ$N) 50--100-hPa layer has the largest RMS residual of any layer in any latitude band, in every member.
Globally, the 50--100-hPa layer contributes 75.4\% of $D^2$ (member median), against 57\% in ERA5 at the same valid times (\cref{fig:precursor}a), and for about 30 days after the crossing the growth of $D$ comes almost entirely from this layer (\cref{tab:hydrostatic_evolution}).
Averaged over the tropics, the residual of this layer is positive, so the predicted geopotential thickness is smaller than the thickness implied by the predicted temperature and humidity.
Expressed as a temperature, this tropical-mean residual is 1.0~K at 60 days before the crossing and 3.9~K at the crossing (member medians, over 18 members at 60 days before the crossing; \cref{tab:hydrostatic_evolution} gives the corresponding RMS values).
The ERA5 values at the same valid times are 1.1 and 1.6~K, and at the crossing every member exceeds the ERA5 value.
This layer lies outside the 850--500-hPa range for which deterministic AI weather models were found to be at least as close to hydrostatic balance as ERA5~\citep{kasteleyn2026physmetrics}.

We next test whether correcting this imbalance prevents rollout failure.
HADRIAN leaves FCN3 predictions unchanged until $D$ first exceeds $\tau$ (\cref{fig:architecture}c; \cref{sec:method}).
From that step on, it corrects every prediction in three stages before the prediction is used as the next input.
The central stage is hydrostatic adjustment.
It changes geopotential on the six levels from 50 to 300~hPa, which include the 50--100-hPa layer where the imbalance is largest.
At each step, it moves geopotential at most a tenth of the way towards hydrostatic balance with the predicted temperature and humidity.
The other two stages are limiters.
In 11 of the 20 uncorrected members, a sustained period of excessively strong zonal-mean westerlies at 50~hPa in the tropics precedes the 100-hPa cooling.
While such a jet persists, the wind limiter reduces the part of this wind that lies outside a range prescribed for each day of the year.
The surface limiter corrects only abnormal values of the 2-m temperature.
FCN3 takes no sea-surface temperature or soil state as input, so its 2-m temperature is not tied to any surface boundary condition.
The limiter compares the mean 2-m temperature of each latitude row, taken separately over sea and over land, with an admissible range for the latitude, season and time of day.
At each step, it reduces the part of the mean that lies outside this range (\cref{app:parameters}).
Each corrected member is identical to its uncorrected control until activation, the step at which $D$ of the control first exceeds $\tau$.
Any later difference between the two is therefore due to the correction.
\Cref{sec:ablations} examines the contribution of each stage.

All 20 corrected members complete two years with finite states and pass all seven physical plausibility checks (\cref{sec:method}; \cref{tab:plausibility}), whereas none of their controls completes the rollout (\cref{fig:imbalance_hadrian}e,f).
After activation, $D$ of the raw FCN3 predictions, evaluated before correction, falls back below $\tau$ and settles near the middle of the ERA5 range, exceeding $\tau$ afterwards on at most 6.8\% of steps (\cref{fig:imbalance_hadrian}c).
Given corrected inputs, the model itself therefore stays close to hydrostatic balance.
T100 stays within 0.5~K of ERA5 in the member mean instead of falling away (\cref{fig:imbalance_hadrian}d).
T2m follows the ERA5 seasonal cycle in both years, with a global-mean cold bias that grows from 0.11~K in the first year to 0.17~K in the second (\cref{fig:imbalance_hadrian}f; \cref{app:climatology}).
All 20 members initialized in 2023 also complete two years and pass the seven checks, whereas none of their controls completes the rollout (\cref{tab:survival_full}). The corrected 2023 members have larger global-mean biases than the 2021 members, at $-0.44$~K for T2m and $-0.53$~K for 850-hPa temperature against $-0.14$ and $-0.18$~K. The comparison models of \cref{sec:spectra} (SFNO, Aurora and AIFS) are likewise 0.2--0.3~K colder relative to ERA5 in T2m for the 2023 initializations than for the 2021 ones, which suggests that the larger bias reflects the verification period and not the correction.

% -----------------------------------------------
% VARIABILITY
% -----------------------------------------------
\FloatBarrier
\subsection{Corrected rollouts retain the mean state and variability}
\label{sec:spectra}
\begin{figure}[h!]
  \centering
  \includegraphics[width=\textwidth]{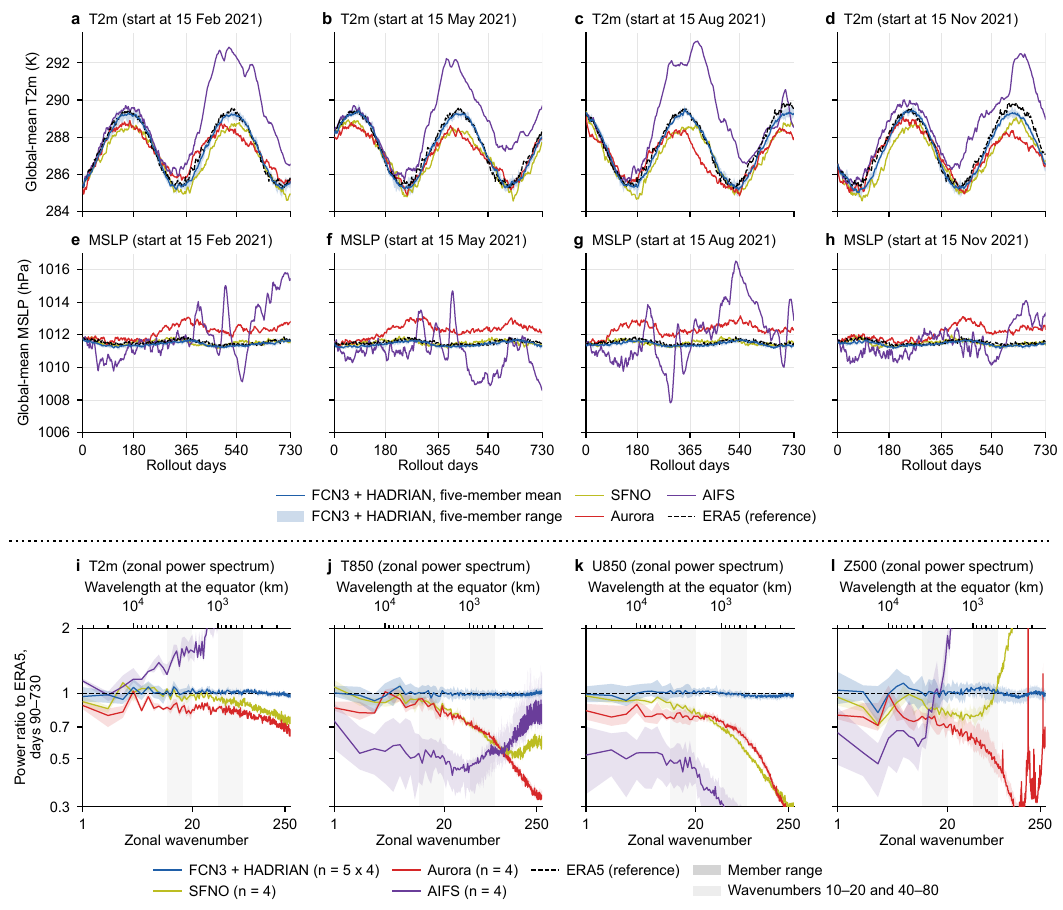}
  \caption{\textbf{FCN3 + HADRIAN retains the seasonal cycle and spectral power close to ERA5 over two years.} Rollouts from the four 2021 initialization dates. (a--d)~Daily 00~UTC global-mean T2m and
  (e--h)~mean sea-level pressure (MSLP) over days 1--730; FCN3 + HADRIAN is shown as the five-member mean and range, and SFNO, Aurora and AIFS as one trajectory per date. (i--l)~Ratio of model to ERA5 zonal power spectra of T2m, 850-hPa temperature (T850), 850-hPa zonal wind (U850) and 500-hPa geopotential (Z500), each averaged over states every ten days from day 90 to day 730; lines show
  medians and shading ranges over the 20 FCN3 + HADRIAN members or the four trajectories of each comparison model. Grey bands mark zonal wavenumbers 10--20 and 40--80.}
  \label{fig:series}
\end{figure}

To test whether the corrected rollouts retain the mean state and variability of the atmosphere, we compare them with ERA5 and, for context, with three released AI weather models that have been evaluated in long rollouts before~\citep{lehmann2026can}, namely  SFNO~\citep{bonev2023spherical}, Aurora~\citep{bodnar2025foundation} and AIFS~\citep{moldovan2026aifs}.
FCN3 + HADRIAN comprises the same 20 members as in \cref{sec:failure}, five per 2021 initialization date; SFNO, Aurora and AIFS are run without correction, with one deterministic trajectory per date.
We call a rollout stable if it stays finite for two years and passes seven physical plausibility checks on its global means, annual drift, meridional temperature structure and seasonal cycle, none of which uses ERA5 (\cref{sec:method}).
SFNO and Aurora are stable in all four rollouts.
The four AIFS trajectories do not blow up, but they become warmer than ERA5 and fail one check: their global-mean T2m first exceeds the upper bound of 292~K on days 296--618 (\cref{fig:series}a--d; \cref{tab:survival,tab:plausibility}).
Aurora and AIFS do not predict total column water vapour (TCWV), so they are assessed on the other six checks.
\begin{table}[!htb]
\centering
\caption{\textbf{Two-year outcomes for the four 2021 initialization dates.} Biases and storm-track variance ratios are means of per-trajectory metrics; Z500 RMSE is computed from the five-member FCN3 + HADRIAN ensemble-mean field at each initialization date, or from the single trajectory of each comparison model, and then averaged over dates. \textbf{Bold} marks the best and \underline{underlining} the second-best value: highest stable fraction, bias closest to zero, lowest RMSE and NH and SH storm-track variance ratios closest to one. Ties share a rank, and stable fractions of zero are not ranked.}
\label{tab:survival}
\footnotesize
\setlength{\tabcolsep}{3pt}
\resizebox{\textwidth}{!}{%
\begin{tabular}{lcccccc}
\toprule
Model & Stable & Activation & T2m bias & T850 bias & Z500 RMSE & Storm-track variance \\
 & & (days) & (K) & (K) & (m$^2$\,s$^{-2}$) & ratio NH / SH \\
\midrule
Uncorrected FCN3 & 0/20 & -- & n/a & n/a & n/a & n/a \\
FCN3 + HADRIAN & \textbf{20/20} & 37.25--342.25 & \textbf{-0.137} & \textbf{-0.179} & \textbf{178.49} & \textbf{0.933} / \textbf{0.944} \\
SFNO & \textbf{4/4} & -- & -0.530 & -0.592 & 358.75 & \underline{0.795} / 0.845 \\
Aurora & \textbf{4/4} & -- & \underline{-0.381} & \underline{-0.543} & \underline{282.16} & 0.694 / \underline{0.850} \\
AIFS & 0/4 & -- & +1.282 & +1.307 & 498.95 & 0.616 / 0.446 \\
\bottomrule
\end{tabular}}
\par\smallskip
\begin{minipage}{\textwidth}\footnotesize
Stable: finite states and all physical plausibility checks passed through day 730 (six of the seven checks for Aurora and AIFS, which do not predict TCWV; \cref{tab:plausibility}).
All metrics cover days 1--730. Temperature biases and the RMSE of the time-mean Z500 field use daily 00~UTC fields; storm-track variance ratios are 2--6-day band-pass Z500 variance ratios to ERA5 from six-hourly fields. All 20 FCN3 + HADRIAN members and all four trajectories of each comparison model are scored, including the four AIFS trajectories, which remain finite but fail the plausibility checks. n/a, not available because no uncorrected FCN3 member completes two years.
\end{minipage}
\end{table}
FCN3 + HADRIAN follows the ERA5 seasonal cycle of global-mean T2m and mean sea-level pressure (MSLP) from every initialization date (\cref{fig:series}a--h).
Over the two years, its global-mean biases in T2m and 850-hPa temperature (T850) are $-0.14$ and $-0.18$~K, the closest to zero among the compared models; the next closest, those of Aurora, are $-0.38$ and $-0.54$~K (\cref{tab:survival}).
The root-mean-square error (RMSE) of its time-mean 500-hPa geopotential (Z500) field is also the lowest, at 178~m$^2$\,s$^{-2}$ against 282~m$^2$\,s$^{-2}$ for Aurora; this value is for the five-member ensemble mean, which is smoother than the single trajectories of the comparison models.
Among the comparison models, the global-mean T2m bias of AIFS reaches about $+1.9$~K in the second year, and Aurora overestimates global-mean MSLP by about 0.7~hPa (\cref{app:climatology}). 
In the 10-year rollout of \cref{sec:longruns}, AIFS is likewise too warm in the second year but then cools at the surface.
The corrected rollouts also retain synoptic-scale variability and small-scale spectral power.
We measure synoptic-scale variability by the storm-track variance ratio, the variance of 2--6-day band-pass-filtered Z500 poleward of $30^\circ$ divided by that of ERA5.
Over the two years, this ratio is 0.933 in the Northern Hemisphere and 0.944 in the Southern Hemisphere for FCN3 + HADRIAN, the closest to one among the compared models; the next closest values are 0.795 (SFNO) and 0.850 (Aurora), respectively (\cref{tab:survival}).
In the second year, which lies largely after activation, the ratios are 0.950 and 0.918, within about 0.03 of their first-year values (\cref{tab:clim_years}).
Over days 90--730, the zonal power spectra of T2m, T850, 850-hPa zonal wind (U850) and Z500 stay close to those of ERA5 up to zonal wavenumber 250, a wavelength of about 160~km at the equator (\cref{fig:series}i--l).
Hence, the stability gained with HADRIAN is not obtained by damping the atmosphere, as storm-track variance stays within 8.2\% of that of ERA5 in both years and small-scale spectral power is retained.

By contrast, SFNO and Aurora, which are stable without correction, lose T2m, T850 and U850 power towards higher wavenumbers, and their U850 power falls to less than half that of ERA5 at the highest wavenumbers (\cref{fig:series}i--k).
AIFS underestimates T850 and U850 power and has more than twice the T2m and Z500 power of ERA5 at wavenumbers above about 30 (\cref{fig:series}i--l).
These three models are deterministic, which favours smoother fields, so part of this contrast reflects the probabilistic design of FCN3 and not the correction.
%

% Keep the end of this section with Fig. 3 so the long-rollout figures fit with their text.
\enlargethispage{\baselineskip}
The agreement with ERA5 persists in the second rollout year, which lies largely after activation.
The RMSE of the time-mean T2m and Z500 fields of FCN3 + HADRIAN is then 0.784~K and 210.0~m$^2$\,s$^{-2}$, against 0.731~K and 210.4~m$^2$\,s$^{-2}$ in the first year, and the lowest among the compared models (\cref{tab:clim_years}).
For seasonal-mean fields in two windows of the first year, the FCN3 + HADRIAN ensemble mean has the lowest RMSE and the five-member ensemble has the lowest continuous ranked probability score (CRPS) among the compared models for all seven variables, although the ensemble is underdispersive for most of them (\cref{app:seasonal_fidelity}).
On average, individual members also have a lower RMSE than every comparison model, except for the 850-hPa meridional wind in the calendar-season window.
At lead times of 6--18 months, the number, intensity and near-surface wind structure of extratropical cyclones over the North Atlantic and Europe in FCN3 + HADRIAN also lie within the year-to-year range of ERA5 (\cref{app:cyclones}).

% -------------------------------------------------
% LONG ROLLOUTS
% -------------------------------------------------
\subsection{Corrected rollouts over 10 and 50 years}
\label{sec:longruns}
To test whether the correction keeps FCN3 stable beyond two years, we extend the rollouts to 10 and 50 years with the configuration of \cref{sec:failure} unchanged.
Five FCN3 + HADRIAN members are initialized at 00~UTC on 1 January 2017, run for 10 years and are then continued to 50 years (\cref{sec:method}).
\begin{figure}[!h]
  \centering
  \includegraphics[width=\textwidth]{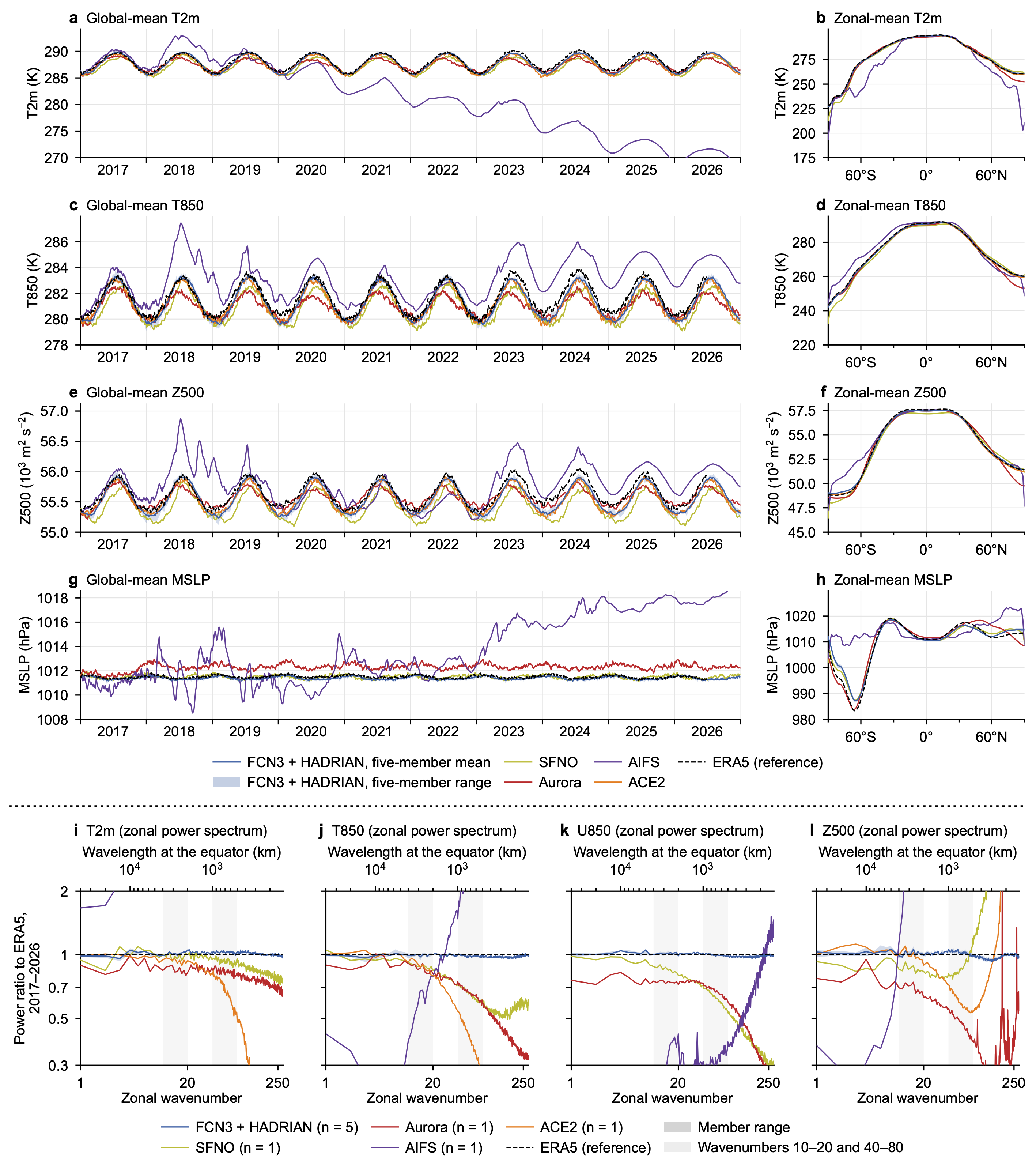}
  \caption{\textbf{FCN3 + HADRIAN retains the seasonal cycle and spectral power close to ERA5 over 10 years.} Rollouts initialized on 1 January 2017. (a,c,e,g)~Global means and (b,d,f,h)~time-mean zonal means of T2m, T850, Z500 and MSLP. (i,j,k,l)~Ratio of model to ERA5 zonal power spectra, averaged over 2017 to 2026. FCN3 + HADRIAN is shown as the five-member mean (median in i,j,k,l) and range, and the other models as one trajectory each. ACE2 is shown for T2m, T850 and Z500 only, and its $1^\circ$ grid limits its power at high wavenumbers.}
  \label{fig:long_2017}
\end{figure}
This start date lies after the training period of FCN3 and after the years 2000 to 2015 from which the thresholds of HADRIAN are derived (\cref{tab:training_periods}).
For context, we also run SFNO, Aurora and AIFS without correction for 10 years from the same date, and the climate emulator ACE2~\citep{wattmeyer2024ace} for 50 years.
No uncorrected controls of the five members have been run.

All five members complete the 10 years without blowing up.
Their global-mean T2m, T850, Z500 and MSLP follow the seasonal cycle of ERA5 throughout (\cref{fig:long_2017}a,c,e,g).
The time-mean zonal means also lie close to those of ERA5, except at the MSLP minimum of the Southern Hemisphere near $65^\circ$S, which is higher than in ERA5 (\cref{fig:long_2017}b,d,f,h).
Averaged in time, the zonal power spectra of T2m, T850, U850 and Z500 stay close to those of ERA5 up to zonal wavenumber 250 (\cref{fig:long_2017}i,j,k,l).
In 2023 to 2025, ERA5 is warmer than the corrected rollouts, which receive no information on the observed warming (\cref{sec:discussion}).

\begin{figure}[!htb]
  \centering
  \includegraphics[width=\textwidth]{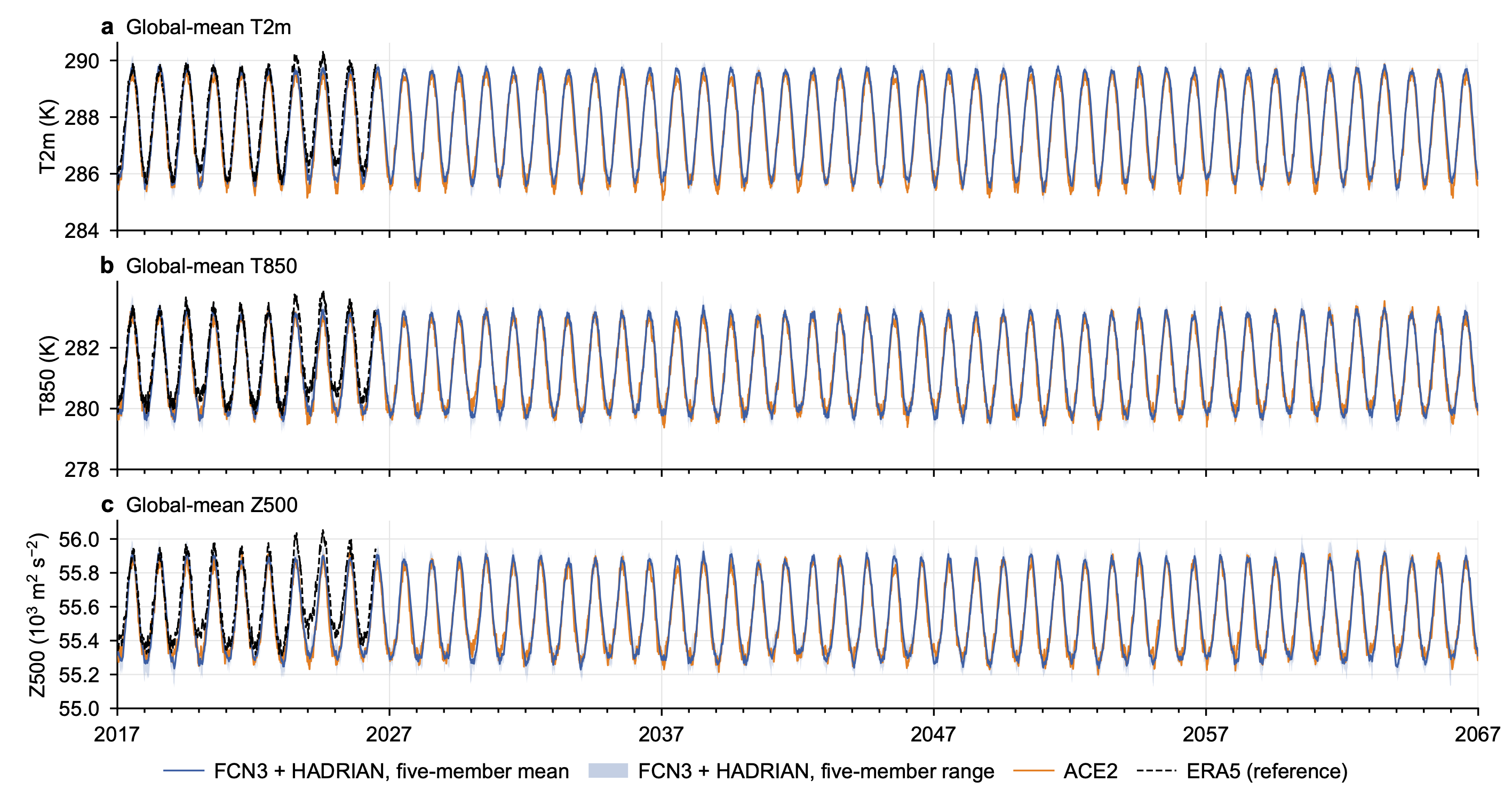}
  \caption{\textbf{Global means of FCN3 + HADRIAN over 50 years.} The five members of \cref{fig:long_2017}, continued to 50 years after a restart at the end of year 10. (a)~Global-mean T2m, (b)~T850 and (c)~Z500 for the FCN3 + HADRIAN five-member mean and range, ACE2 and ERA5, which ends in 2026.}
  \label{fig:long_50y}
\end{figure}

Continued to 50 years, the five members again do not blow up, and their global-mean T2m, T850 and Z500 show no visible drift (\cref{fig:long_50y}); the global-mean T2m and T850 of FCN3 + HADRIAN stay within about 0.5~K of those of ACE2.

Over 10 years, the comparison models behave as in the two-year rollouts or depart further from ERA5.
SFNO and Aurora remain finite and keep their seasonal cycle, with cold offsets in global-mean temperature, a global-mean MSLP that is about 1~hPa too high in Aurora and a loss of T2m, T850 and U850 power towards higher wavenumbers (\cref{fig:long_2017}a,c,g,i,j,k).
AIFS is too warm in the second year, as in \cref{sec:spectra}, but then cools at the surface: by the tenth year, its global-mean T2m is more than 10~K below ERA5, while its T850 remains warmer than ERA5 and its global-mean MSLP rises (\cref{fig:long_2017}a,c,g).
ACE2 stays close to ERA5 in the global and zonal means of T2m, T850 and Z500; it runs on a $1^\circ$ grid, which holds zonal wavenumbers up to 180, so its spectra are not comparable with those of the $0.25^\circ$ models at the highest wavenumbers.
Over 10 years, as over 2, FCN3 + HADRIAN therefore avoids blow-up without damping atmospheric variability.

% -------------------------------------------------
% ABLATION
% -------------------------------------------------
\FloatBarrier
\subsection{Ablation study}
\label{sec:ablations}
\begin{figure}[!htb]
  \centering
  \includegraphics[width=\textwidth]{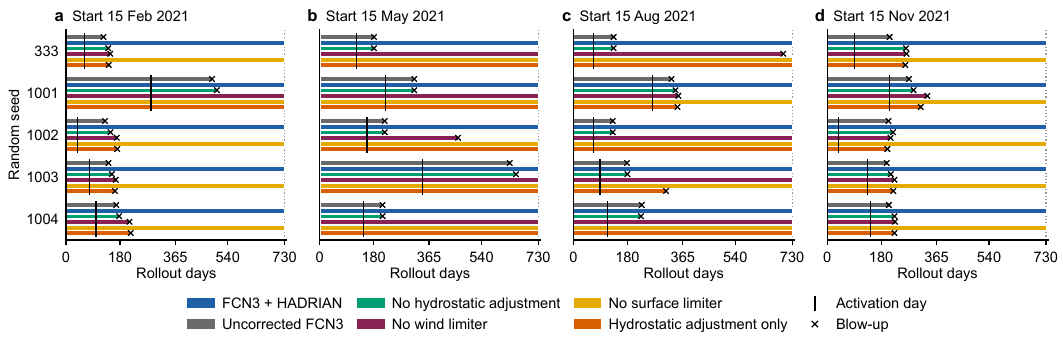}
  \caption{\textbf{Rollout length of each member in the ablation configurations.} Two-year rollouts from the four 2021 initialization dates, (a)~15 February, (b)~15 May, (c)~15 August and (d)~15 November, five seeds each (20 members per configuration). Bars show the last day of each member's rollout (rows: seeds) for uncorrected FCN3, FCN3 + HADRIAN, the configurations with one stage removed and the configuration with hydrostatic adjustment only; crosses mark blow-up. Black vertical markers show the activation day of the FCN3 + HADRIAN member with the same initialization date and random seed; before it, every configuration that keeps the activation criterion is identical to uncorrected FCN3. The number of stable members and the days to failure of each configuration are given in \cref{tab:ablation}.}
  \label{fig:ablation}
\end{figure}

To identify which of the three stages of HADRIAN are required for stability, we repeat the two-year 2021 rollouts of the same 20 members with one stage removed at a time, the correction being activated, as before, when $D$ first exceeds $\tau$ (\cref{app:ablation_details}).
Removing hydrostatic adjustment or the wind limiter leads to rollout failure, whereas removing the surface limiter does not (\cref{fig:ablation}).
Without hydrostatic adjustment, none of the 20 members completes two years, and without the wind limiter only eight members complete the two-year rollout.
All 20 members remain stable without the surface limiter (\cref{tab:ablation}).

Hydrostatic adjustment is a necessary condition for stability in these experiments.
Without it, all 20 members fail even though the other two stages continue to operate.
The excessive tropical 50-hPa westerly jet forms in all 20 members, against 12 of the 20 FCN3 + HADRIAN members, so that the wind limiter is switched on in every member and remains on until the end of the rollout.
Neither of these two stages, however, reduces the hydrostatic residual.
After activation, $D$ of the raw predictions remains above $\tau$ for the rest of the rollout in most members, as in uncorrected FCN3, whereas it exceeds $\tau$ on at most 6.8\% of the steps in the members that remain stable.

The members without hydrostatic adjustment fail in the same way as uncorrected FCN3, with 100-hPa cooling and an increase in column water vapour within weeks of activation (\cref{fig:app_abl_routes}a,e).
They also fail at nearly the same time, a median of 81 days after activation (range 58--313 days) against 66 days (58--293 days) for uncorrected FCN3.
Relative to the uncorrected control with the same initialization date and random seed, failure occurs a median of 12.5 days later.
The other two stages therefore cannot replace hydrostatic adjustment, which is the only stage that acts on the hydrostatic imbalance.
Hydrostatic adjustment is accordingly the stage around which the adaptations of HADRIAN to other AI weather models are built (\cref{sec:crossmodel}).

Hydrostatic adjustment is not, however, a sufficient condition.
The wind limiter is required in addition whenever the excessive tropical 50-hPa westerly jet forms.
FCN3 + HADRIAN switches the wind limiter on in 12 of the 20 members, and all 12 fail without it, 76--120 days after the step at which it would have been switched on.
In the other eight members it is never switched on, so the rollouts without it are identical to those of FCN3 + HADRIAN and do not test this stage.
The 12 members that fail do so in the same way as uncorrected FCN3 (\cref{fig:app_abl_routes}b,f), and their $D$ exceeds $\tau$ on a median of 94.2\% of the steps after activation, although hydrostatic adjustment remains active.
In these 12 members, hydrostatic adjustment and the surface limiter therefore do not keep the rollout stable without the wind limiter.
With hydrostatic adjustment only, nine members complete two years, seven of which also remain stable without the wind limiter (\cref{fig:ablation}; \cref{app:ablation_details}).

The surface limiter is not required for stability, but it limits a warming of Northern Hemisphere land.
In FCN3 + HADRIAN it changes the state on a median of 34\% of the steps after activation (member range 17--55\%).
Without it, Northern Hemisphere land T2m is 0.23~K and global-mean T2m 0.05~K warmer than in FCN3 + HADRIAN, global-mean T850 is 0.03~K warmer and global-mean TCWV $0.03\,\mathrm{kg\,m^{-2}}$ higher, and the RMSE of the time-mean T2m and Z500 fields rises by 0.05~K and $6.3\,\mathrm{m^2\,s^{-2}}$ (\cref{tab:ablation}; medians over the 20 members, days 1--730).
Tropical-mean T2m and the storm-track variance ratios are unchanged to two decimal places.

% -------------------------------------------------
% OTHER MODELS
% -------------------------------------------------
\FloatBarrier
\subsection{Physical corrections applied to other AI weather models}
\label{sec:crossmodel}
Hydrostatic imbalance also develops in five other AI weather models, namely GraphCast~\citep{lam2023learning}, GenCast~\citep{price2025probabilisti}, Aurora~\citep{bodnar2025foundation}, AIFS~\citep{moldovan2026aifs} and FuXi~\citep{chen2023fuxi}.
Without correction, $D$ crosses $\tau$ in all five and stays above it (\cref{app:crossmodel_hydrostatic}).
To each we apply a preliminary adaptation of HADRIAN built around hydrostatic adjustment, which, as for FCN3, acts on the predicted states without training or changing the model weights.
Each adaptation is matched to the variables the model predicts and to its imbalance characteristics.
Four of the five adaptations also include model-specific limits, some of which hold a variable within a day-of-year or climatological range, whereas that of FuXi consists of hydrostatic adjustment alone.
The adaptations are defined in \cref{app:crossmodel_setup} and summarized in \cref{tab:crossmodel_hosts}.
With these adaptations, $D$ stays close to the ERA5 range in GraphCast, Aurora and AIFS and, for about 20 months, in GenCast, but rises above $\tau$ again in FuXi (\cref{fig:crossmodel_residual}).
In a preliminary ablation from the same initialization date, $D$ remains above the ERA5 maximum throughout the second year without hydrostatic adjustment in GraphCast, GenCast and AIFS, whereas hydrostatic adjustment alone keeps it near the ERA5 range in GraphCast and AIFS; in Aurora, the humidity limit alone keeps $D$ near that range (\cref{app:crossmodel_ablation}).
The adaptations improve the stability and the seasonal cycle of GraphCast, GenCast and AIFS and the long-term temperature evolution of Aurora, which is stable without correction, but do not stabilize FuXi.

\begin{figure}[!htb]%[p]
  \centering
  \includegraphics[width=\textwidth,pagebox=cropbox]{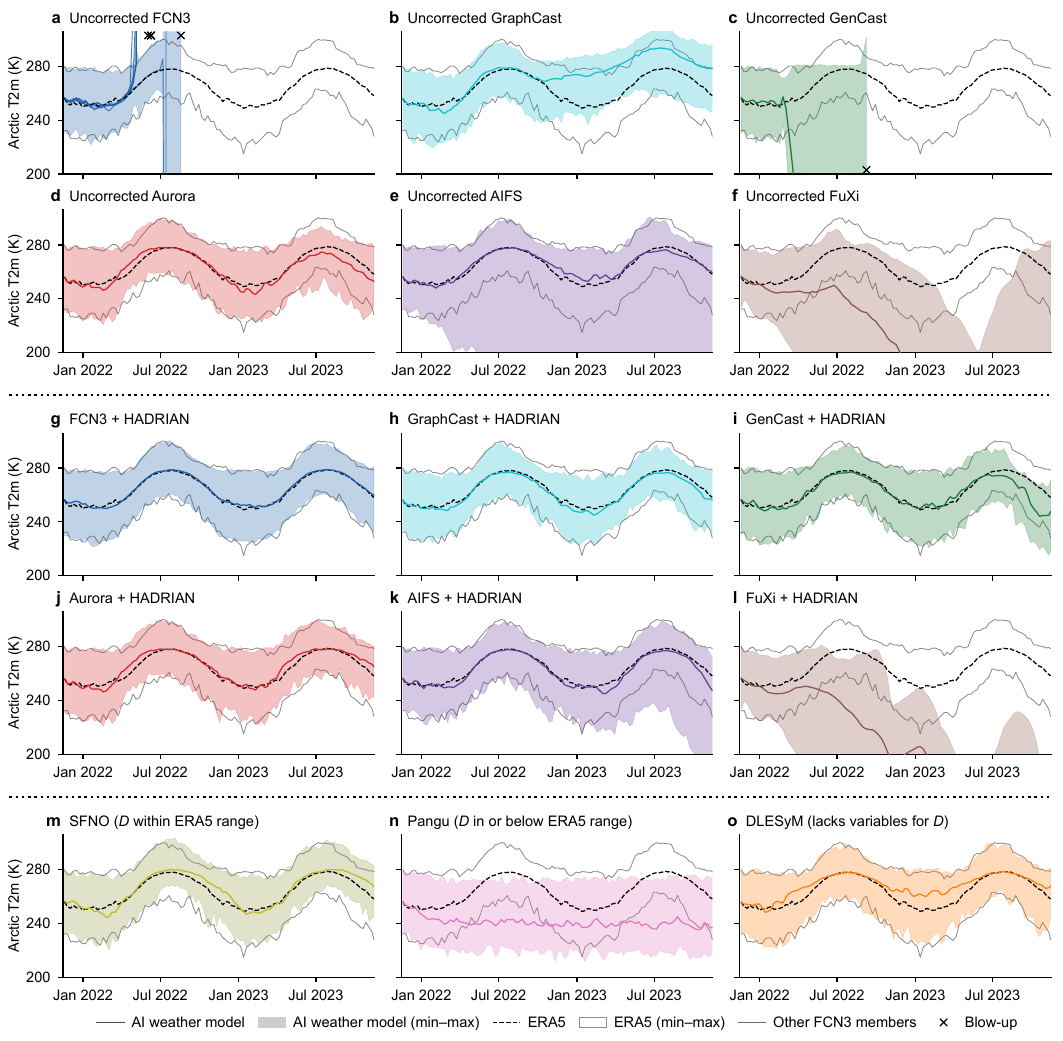}
  \caption{\textbf{Arctic T2m with and without HADRIAN.} Weekly means poleward of $66.5^\circ$N for two-year rollouts from 15 November 2021. (a--f)~Six models without HADRIAN. (g--l)~The same models with HADRIAN. (m--o)~SFNO, Pangu and DLESyM, run without HADRIAN. Solid curves show area-weighted regional means and dashed black curves ERA5; coloured bands (models) and grey lines (ERA5) mark the weekly-mean regional minimum and maximum. FCN3 + HADRIAN is shown as the five-member mean, and uncorrected FCN3 members individually. Crosses mark blow-up. Details are given in \cref{app:crossmodel_setup}.}
  \label{fig:crossmodel_arctic}
\end{figure}

For each of these five models, we compare one two-year rollout with HADRIAN and one without, both initialized on 15 November 2021 (\cref{fig:crossmodel_arctic}a--l; \cref{app:crossmodel_setup}).
Three further models, namely SFNO~\citep{bonev2023spherical}, Pangu~\citep{bi2023accurate} and DLESyM~\citep{cresswellclay2025deep}, are run without HADRIAN, because their residual either never reaches $\tau$ or cannot be computed (\cref{fig:crossmodel_arctic}m--o).

In GraphCast, HADRIAN removes the pronounced Arctic warm drift and keeps the T2m extremes within their physical range throughout the two-year rollout, whereas those of the uncorrected rollout first leave it on day 414, although the global mean of the corrected rollout remains colder than ERA5 (\cref{fig:crossmodel_global}b).
Uncorrected GenCast fails in its tenth month, whereas GenCast + HADRIAN follows the seasonal cycle for about 20 months before cooling sharply (\cref{app:crossmodel_gencast}).
In AIFS, HADRIAN delays the day on which extreme values first leave their physical ranges from day 38 to day 446.
It also suppresses the strong late Antarctic warming of the uncorrected rollout and moderates its later global warm bias (\cref{fig:crossmodel_global,fig:crossmodel_antarctic}).
The corrected Arctic mean, however, falls below ERA5 in the final months (\cref{fig:crossmodel_arctic}k).
In Aurora, which does not fail without correction, HADRIAN restores hydrostatic balance in the 50--100-hPa layer (\cref{app:crossmodel_hydrostatic}) and reduces the second-year Arctic and global cold biases.

FuXi is the exception, because its rollout failure does not originate from a single balance relation.
In the long rollouts of \citet{lehmann2026can}, FuXi amplifies small-scale energy and blows up, and its minimum Arctic temperature diverges rapidly from ERA5.
In our rollouts, its successive loss of hydrostatic and geostrophic balance is accompanied by surface and mass drift.
Restoring hydrostatic balance alone therefore cannot stabilize the model, and FuXi + HADRIAN departs from ERA5 earlier than uncorrected FuXi (\cref{fig:crossmodel_global}f; \cref{app:crossmodel_fuxi}).

In Pangu, rollout failure takes the form of a loss of seasonality and variability instead of a blow-up.
\citet{lehmann2026can} report the same behaviour and consider the absence of a time-encoding input the likely cause.
Pangu cools throughout the column, and its Arctic T2m loses most of its seasonal cycle (\cref{fig:crossmodel_arctic}n; \cref{app:crossmodel_reference}).
Because geopotential thickness and temperature decrease together, its residual stays at or below ERA5 levels and HADRIAN would never activate.
SFNO remains stable with its residual within the ERA5 range, but underestimates small-scale T2m, T850 and U850 power and synoptic-scale variability (\cref{fig:series}i--k; \cref{tab:survival}).
DLESyM does not predict the fields needed to compute the residual.

%%%%%%%%%%%%%%%%%%%%%%%%%%%%%%%%%%%%%%%%%
% DISCUSSION
%%%%%%%%%%%%%%%%%%%%%%%%%%%%%%%%%%%%%%%%%
\FloatBarrier
\section{Discussion}
\label{sec:discussion}
Our results show that, in FCN3, rollout failure is preceded by hydrostatic imbalance among its own predicted variables.
In the 20 members initialized in 2021, the hydrostatic residual exceeds its fixed threshold a median of about 66 days before failure, when the global-mean 100-hPa and 2-m temperatures are still within 0.2~K of ERA5 in the member median.
The imbalance develops first in the tropical 50--100-hPa layer, which lies outside the 850--500-hPa range in which deterministic AI weather models were found to be at least as close to hydrostatic balance as ERA5 over the first ten days~\citep{kasteleyn2026physmetrics}.
The ablation experiments indicate that this imbalance contributes to the failure, since all 20 members fail when hydrostatic adjustment is removed, although the other two stages remain active.
In addition, once the corrected states are fed back, the hydrostatic residual of the raw FCN3 predictions returns to the ERA5 range, which suggests that FCN3 departs from hydrostatic balance mainly when its own input is already unbalanced.

In HADRIAN, this residual determines the step from which the predicted states are corrected; FCN3 runs as released until then.
With the correction, all 40 two-year members remain stable, whereas none of their uncorrected controls completes the rollout.
This includes the 20 members initialized in 2023.
Without the wind limiter, 12 of the 20 members initialized in 2021 fail although hydrostatic adjustment remains active, whereas all members remain stable without the surface limiter, whose removal mainly warms Northern Hemisphere land by about 0.2~K.
All constants of HADRIAN for FCN3 are fixed, and its threshold, admissible ranges and target residual are statistics of the climatologies of 2000--2015.
The correction starts, and the two limiters act, only when the predicted state reaches values that almost never occurred in those years (\cref{tab:threshold_ranges}).
The correction requires no retraining and adds about 8\% to the cost of each step.

In the diagnostics that we examined, the corrected rollouts remain stable without a loss of atmospheric variability.
During the second rollout year, which lies largely after activation, the storm-track variance stays within 8.2\% of that of ERA5 in both hemispheres.
The zonal power spectra of the corrected rollouts also stay close to those of ERA5 up to zonal wavenumber 250.
Five further members initialized on 1 January 2017 run for 50 years without blowing up; over the years for which ERA5 is available, their global means, zonal means and zonal power spectra stay close to those of ERA5 (\cref{sec:longruns}).
These results indicate that, for a released model such as FCN3, stable long rollouts can be obtained at inference time, without the retraining, physical constraints or dedicated architectures used so far~\citep{wattmeyer2024ace,subramaniam2025imposing,bonev2023spherical}.

In the seven other models for which it can be computed, the hydrostatic residual exceeds the threshold in five and stays below it in SFNO and Pangu.
Adaptations of HADRIAN improve the stability and the seasonal cycle of GraphCast, GenCast and AIFS and reduce the cold biases of Aurora (\cref{fig:crossmodel_arctic}; \cref{fig:crossmodel_global,fig:crossmodel_antarctic}).
In these four models, the corrected rollouts keep $D$ close to the ERA5 range for most of the two years, whereas in FuXi, $D$ rises above $\tau$ again.
The adaptation to FuXi brings no improvement, since its failure also involves geostrophic balance and surface and mass drift.
The warm winter Arctic stratosphere of SFNO and the whole-column cooling of Pangu keep geopotential and temperature mutually consistent, so the residual does not rise (\cref{app:crossmodel_reference}).
HADRIAN is therefore limited to errors that involve a departure from hydrostatic balance.

The full evaluation concerns one model, with four initialization dates in each of two years, five members per date and overlapping verification windows.
In the corrected rollouts, the global-mean cold bias grows from the first to the second year.
The three stages are applied in sequence and do not enforce all balances jointly. 
The admissible ranges and the target residual are fixed statistics of 2000 to 2015, and FCN3 receives no sea-surface temperature, so the corrected rollouts represent a stationary climate and cannot follow a changing one; in the rollouts from 2017, they stay below the ERA5 global-mean temperatures of 2023 to 2025 (\cref{fig:long_2017}a,c). 
The origin of the imbalance in the tropical lower stratosphere and of the excessive tropical 50-hPa westerly jet remains to be explained.
The cross-model tests use one rollout per model and configuration, and the model-specific limits of the adaptations are preliminary (\cref{tab:crossmodel_hosts}).
A preliminary ablation from one initialization date shows that the stage that keeps $D$ close to the ERA5 range differs between these models (\cref{app:crossmodel_ablation}), and the mechanism that links hydrostatic balance to their stability and seasonal cycle remains to be explained.
GenCast + HADRIAN also cools sharply after about 20 months.
Forecast skill at subseasonal-to-seasonal lead times was not evaluated, since the seasonal-mean comparisons concern climatological fidelity.

Activation criteria that combine several physical relations, such as geostrophic balance and the global mass and energy budgets, with diagnostics of mean-state drift could identify rollout failures that leave hydrostatic balance intact.
With more members and initialization dates, the preliminary ablation across models could be extended to test whether $D$ follows the stability and seasonal cycle of a rollout whichever stage of the correction lowers it; if it does, $D$ could guide the design of corrections for other AI weather models.
For models that accept prescribed ocean and sea-ice boundary conditions, corrected rollouts could be tested against observed climate variability and forced change.
Because HADRIAN acts only on predicted states, it can be combined with constraints imposed during training~\citep{beucler2021enforcing,subramaniam2025imposing}, which would reduce how often imbalance arises.
This study indicates that the physical consistency of AI weather models can be monitored during long rollouts and that, in FCN3, hydrostatic balance can be restored once it is lost.
We regard this as a step towards using the same models seamlessly from weather to climate timescales~\citep{palmer2008seamless}.

%%%%%%%%%%%%%%%%%%%%%%%%%%%%%%%%%%%%%%%%%
% METHODS
%%%%%%%%%%%%%%%%%%%%%%%%%%%%%%%%%%%%%%%%%
\FloatBarrier
\section{Methods}
\refstepcounter{methods}\label{sec:method}
\subsection{Overview of HADRIAN}
\label{sec:hadrian_overview}
HADRIAN is a training-free physical corrector whose central stage is hydrostatic adjustment.
It addresses the hydrostatic imbalance that precedes rollout failure in uncorrected FCN3: in every member, the hydrostatic residual of the predicted states exceeds a fixed threshold before failure, and the imbalance is largest in the tropical 50--100-hPa layer (\cref{sec:failure}).
This residual is the difference between the predicted geopotential thickness of a layer and the thickness implied by the predicted temperature and humidity (\cref{fig:architecture}b).
HADRIAN uses it in two ways.
The global root-mean-square residual of each raw prediction determines the step at which correction starts (\cref{fig:architecture}a); from that step on, hydrostatic adjustment changes geopotential so that the residual of each layer moves towards its target value (\cref{fig:architecture}b).

HADRIAN acts between each FCN3 prediction and its use as the next input (\cref{fig:architecture}c). Until the residual first exceeds the threshold, predictions are passed on unchanged, so forecasts up to that step are those of the released model; from the activation step onwards, every predicted state is corrected before it is fed back.

Once activated, three stages correct each predicted state at the same valid time, in the order stratospheric wind limiter ($\mathcal W$), surface limiter ($\mathcal S$) and hydrostatic adjustment ($\mathcal H$).
The wind limiter and the surface limiter change the state only when a zonal mean of their variable lies outside its admissible range and otherwise return the state unchanged; hydrostatic adjustment moves every column towards its target residual at each step after activation.
Each stage changes a different variable: the wind limiter the 50-hPa zonal wind, the surface limiter the 2-m temperature and hydrostatic adjustment the geopotential at 50--300 hPa.
The temperature, humidity and geopotential on which hydrostatic adjustment acts are therefore those of the raw prediction.

The other two stages address errors that hydrostatic adjustment cannot remove.
The wind limiter limits the excessive tropical 50-hPa westerly jet that precedes rapid 100-hPa cooling in 11 of the 20 uncorrected members (\cref{sec:failure}). The surface limiter corrects only abnormal values of the zonal-mean 2-m temperature, those outside its admissible range (\cref{fig:architecture}c).

At each step, the corrector reads only the prediction, its valid time and a fixed set of constants (\cref{tab:parameters}).
The threshold, the admissible ranges of the two limiters and the target residual of hydrostatic adjustment are statistics of the climatologies of 2000--2015.
The correction is activated, and the two limiters and the special case of hydrostatic adjustment act, only when the predicted state reaches values that almost never occurred in those years (\cref{tab:threshold_ranges}).
The corrector reads no reanalysis or observations during the rollout and leaves the model weights, input noise and six-hour step unchanged. On one NVIDIA H200 SXM GPU, a six-hour step takes 0.969~s for uncorrected FCN3 and 1.049~s for FCN3 + HADRIAN with the correction applied, so the correction adds 80~ms (8.2\%) per step.

Hydrostatic adjustment is also the central stage of preliminary adaptations of HADRIAN to GraphCast, GenCast, Aurora, AIFS and FuXi (\cref{sec:crossmodel}). Their activation criteria and other components are adapted to the variables each model predicts and to its imbalance characteristics. The approach requires predicted geopotential, temperature and specific humidity on pressure levels, from which the residual of \cref{eq:hydrostatic_residual} is computed, and it can repair only rollout failures that begin with hydrostatic imbalance. These adaptations are preliminary; their setup and outcomes are given in \cref{app:crossmodel_temperature} (\cref{app:crossmodel_setup}, \cref{tab:crossmodel_hosts}). The sections below describe the FCN3 configuration.

\subsection{Hydrostatic residual and activation}
Activation is decided by comparing a global measure of the hydrostatic imbalance of each raw FCN3 prediction with a fixed threshold. Let $y_n=F_\theta(x_n;\xi_n)$ denote the six-hour prediction of the frozen FCN3 model~\citep{bonev2025fourcastnet} from the state $x_n$ with stochastic input $\xi_n$. The state comprises 72 channels on the $0.25^\circ$ grid: zonal and meridional wind, geopotential, temperature and specific humidity on 13 pressure levels, and seven single-level variables, among them 2-m temperature, mean sea-level pressure and total column water vapour.

The hydrostatic residual $r_{k,i}$ and its global RMS $D$ are defined in \cref{eq:hydrostatic_residual} and \cref{eq:residual_statistic}. \Cref{eq:hydrostatic_residual} is the discrete residual of the hypsometric equation, which relates the geopotential thickness of a layer to its mean virtual temperature; the layer-mean virtual temperature is approximated by the mean of its values at the two bounding levels, and $R_v/R_d-1\simeq0.608$. In \cref{eq:residual_statistic}, $N$ counts all points of the $0.25^\circ$ grid.

At each step, HADRIAN computes $D(y_n)$ on the native grid and compares it with the fixed threshold $\tau=157.357\,\mathrm{m^2\,s^{-2}}$. Because the vertical discretization over 13 levels and analysis noise leave a residual even in reanalysis, $D$ of the daily ERA5 states over 2000--2015 ranges from 83.2 to $144.7\,\mathrm{m^2\,s^{-2}}$ (\cref{app:precursor}). The threshold lies above this range, at 1.09 times its largest value; its calibration is given in \cref{app:parameters}.

The first step with $D(y_n)>\tau$ is the activation step, and the correction then stays on for the rest of the rollout; a non-finite $D(y_n)$ is treated as exceeding $\tau$.
Starting from $a_{-1}=0$, the activation state and the next input are
\begin{equation}
a_n=a_{n-1}\lor\mathbf{1}\{D(y_n)>\tau\},\qquad
 x_{n+1}=\begin{cases}
 y_n, & a_n=0,\\
 (\mathcal H_n\circ\mathcal S_n\circ\mathcal W_n)(y_n), & a_n=1,
 \end{cases}
\label{eq:hadrian_update}
\end{equation}
where the subscript $n$ marks the dependence of each stage on the valid time of $y_n$ and, for $\mathcal W_n$, on a switching state carried from step to step. Before activation, the prediction is passed on unmodified, so each FCN3 + HADRIAN member is bitwise identical to its uncorrected control until its activation step.

\subsection{Hydrostatic adjustment (\texorpdfstring{$\mathcal H$}{H})}
Hydrostatic adjustment moves the geopotential $\Phi$ of the upper part of each predicted column towards a target profile $\widetilde\Phi$. The hydrostatic residual thereby moves from $r_k(\Phi)$ towards the target residual $r_k(\widetilde\Phi)$. \Cref{fig:architecture}b shows a layer before the adjustment and after the adjustment is complete, when the geopotential equals $\widetilde\Phi$. A single step makes at most a tenth of this change (\cref{eq:hydrostatic_adjustment}). The target residual is the non-zero residual that reanalyses such as ERA5 have on the same pressure levels. The target profile is computed from the temperature and humidity of the same column at the same step. The stage leaves $T$ and $q$ unchanged and adjusts geopotential on the six levels from 50 to 300 hPa ($k=1,\dots,6$), which contain the tropical 50--100-hPa layer where the imbalance before threshold crossing is largest (\cref{fig:precursor}a). Geopotential at 400--1000 hPa is left unchanged. Because $T$ and $q$ are unchanged, the residual is written below as a function of geopotential only.

The residual is not reduced to zero because ERA5 itself has a non-zero residual: $D$ of its daily states over 2000--2015 lies between 83.2 and $144.7\,\mathrm{m^2\,s^{-2}}$ (grey band in \cref{fig:imbalance_hadrian}a). An independent evaluation of ERA5 with the same two-level approximation also finds a non-zero residual~\citep{subramaniam2025imposing}. Because \cref{eq:hydrostatic_residual} approximates the layer-mean virtual temperature by the mean of its values at the two bounding levels, a hydrostatically balanced atmosphere sampled on the 13 levels has a non-zero residual wherever temperature does not vary linearly with $\ln p$. This residual is largest in the layers adjoining the tropical tropopause and follows the annual cycle of the tropopause, so a relaxation towards zero would change the geopotential of even a balanced column, always in the same direction. The target residual $r_{k,i}(\widetilde\Phi)$ is therefore the residual that a balanced atmosphere has in the same layer at the location and time of year of the column. Its values are estimated from the 2000--2015 climatology (\cref{app:parameters}).

The target residual enters the target profile only as a fixed offset in the thickness of each layer. The thickness terms $b_{k,i}$ of \cref{eq:hydrostatic_residual} are computed at every step from the predicted $T$ and $q$ and are not altered by the adjustment, so the target thickness of each layer is $b_{k,i}-r_{k,i}(\widetilde\Phi)$. Hydrostatic adjustment thus reduces the difference between the predicted geopotential thickness of each layer and the thickness implied by the temperature and humidity that the model itself predicts (\cref{fig:architecture}b). Profiles with this residual differ from one another only by a constant added at all six levels. Within each column $i$, we construct the target profile by integrating the hydrostatic relation downwards from 50~hPa and fix the constant by keeping the geopotential at 300~hPa unchanged:
\begin{equation}
H_{1,i}=0,\qquad H_{k+1,i}=H_{k,i}-b_{k,i}+r_{k,i}(\widetilde\Phi)\ \ (k<6),\qquad
 \widetilde\Phi_{k,i}=H_{k,i}+\Phi_{6,i}-H_{6,i}.
\label{eq:hydrostatic_target}
\end{equation}
By construction, $\widetilde\Phi$ has the target residual in the five layers between 50 and 300~hPa. Rather than replacing $\Phi$ by $\widetilde\Phi$ in one step, we move the geopotential on the six levels towards $\widetilde\Phi$ by a fixed fraction per step, with a cap on the change in each column:
\begin{equation}
\begin{aligned}
\Phi'_{k,i}&=\Phi_{k,i}+\beta s_i\bigl(\widetilde\Phi_{k,i}-\Phi_{k,i}\bigr),\\
s_i&=\min\!\left(1,\frac{c}{\max_{k\le6}|\widetilde\Phi_{k,i}-\Phi_{k,i}|}\right),\qquad
\beta=0.1,\quad c=250\,\mathrm{m^2\,s^{-2}},
\end{aligned}
\label{eq:hydrostatic_adjustment}
\end{equation}
with $s_i=1$ if the column is already at its target profile. The relaxation factor $\beta$ sets the fraction of the departure from the target profile removed per step, and $c$ the largest departure within a column that is corrected at that rate. Because $T$ and $q$ are unchanged,
\begin{equation}
\begin{aligned}
r_{k,i}(\Phi')-r_{k,i}(\widetilde\Phi)&=(1-\beta s_i)\bigl[r_{k,i}(\Phi)-r_{k,i}(\widetilde\Phi)\bigr]\qquad(k=1,\dots,5),\\
\lvert\Phi'_{k,i}-\Phi_{k,i}\rvert&\le\beta c=25\,\mathrm{m^2\,s^{-2}}.
\end{aligned}
\label{eq:hydrostatic_bound}
\end{equation}
The first relation holds because the residual is affine in geopotential. The geopotential at 300~hPa is unchanged, so the residuals of the 300--400-hPa layer and of all layers below are also unchanged. Both relations apply to a single step; the residual over the full rollouts is shown in \cref{fig:imbalance_hadrian}c.

\subsection{Stratospheric wind limiter and surface limiter}\label{sec:other_stages}
\subsubsection{Stratospheric wind limiter (\texorpdfstring{$\mathcal W$}{W})}
The wind limiter reduces the part of the tropical zonal-mean 50-hPa zonal wind that lies outside an admissible range (\cref{fig:architecture}c) while an excessive westerly jet persists (\cref{sec:failure}), and changes no other variable. At latitude $\varphi$ and day of year $d$, let $\bar u_{50}(\varphi)$ be the zonal-mean wind and $[\ell_d,h_d]$ the admissible range; the range and the rule that switches the wind limiter on and off are given in \cref{app:parameters}. While the wind limiter is on, the 50-hPa zonal wind at longitude $\lambda$ is updated as
\begin{equation}
e_\varphi=\operatorname{clip}(\bar u_{50}(\varphi);\ell_d,h_d)-\bar u_{50}(\varphi),\qquad
 u'_{50}(\varphi,\lambda)=u_{50}(\varphi,\lambda)+w(\varphi)\operatorname{clip}(0.1e_\varphi;-2,2),
\label{eq:wind_update}
\end{equation}
where $w=1$ within $20^\circ$ of the equator and tapers with a cosine to zero at $30^\circ$. Each step removes 10\% of the excess of the zonal-mean wind over the range, at most $2\,\mathrm{m\,s^{-1}}$, and adds the same increment at all longitudes of a latitude row, so the deviations from the zonal mean along each row are unchanged.

\subsubsection{Surface limiter (\texorpdfstring{$\mathcal S$}{S})}
The surface limiter reduces the part of the zonal-mean 2-m temperature that lies outside an admissible range (\cref{fig:architecture}c), and changes no other variable. The zonal mean is taken along each latitude row, separately over sea and over land points, and the range depends on the latitude, surface type, season and time of day. For the sea (or land) points of the row at latitude $\varphi$, let the row mean $\overline T_\varphi$ be their mean 2-m temperature. Its admissible range is $\mu_\varphi\pm4.5\sigma_\varphi$, where $\mu_\varphi$ is the climatological mean seasonal cycle of $\overline T_\varphi$ at the UTC hour of the valid time and $\sigma_\varphi$ the root-mean-square deviation from it, both smooth functions of the day of year (\cref{app:parameters}). Because land is unevenly distributed in longitude, the row mean over land has a diurnal cycle in UTC; separate values of $\mu_\varphi$ and $\sigma_\varphi$ for each UTC hour keep this cycle within the range. The increment added to every sea (or land) point of the row is
\begin{equation}
e_\varphi=\operatorname{clip}\bigl(\overline T_\varphi;\ \mu_\varphi-4.5\sigma_\varphi,\ \mu_\varphi+4.5\sigma_\varphi\bigr)-\overline T_\varphi,\qquad
 \delta_\varphi=\frac15\sum_{j=-2}^{2}\operatorname{clip}\bigl(0.5\,e_{\varphi_j};-1,1\bigr),
\label{eq:surface_update}
\end{equation}
in kelvin, where $\varphi_j$ is the latitude of the row $j$ rows from that at $\varphi$. A row mean outside its range contributes to the increment half of its excess over the range, at most 1~K, and the average over five adjacent rows spreads the increment to neighbouring rows. Along each row, the deviations of the sea points and of the land points from their row means are unchanged, and the state is returned unchanged when every row mean lies within its range.

\subsection{Experimental design}
All FCN3 experiments start from ERA5~\citep{hersbach2020era5} on the $0.25^\circ$ grid and advance in six-hour steps, each prediction serving as the next input, without data assimilation. 
Every FCN3 + HADRIAN member of the two-year experiments has an uncorrected control with the same initial state and random seed, which is identical to it until activation, so each pair isolates the effect of the correction.

The two-year experiments start at 00~UTC on 15 February, 15 May, 15 August and 15 November of 2021 and of 2023, with five random seeds per date, and run for 730 days; the same set of five seeds is used for every initialization date, and the 2023 rollouts use the same fixed configuration. The periods of the training, validation and test data of FCN3 and of the comparison models are listed in \cref{tab:training_periods}.

For comparison, three released AI weather models that have been evaluated in long rollouts~\citep{lehmann2026can} are run without correction: SFNO~\citep{bonev2023spherical}, pretrained Aurora~\citep{bodnar2025foundation} and AIFS Single 1.1~\citep{moldovan2026aifs}, one deterministic trajectory from each 2021 initialization date (\cref{app:evaluation}).

The long rollouts start at 00~UTC on 1 January 2017, with five random seeds, and use the same fixed configuration as the two-year experiments.
No uncorrected controls were run for these members.
They run for 10 years and are then continued to 50 years from a restart at the end of year 10.
The comparison models start from the same date. 
SFNO, pretrained Aurora and AIFS Single 1.1 are run without correction for 10 years, one trajectory each. 
ACE2~\citep{wattmeyer2024ace}, a climate emulator on a $1^\circ$ grid with eight layers that needs prescribed boundary conditions, is run for 50 years.

\begin{table}[!htb]
\centering
\caption{\textbf{Periods of the training, validation and test data of FCN3 and the comparison models.} Years as stated in the papers that describe the models~\citep{bonev2025fourcastnet, bonev2023spherical, bodnar2025foundation, moldovan2026aifs, wattmeyer2024ace}; the data are ERA5 unless another dataset is named. IFS, Integrated Forecasting System of the European Centre for Medium-Range Weather Forecasts; HRES, high-resolution configuration of the IFS; GFS, Global Forecast System.}
\label{tab:training_periods}
\footnotesize
\renewcommand{\arraystretch}{1.15}
\begin{tabularx}{\textwidth}{@{}>{\raggedright\arraybackslash}p{0.16\textwidth}>{\raggedright\arraybackslash}X>{\raggedright\arraybackslash}p{0.13\textwidth}>{\raggedright\arraybackslash}p{0.22\textwidth}@{}}
\toprule
Model & Training & Validation & Test \\
\midrule
FCN3 & 1980--2016, fine-tuned on 2012--2016 & 2018--2021 & 2017 \\
SFNO & 1979--2015 & 2016--2017 & 2018 \\
Pretrained Aurora & 1979--2020, with other datasets & IFS HRES 2020 & 2021; IFS HRES 2022 \\
AIFS Single 1.1 & 1979--2022, fine-tuned on IFS operational analyses of 2016--2022 & not stated & 2023, against IFS operational analyses \\ ACE2 & 1940 to 1995, 2011 to 2019 and 2021 to 2022 & 1996 to 2000 & 2001 to 2010; 2020 \\
\bottomrule
\end{tabularx}
\par\smallskip
\begin{minipage}{\textwidth}\footnotesize
FCN3 calls 2017 its test data and 2018--2021 its out-of-sample validation data, which is used only for validation and on which its reported scores (2020) are computed; the card of the released FCN3 checkpoint gives 1980--2015 for training, 2016--2017 for testing and 2018--2019 for evaluation. The SFNO years are those of the SFNO paper; the documentation of the released checkpoint states no period. Pretrained Aurora was also trained on 1-day IFS HRES forecasts of 2016--2020, 1-day GFS forecasts and GFS analyses of 2015--2020 and the CMCC-CM2-VHR4 and ECMWF-IFS-HR climate simulations of 1950--2014.
\end{minipage}
\end{table}

The ablation experiments, which repeat the two-year 2021 rollouts with one stage removed at a time, are described in \cref{app:ablation_details}.

\subsection{Stability criteria}
A rollout fails when it blows up, drifts or loses its seasonality and variability \citep{lehmann2026can}. Blow-up ends with a numerical error at the first non-finite state; failure times, and the crosses in the figures, refer to this step. Drift and loss of seasonality leave the rollout finite and are not assigned a failure time; a rollout that drifts fails the physical plausibility checks below.

A member is stable if it completes the target duration with finite states and passes seven physical plausibility checks, none of which uses ERA5: global-mean T2m within 283--292~K, TCWV within $15$--$40\,\mathrm{kg\,m^{-2}}$ and MSLP within 1000--1025~hPa; annual drift of global-mean MSLP within $\pm500$ Pa and of a total-energy proxy, the dry static and kinetic energy on the 13 levels, within $\pm5\,\mathrm{W\,m^{-2}}$; tropical-mean ($20^\circ$S--$20^\circ$N) T2m no higher than the extratropical mean of either hemisphere at most 1\% of the time; and a phase difference of $182.5\pm45$ days between the annual cycles of the two extratropical means. Aurora and AIFS do not predict TCWV and are therefore assessed on the other six checks; \cref{tab:plausibility} gives the outcome of each check. Stable fractions and quality metrics include all rollouts, among them the AIFS trajectories that remain finite but fail the checks; quality metrics of uncorrected FCN3 are reported as n/a because none of its members completes the rollout.

\subsection{Evaluation metrics}
ERA5 is sampled at the same valid times as the models, and two-year scores cover days 1--730, excluding the initial state, unless stated otherwise. T850 and U850 denote the 850-hPa temperature and zonal wind, and MSLP mean sea-level pressure.

The bias of a trajectory is the cosine-latitude-weighted global mean of its time-mean departure from ERA5; table entries are means of these per-trajectory biases. The Z500 RMSE is the RMSE of the time-mean field, computed from the five-member FCN3 + HADRIAN ensemble-mean field at each initialization date, or from the single trajectory of a comparison model, and averaged over the four dates.

Synoptic-scale variability is measured by the storm-track variance ratio: the time mean is removed from six-hourly Z500, periods of 2--6 days are retained with a rectangular FFT filter, and the cosine-latitude-weighted variance poleward of $30^\circ$ in each hemisphere is divided by that of ERA5 over the same days. The same definition is applied to every window for which a storm-track variance ratio is reported (\cref{tab:survival,tab:survival_full,tab:ablation,tab:seasonal_late_diagnostics,tab:clim_years}). Zonal power spectra are computed after removing the zonal mean and weighted by cosine latitude; model and ERA5 spectra are averaged in time within each trajectory before their ratio is taken.

Seasonal-mean fields are scored in two windows per initialization by the RMSE of the FCN3 + HADRIAN ensemble-mean field and by the continuous ranked probability score (CRPS) of the five members, both against ERA5 for seven variables (\cref{app:seasonal_fidelity}). The grids, samples and aggregation of the other scores are given in \cref{app:evaluation}.

\label{maintext:end}

\subsection*{AI use statement}
Generative AI tools assisted with research planning, interpretation of experimental results, language editing and proofreading of the draft, literature discovery and organization, and figure-code preparation. 
The authors checked all numerical claims against the experimental records and all cited sources against the original publications.

The authors retain responsibility for the scientific claims, analyses and final manuscript.

\subsection*{Data availability}
ERA5~\citep{hersbach2020era5} is available from the Copernicus Climate Change Service Climate Data Store (\url{https://cds.climate.copernicus.eu}).
All AI weather models were run from the pretrained weights released by their developers, cited in the text; FCN3, SFNO and Aurora were run through Earth2Studio (\url{https://github.com/NVIDIA/earth2studio}).
The rollout outputs and the diagnostics behind the tables and figures will be released publicly upon publication. Until then, they are available from the corresponding author upon reasonable request.

\subsection*{Code availability}
The HADRIAN package for FCN3, comprising the activation criterion, the three correction stages, their parameter files and tests, will be released publicly upon publication. Until then, it is available from the corresponding author upon reasonable request.

\subsection*{Acknowledgments}
We thank ECMWF and the Copernicus Climate Change Service for ERA5, and the developers of the AI weather models evaluated here for releasing their pretrained weights.

\subsection*{Competing interests}
The authors declare no competing interests.

% Bibliography
\setlength{\bibsep}{0.9em}
\bibliography{references}

\appendix
\setcounter{table}{0}
\setcounter{figure}{0}
\renewcommand{\thetable}{S\arabic{table}}
\renewcommand{\thefigure}{S\arabic{figure}}
\renewcommand{\theHtable}{S\arabic{table}}
\renewcommand{\theHfigure}{S\arabic{figure}}

\clearpage
\section*{Appendix}
\crefalias{section}{appendix}
% Appendices A--B: correction parameters and evaluation.
\section{FCN3 Correction Parameters}\label{app:parameters}

\begin{table}[!htb]
\centering
\caption{\textbf{Constants of the activation criterion and the three correction stages of HADRIAN for FCN3.} The update rules are given in \cref{sec:method}.}
\label{tab:parameters}
\footnotesize
\renewcommand{\arraystretch}{1.15}
\begin{tabularx}{\textwidth}{@{}>{\raggedright\arraybackslash}p{0.22\textwidth}>{\raggedright\arraybackslash}X@{}}
\toprule
Stage & Constants \\
\midrule
Activation & $\tau=157.357\,\mathrm{m^2\,s^{-2}}$, the 99th percentile of the calibration below; the 95th and 99.9th percentiles are 144.9 and $164.3\,\mathrm{m^2\,s^{-2}}$ \\
Wind limiter $\mathcal W$ & after activation, switched on after 8 consecutive steps (two days) with $U_n\ge80\,\mathrm{m\,s^{-1}}$ and off after 40 consecutive steps (ten days) with $U_n<60\,\mathrm{m\,s^{-1}}$; 10\% of the excess per step, at most $2\,\mathrm{m\,s^{-1}}$; full weight within $20^\circ$ of the equator, zero beyond $30^\circ$; admissible range $[\ell_d,h_d]$ for each day of the year, with $h_d\le21.6\,\mathrm{m\,s^{-1}}$ \\
Surface limiter $\mathcal S$ & admissible range $\mu_\varphi\pm4.5\sigma_\varphi$ for each latitude row, surface type (sea, land), UTC hour and day of year; increment $\delta_\varphi$ of 50\% of the excess, at most 1~K, averaged over five adjacent rows \\
Hydrostatic adjustment $\mathcal H$ & 50--300~hPa (six levels), $\beta=0.1$, $c=250\,\mathrm{m^2\,s^{-2}}$; target residual $r_{k,i}(\widetilde\Phi)$ equal to the climatological mean of the residual; zero in the special case below \\
\bottomrule
\end{tabularx}
\end{table}

The threshold $\tau$ is the 99th percentile of the cosine-latitude-weighted RMS residual of 11,520 six-hourly ERA5 states from 2000--2015 (192 fifteen-day windows starting on the first of each month), each estimated from 8,192 area-stratified columns of the native grid; geopotential, temperature and specific humidity are all taken from ERA5. In the switching rule of the wind limiter, $U_n$ is the maximum of the zonal-mean 50-hPa zonal wind within $20^\circ$ of the equator. The admissible range $[\ell_d,h_d]$ of the wind limiter is derived from the daily 00~UTC ERA5 states of 2000--2015. For each state, the maximum and minimum of the zonal-mean 50-hPa zonal wind over the latitudes within $20^\circ$ of the equator are taken; $h_d$ and $\ell_d$ are the largest maximum and the smallest minimum among the 16 years on each day of the year, smoothed with a 15-day running mean and widened by $5\,\mathrm{m\,s^{-1}}$ at each end. Over the year, $h_d$ lies between 17.2 and $21.6\,\mathrm{m\,s^{-1}}$ and $\ell_d$ between $-33.1$ and $-21.9\,\mathrm{m\,s^{-1}}$. The centre $\mu_\varphi$ and width $\sigma_\varphi$ of the surface-limiter range are given for each UTC hour from the 2000--2015 climatology of the row-mean 2-m temperature and are each represented by an annual mean and annual and semi-annual harmonics. The target residual $r_{k,i}(\widetilde\Phi)$ of hydrostatic adjustment is the climatological mean of the residual of \cref{eq:hydrostatic_residual}, which represents the residual that a balanced atmosphere has under the two-level approximation of that equation. It consists of the calendar-month means of this residual in the five layers between 50 and 300~hPa from the 2000--2015 climatology, interpolated linearly in time to the valid time.

\paragraph{Special case of hydrostatic adjustment.} Hydrostatic adjustment has a special case for states in which the tropical 50--100-hPa residual is far above its climatological mean (\cref{app:precursor}). In these states, the target residual is zero instead of the climatological mean. With a target of zero, the adjustment removes more of the residual over the following steps. The change made at a single step remains limited: \cref{eq:hydrostatic_adjustment} and \cref{eq:hydrostatic_bound} apply unchanged, with $\beta=0.1$ and $c=250\,\mathrm{m^2\,s^{-2}}$, and the residual of the 300--400-hPa layer changes by at most $25\,\mathrm{m^2\,s^{-2}}$ per step. Let $A_n$ be the cosine-latitude-weighted mean over $20^\circ$S--$20^\circ$N of the departure of $r_1(y_n)$ from its climatological mean. The special case begins after eight consecutive steps with $A_n>350\,\mathrm{m^2\,s^{-2}}$ and ends after eight consecutive steps with $A_n<0$, with the steps counted from activation. At activation and at every step outside the special case, the target residual is the climatological mean. In the special case, $H_{k,i}$ follows \cref{eq:hydrostatic_target} with $r_{k,i}(\widetilde\Phi)=0$. The constant is chosen to keep the mean geopotential of the six adjusted levels unchanged, instead of the geopotential at 300~hPa: $\widetilde\Phi_{k,i}=H_{k,i}+\tfrac16\sum_{\ell=1}^{6}(\Phi_{\ell,i}-H_{\ell,i})$. This choice gives the hydrostatic profile closest to the predicted one in the least-squares sense.

\Cref{tab:threshold_ranges} compares the thresholds and admissible ranges of HADRIAN with the values that occurred in 2000--2015.

\begin{table}[!htb]
\centering
\caption{\textbf{Thresholds and admissible ranges of HADRIAN compared with the 2000--2015 climatologies.} The climatologies are computed from ERA5: all 23,376 six-hourly states from 00~UTC on 1 January 2000 to 18~UTC on 31 December 2015, except for $D$, which uses one state per day. $U_n$ and $A_n$ are defined above.}
\label{tab:threshold_ranges}
\footnotesize
\renewcommand{\arraystretch}{1.15}
\begin{tabularx}{\textwidth}{@{}>{\raggedright\arraybackslash}p{0.30\textwidth}>{\raggedright\arraybackslash}p{0.22\textwidth}>{\raggedright\arraybackslash}X@{}}
\toprule
Constant & Value & Climatology, 2000--2015 \\
\midrule
Activation threshold & $\tau=157.357\,\mathrm{m^2\,s^{-2}}$ & $D$ between 83.2 and $144.7\,\mathrm{m^2\,s^{-2}}$ \\
Switch-on threshold of the wind limiter & $U_n\ge80\,\mathrm{m\,s^{-1}}$ & $U_n$ at most $17.4\,\mathrm{m\,s^{-1}}$ \\
Admissible range of the wind limiter & $[\ell_d,h_d]$ & zonal-mean 50-hPa zonal wind within $30^\circ$ of the equator inside the range in all states \\
Admissible range of the surface limiter & $\mu_\varphi\pm4.5\sigma_\varphi$ & 0.0012\% of the row means over sea and 0.0013\% of those over land outside the range \\
Threshold for starting the special case of hydrostatic adjustment & $A_n>350\,\mathrm{m^2\,s^{-2}}$ & $A_n$ between $-310$ and $+267\,\mathrm{m^2\,s^{-2}}$ \\
\bottomrule
\end{tabularx}
\end{table}

\section{Precursor Analysis and Evaluation Details}\label{app:evaluation}

\subsection{Hydrostatic Imbalance before Rollout Failure}\label{app:precursor}

\begin{figure}[!htb]
  \centering
  \includegraphics[width=\linewidth]{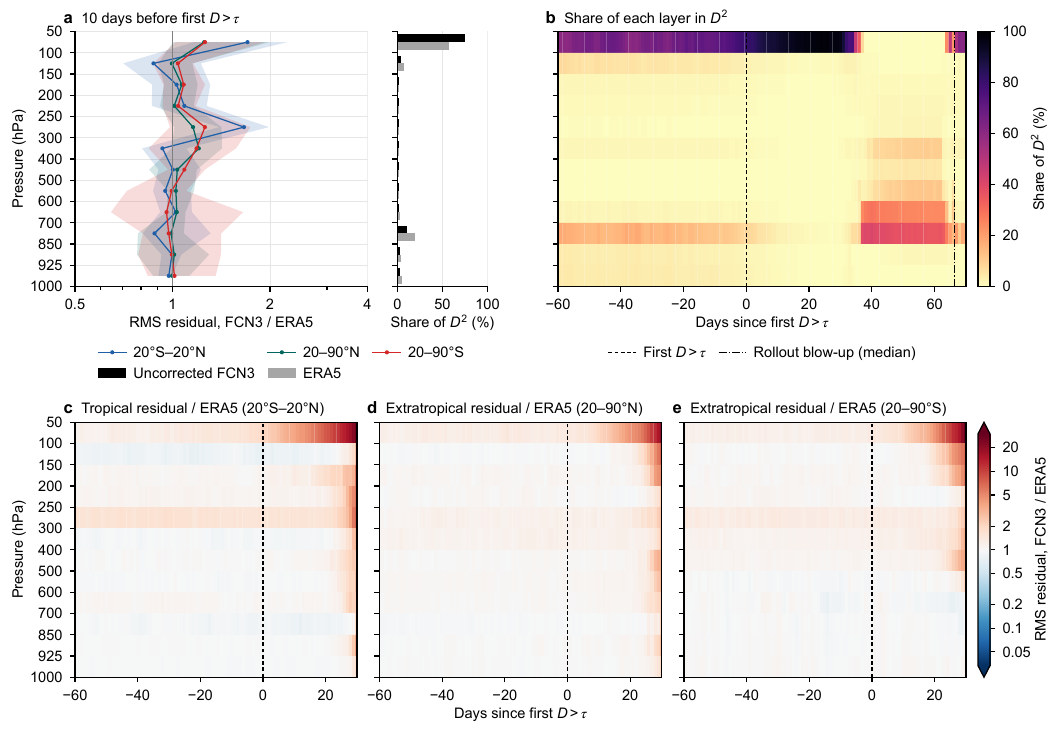}
  \caption{\textbf{Hydrostatic imbalance before rollout failure in 20 uncorrected FCN3 members.} (a)~Ratio of the RMS residual of each layer to that of ERA5 at the same valid times over the 40 six-hourly states (10 days) before the first $D>\tau$ of each member, for the tropics ($20^\circ$S--$20^\circ$N) and the two extratropical bands (poleward of $20^\circ$); lines show member medians, shading member ranges and the grey line a ratio of one. The right-hand panel gives the share of each layer in the global $D^2$ for FCN3 and ERA5 over the same states. (b)~Median over members of the share of each layer in the global $D^2$ against days since the first $D>\tau$; the dashed line marks the first $D>\tau$ and the dash-dotted line the median time of rollout blow-up. (c--e)~Median over members of the ratio of the RMS residual of each layer to that of ERA5 against days since the first $D>\tau$, for the tropics~(c) and the northern~(d) and southern~(e) extratropics; the colour scale is logarithmic and centred on one. Pressure levels on the vertical axes mark the layer boundaries.}
  \label{fig:precursor}
\end{figure}

\Cref{tab:hydrostatic_evolution,fig:precursor} resolve the imbalance of the 20 uncorrected members by layer, latitude band and time relative to each member's first crossing of $\tau$. Because \cref{eq:hydrostatic_residual} approximates the layer-mean virtual temperature by the mean of its values at the two bounding levels, it leaves a residual even in a hydrostatic column wherever temperature does not vary linearly with $\ln p$; in ERA5 over 2000--2015, $D$ lies between 83.2 and $144.7\,\mathrm{m^2\,s^{-2}}$. \Cref{fig:precursor} therefore shows the residual of each layer relative to that of ERA5 at the same valid times, so that a ratio of one means that the prediction is as balanced as the reanalysis.

The imbalance grows first in the tropical 50--100-hPa layer and reaches the rest of the column only after the crossing. Over the ten days before the crossing, this layer has the largest RMS residual of all band--layer combinations in every member and the largest ratio to ERA5 in 8 of the 20 members (\cref{fig:precursor}a); in 10 of the other 12, the largest ratio is in the 250--300-hPa layer, which holds only 0.3\% of $D^2$, and in the remaining two it is in the 50--100-hPa or the 600--700-hPa layer of the southern extratropics. Expressed as a temperature, the median tropical RMS residual of this layer is 1.9~K sixty days before the crossing and 4.3~K at the crossing (\cref{tab:hydrostatic_evolution}), against 1.9--2.3~K in ERA5 at the same valid times; the median tropical mean of the residual is 1.0 and 3.9~K at these times, against 1.1 and 1.6~K in ERA5. The ratio of the tropical 50--100-hPa layer exceeds 2 at the crossing, whereas that of the 50--100-hPa layer exceeds 2 only 10 days after the crossing in the northern extratropics and 11 days after it in the southern extratropics, and that of every other layer only 21--33 days after it, when $D$ is already about 3 to 22 times $\tau$ (\cref{fig:precursor}c--e).

Rollout failure then proceeds in two stages (\cref{fig:precursor}b, \cref{tab:hydrostatic_evolution}). At the crossing, the departure of global-mean T100 from ERA5 ranges from $-2.6$ to $+0.7$~K across the 20 members (median $-0.11$~K). For about 30 days after the crossing, the 50--100-hPa layer holds up to 98\% of $D^2$, the other layers stay near their ERA5 levels, and global-mean T100 begins to fall; global-mean T2m first departs from ERA5 by more than 10~K a median of 29 days after the crossing (range 21--254.75 days). Between 34 and 38 days after the crossing, the bulk of the residual moves to the 600--850-hPa layers; thereafter, the 300--850-hPa layers together hold more than 85\% of $D^2$ until the rollout blows up, a median of 66.25 days after the crossing (range 58.25--292.5 days).

With each of the three thresholds of \cref{tab:parameters}, the first crossing precedes failure in all 20 members, but the lead to failure depends on the threshold. Its median is about 64 days (range 57.5--291.5 days) with the 99.9th percentile of the calibration as threshold and about 94 days (range 60.25--444 days) with the 95th percentile, which lies only $0.2\,\mathrm{m^2\,s^{-2}}$ above the largest ERA5 value of $D$ over 2000--2015, against about 66 days with $\tau$.

$D$ returns below $\tau$ after the first crossing in 10 members. In the two members that fail in the second year (the dash-dotted lines in \cref{fig:imbalance_hadrian}a), initialized on 15 May with seed 1003 and on 15 February with seed 1001, $D$ first exceeds $\tau$ on days 342.25 and 283.75 but exceeds it continuously only from days 551.25 and 423.75 until failure on days 634.75 and 488. In the members initialized on 15 February and 15 November with seed 1002, $D$ first exceeds $\tau$ on days 37.25 and 38 and exceeds it continuously from days 70 and 142; in the members initialized on 15 August with seed 1004 and on 15 November with seed 333, it first exceeds $\tau$ on days 113.5 and 91.25 and exceeds it continuously from days 138 and 144.75. In the other four members, $D$ falls below $\tau$ on at most ten six-hourly steps after the first crossing.

\begin{table}[!htb]
\centering
\caption{\textbf{Hydrostatic imbalance and temperature departures around the first threshold crossing in the 20 uncorrected members.}}
\label{tab:hydrostatic_evolution}
\small
\setlength{\tabcolsep}{4pt}
\renewcommand{\arraystretch}{1.12}
\begin{tabularx}{\linewidth}{@{}>{\centering\arraybackslash}p{0.18\linewidth}*{5}{>{\centering\arraybackslash}X}@{}}
\toprule
Days since first crossing & $D/\tau$ & Tropical 50--100-hPa RMS residual (K) & 50--100-hPa contribution to global $D^2$ (\%) & Global-mean T100 departure (K) & Global-mean T2m departure (K) \\
\midrule
$-60$ & 0.68 & 1.9 & 56.6 & $-0.07$ & $0.00$ \\
$-10$ & 0.78 & 2.9 & 69.7 & $-0.18$ & $+0.03$ \\
0 & 0.99 & 4.3 & 80.8 & $-0.11$ & $+0.06$ \\
$+10$ & 1.39 & 6.6 & 91.2 & $-0.07$ & $-0.01$ \\
$+20$ & 2.89 & 14.4 & 96.9 & $-2.70$ & $+0.20$ \\
\bottomrule
\end{tabularx}
\par\smallskip
\begin{minipage}{\linewidth}\footnotesize
Medians over the 20 uncorrected members of daily 00~UTC values; the row for day $-60$ uses the 18 members that first cross $\tau$ after day 60. Day 0 is the last 00~UTC state at or before each member's first crossing of $\tau$. The tropical ($20^\circ$S--$20^\circ$N) 50--100-hPa residual is the RMS over the band of the layer residual divided by $R_d\ln 2=199\,\mathrm{m^2\,s^{-2}\,K^{-1}}$, the change in its hydrostatic thickness per kelvin of layer-mean virtual temperature. Temperature departures are cosine-latitude-weighted global means of FCN3 minus ERA5.
\end{minipage}
\end{table}

\subsection{Evaluation Windows and Aggregation}\label{app:evaluation_windows}

FCN3, SFNO and pretrained Aurora are run through the Nvidia Earth2Studio platform, and all AIFS Single 1.1 trajectories use one inference configuration.

\paragraph{Grids.}
The $1.5^\circ$ fields are obtained from the $0.25^\circ$ state fed back as the next input by first-order conservative regridding with the WeatherBench 2 algorithm~\citep{rasp2023weatherbench}, in which each $1.5^\circ$ cell is the area-weighted mean of the $0.25^\circ$ cells it overlaps; ERA5 is coarsened in the same way and compared at the same valid times. Models whose native grid is not the $0.25^\circ$ latitude--longitude grid are first brought to it: AIFS by interpolation from N320, Aurora by extending its 720 latitude rows to 721 with the $90^\circ$S row copied from $89.75^\circ$S, and DLESyM by bilinear interpolation from HEALPix64 (\cref{tab:grids}).

\begin{table}[!htb]
\centering
\caption{\textbf{Grids of the fields used in the evaluation.}}
\label{tab:grids}
\footnotesize
\renewcommand{\arraystretch}{1.15}
\begin{tabularx}{\textwidth}{@{}>{\raggedright\arraybackslash}p{0.17\textwidth}>{\raggedright\arraybackslash}p{0.15\textwidth}>{\raggedright\arraybackslash}X>{\raggedright\arraybackslash}p{0.22\textwidth}@{}}
\toprule
Fields & Grid & Content & Used for \\
\midrule
Native model output & $0.25^\circ$ ($721\times1440$) & T2m, TCWV, MSLP, 10-m wind components, T850, U850, V850 and Z500, six-hourly & all scores of these nine variables \\
WeatherBench 2 grid & $1.5^\circ$ ($240\times121$) & all 72 channels, six-hourly & all other quantities, such as the 100-hPa temperature \\
Hydrostatic residual records & native $0.25^\circ$ & per-step layer and latitude-band residual statistics, computed as for the activation criterion (\cref{sec:method}), with the ERA5 residual on the same grid at the same valid times & \cref{app:precursor,tab:hydrostatic_evolution,fig:precursor} \\
\bottomrule
\end{tabularx}
\end{table}

\paragraph{Figure~\ref*{fig:series}.}
Global means are cosine-latitude-weighted means of the daily 00~UTC states on days 1--730. Zonal power spectra are computed on the $0.25^\circ$ fields at days 90, 100, $\ldots$, 730, and band values average the median ratio over the wavenumbers in each band; the four two-year windows overlap.

\paragraph{Synoptic-scale variability.}
The two-year storm-track variance ratios use the same 2,920 six-hourly samples on days 1--730 for all models.

\begin{table}[!htb]
\centering
\caption{\textbf{Two-year outcomes by initialization year.} Metrics are defined as in \cref{tab:survival}: biases and storm-track variance ratios are means of per-trajectory metrics, and Z500 RMSE uses the five-member FCN3 + HADRIAN ensemble-mean field at each initialization date, or the single trajectory of each comparison model. Each year has four initialization dates, with five seeds per date for FCN3 and one trajectory per date for SFNO, Aurora and AIFS; all 20 FCN3 + HADRIAN members and all four trajectories of each comparison model are scored. As in 2021, the AIFS trajectories initialized in 2023 remain finite but fail the plausibility check on global-mean T2m. The 2021 rows repeat \cref{tab:survival}. n/a, not available because no uncorrected member completes two years.}
\label{tab:survival_full}
\footnotesize
\setlength{\tabcolsep}{2pt}
\begin{tabular*}{\textwidth}{@{\extracolsep{\fill}}lcccccc@{}}
\toprule
Initialization year, model & Stable & Activation & T2m bias & T850 bias & Z500 RMSE & Storm-track variance \\
 & & (days) & (K) & (K) & (m$^2$\,s$^{-2}$) & ratio NH / SH \\
\midrule
2021 uncorrected FCN3 & 0/20 & -- & n/a & n/a & n/a & n/a \\
2021 FCN3 + HADRIAN & 20/20 & 37.25--342.25 & -0.137 & -0.179 & 178.49 & 0.933 / 0.944 \\
2021 SFNO & 4/4 & -- & -0.530 & -0.592 & 358.75 & 0.795 / 0.845 \\
2021 Aurora & 4/4 & -- & -0.381 & -0.543 & 282.16 & 0.694 / 0.850 \\
2021 AIFS & 0/4 & -- & +1.282 & +1.307 & 498.95 & 0.616 / 0.446 \\
\midrule
2023 uncorrected FCN3 & 0/20 & -- & n/a & n/a & n/a & n/a \\
2023 FCN3 + HADRIAN & 20/20 & 49.25--387.5 & -0.440 & -0.526 & 203.01 & 0.968 / 0.955 \\
2023 SFNO & 4/4 & -- & -0.871 & -0.975 & 397.81 & 0.819 / 0.863 \\
2023 Aurora & 4/4 & -- & -0.664 & -0.872 & 341.64 & 0.669 / 0.860 \\
2023 AIFS & 0/4 & -- & +1.047 & +1.216 & 518.62 & 0.600 / 0.432 \\
\bottomrule
\end{tabular*}
\end{table}

\begin{table}[!htb]
\centering
\caption{\textbf{Physical plausibility checks of the two-year rollouts from the four 2021 initialization dates.} Entries give the number of trajectories that pass the check out of the number checked and, in parentheses, the range over these trajectories of the checked quantity (daily global mean, annual drift, percentage of days or phase difference). The checks and thresholds are defined in \cref{sec:method}. They use cosine-latitude-weighted global and regional means of the daily 00~UTC states from day 1 to the end of the rollout, except that the drift of the total-energy proxy uses all six-hourly states; drifts are least-squares linear trends, and the phase of each annual cycle is that of the annual harmonic fitted to the extratropical-mean T2m of each hemisphere. n/a, not checked because Aurora and AIFS do not predict TCWV. Uncorrected FCN3 is not listed because none of its members completes two years.}
\label{tab:plausibility}
\footnotesize
\setlength{\tabcolsep}{3pt}
\resizebox{\textwidth}{!}{%
\begin{tabular}{llcccc}
\toprule
Check & Threshold & FCN3 + HADRIAN & SFNO & Aurora & AIFS \\
\midrule
Global-mean T2m (K) & 283--292 & 20/20 (284.7--289.8) & 4/4 (284.6--289.4) & 4/4 (284.8--289.1) & 0/4 (285.4--293.2) \\
Global-mean TCWV (kg\,m$^{-2}$) & 15--40 & 20/20 (22.1--27.9) & 4/4 (22.4--26.8) & n/a & n/a \\
Global-mean MSLP (hPa) & 1000--1025 & 20/20 (1011.0--1011.9) & 4/4 (1011.1--1012.0) & 4/4 (1011.2--1013.2) & 4/4 (1007.8--1016.5) \\
Annual drift of global-mean MSLP (Pa) & $\pm500$ & 20/20 ($-6.9$ to $+12.7$) & 4/4 ($-6.9$ to $-0.1$) & 4/4 ($+23.2$ to $+48.4$) & 4/4 ($-55.6$ to $+181.4$) \\
Drift of the total-energy proxy (W\,m$^{-2}$) & $\pm5$ & 20/20 ($-0.22$ to $+0.20$) & 4/4 ($-0.28$ to $+0.20$) & 4/4 ($-0.18$ to $+0.11$) & 4/4 ($+0.09$ to $+0.80$) \\
Tropical-mean T2m $\le$ extratropical mean (\% of days) & $\le1$ & 20/20 (0) & 4/4 (0) & 4/4 (0) & 4/4 (0) \\
Phase difference of the annual cycles (days) & $182.5\pm45$ & 20/20 (179.1--182.2) & 4/4 (175.6--178.5) & 4/4 (177.6--182.0) & 4/4 (162.8--175.6) \\
\bottomrule
\end{tabular}}
\end{table}

% Appendix C: reserved for detailed ablation experiments.
% \clearpage
\section{Detailed Ablation Experiments}\label{app:ablation_details}

\begin{figure}[!htb]
  \centering
  \includegraphics[width=\textwidth]{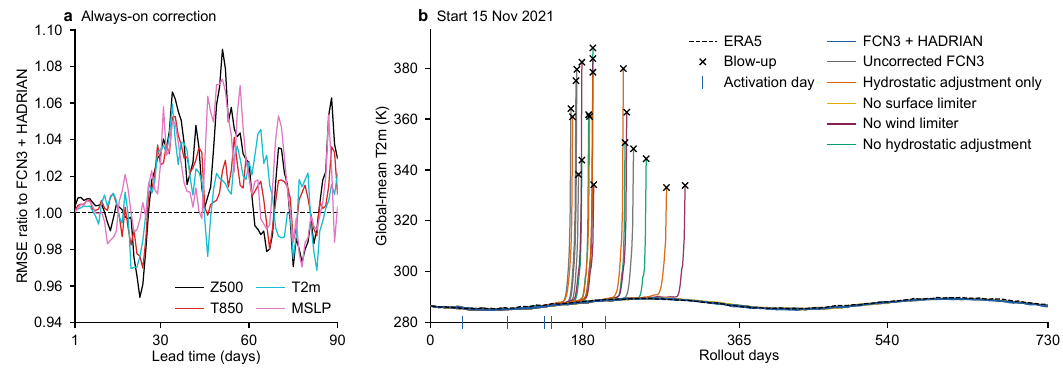}
  \caption{\textbf{Short-range errors of always-on correction, and global-mean T2m of the ablation configurations for the 15 November 2021 initialization.} (a)~Ratio of the RMSE of always-on correction to that of FCN3 + HADRIAN, which equals uncorrected FCN3 before activation, for Z500, T850, T2m and MSLP at daily lead times 1--90; mean of the per-member ratios over the 20 members from the four 2021 initialization dates; cosine-latitude-weighted RMSE against ERA5 at 00~UTC. (b)~Daily global-mean T2m of the five members initialized on 15 November 2021 for FCN3 + HADRIAN, uncorrected FCN3, the configurations with one stage removed and the configuration with hydrostatic adjustment only. Crosses mark blow-up. Before activation, all configurations that keep the activation criterion are identical to uncorrected FCN3.}
  \label{fig:app_abl_alwayson}
\end{figure}

\begin{figure}[!htb]
  \centering
  \includegraphics[width=\textwidth]{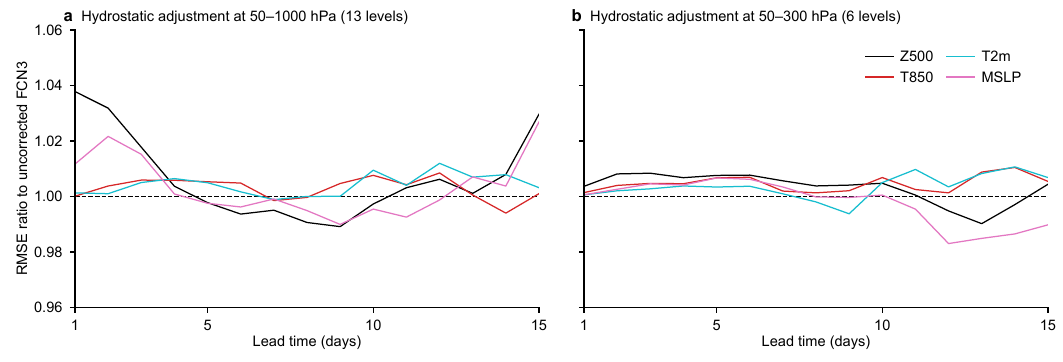}
  \caption{\textbf{Short-range errors of always-on correction with hydrostatic adjustment at all 13 levels and at 50--300~hPa.} Ratio of the RMSE of always-on correction to that of uncorrected FCN3 for Z500, T850, T2m and MSLP at daily lead times 1--15; mean of the per-member ratios over the 20 members from the four 2021 initialization dates. Hydrostatic adjustment acts (a)~at all 13 levels (50--1000~hPa) and (b)~at 50--300~hPa, as in FCN3 + HADRIAN. Global cosine-latitude-weighted RMSE of daily 00~UTC fields against ERA5; dashed line, equal RMSE.}
  \label{fig:app_abl_levels}
\end{figure}

The ablation experiments of \cref{sec:ablations} rerun the 20 two-year 2021 FCN3 + HADRIAN members, with the same initialization dates and random seeds, in three configurations, each with one stage removed and the activation criterion kept. A fourth configuration, run for the same 20 members, keeps hydrostatic adjustment only and removes both the wind limiter and the surface limiter. A member in which the removed stage never acts remains identical to its FCN3 + HADRIAN member; it is counted but does not test that stage. Two further configurations apply the correction from the first step (always-on), with hydrostatic adjustment at 50--300~hPa as in FCN3 + HADRIAN or at all 13 levels; they are run for 90 days and used only to compare short-range errors (\cref{fig:app_abl_alwayson}a, \cref{fig:app_abl_levels}).

\begin{table}[!htb]
\centering
\caption{\textbf{Stability, mean state and variability of each configuration.}}
\label{tab:ablation}
\footnotesize
\setlength{\tabcolsep}{3pt}
\resizebox{\textwidth}{!}{%
\begin{tabular}{lcccccccccc}
\toprule
Configuration & Stable & Days to & $\Delta$T2m & $\Delta$T2m & $\Delta$T2m & $\Delta$T850 & $\Delta$TCWV & $\Delta$T2m & $\Delta$Z500 & $\Delta$Storm-track variance \\
 & & failure & global & tropics & NH land & global & global & RMSE & RMSE & ratio NH / SH \\
\midrule
FCN3 + HADRIAN & 20/20 & -- & -- & -- & -- & -- & -- & -- & -- & -- \\
Uncorrected FCN3 & 0/20 & 66 (58--293) & n/a & n/a & n/a & n/a & n/a & n/a & n/a & n/a \\
\midrule
No hydrostatic adjustment & 0/20 & 81 (58--313) & n/a & n/a & n/a & n/a & n/a & n/a & n/a & n/a \\
No wind limiter & 8/20$^{\mathrm{a}}$ & 120 (83--633) & n/a & n/a & n/a & n/a & n/a & n/a & n/a & n/a \\
No surface limiter & 20/20 & -- & $+0.05$ & $0.00$ & $+0.23$ & $+0.03$ & $+0.03$ & $+0.05$ & $+6.3$ & 0.00 / 0.00 \\
Hydrostatic adjustment only & 9/20$^{\mathrm{b}}$ & 106 (82--221) & n/a & n/a & n/a & n/a & n/a & n/a & n/a & n/a \\
\bottomrule
\end{tabular}}
\par\smallskip
\begin{minipage}{\textwidth}\footnotesize
Activation and $\Delta$ refer to the FCN3 + HADRIAN member with the same initialization date and random seed. Stable: members that complete 730 days and pass the plausibility checks, as in \cref{tab:survival}. Days to failure: median (range) of the days from activation to the first non-finite state. $\Delta$: median over the 20 members of the score of each member minus that of its FCN3 + HADRIAN member over days 1--730, with biases and the RMSE of the time-mean field computed from daily 00~UTC fields against ERA5 and storm-track variance ratios from six-hourly Z500 fields (\cref{sec:method}). Before activation (days 37.25--342.25), all configurations are identical to FCN3 + HADRIAN. Tropics, $|\varphi|\le20^\circ$; NH land, $\varphi\ge30^\circ$N with land fraction $\ge0.5$. Units of $\Delta$: K for T2m and T850, $\mathrm{kg\,m^{-2}}$ for TCWV and $\mathrm{m^2\,s^{-2}}$ for Z500. n/a, not available because members fail. $^{\mathrm{a}}$In these eight members, the wind limiter of FCN3 + HADRIAN is never switched on. $^{\mathrm{b}}$Seven of these nine members also remain stable without the wind limiter.
\end{minipage}
\end{table}

\begin{figure}[!htb]
  \centering
  \includegraphics[width=\textwidth]{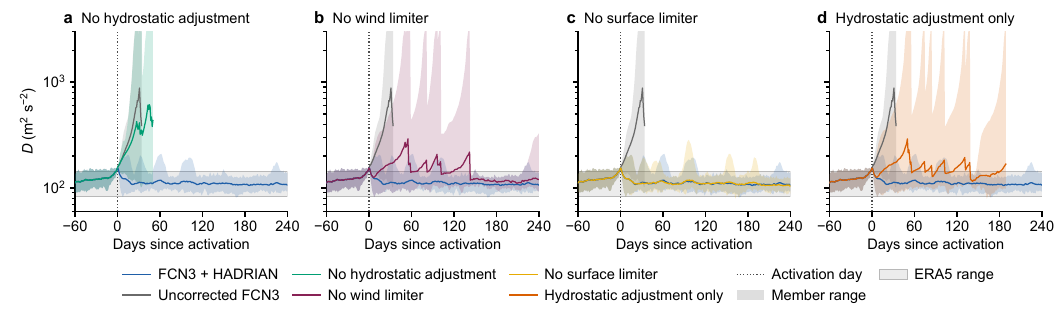}
  \caption{\textbf{Global hydrostatic residual $D$ after activation in the ablation configurations.} $D$ of the raw prediction against days since activation for the configurations (a)~without hydrostatic adjustment, (b)~without the wind limiter, (c)~without the surface limiter and (d)~with hydrostatic adjustment only, each shown with FCN3 + HADRIAN and uncorrected FCN3; 20 members from the four 2021 initialization dates. Daily geometric means of $D$, as in \cref{fig:crossmodel_residual}; lines show the geometric mean and shading the range over the members whose $D$ is finite and at most $10^4\,\mathrm{m^2\,s^{-2}}$, drawn where this holds for at least half of the members. The dotted vertical line marks activation, and the grey band shows the ERA5 range of $D$. Before day 0, all configurations that keep the activation criterion are identical to uncorrected FCN3.}
  \label{fig:app_abl_residual}
\end{figure}

\paragraph{Failure without hydrostatic adjustment or the wind limiter.}
Without hydrostatic adjustment, members fail at nearly the same time as their uncorrected controls: between 3.0 days earlier and 55 days later (median 12.5 days later), and within one day in 6 of the 20 members (\cref{fig:ablation}). In these members, the wind limiter is first switched on a median of 29 days after activation (range 2.75--278 days) and is on for a median of 57\% of the steps after activation, against 2.7\% in FCN3 + HADRIAN. In both failing configurations, $D$ of the raw predictions exceeds $\tau$ on most steps after activation, as in uncorrected FCN3 (\cref{fig:app_abl_residual}a,b). The median share of these steps is 99.8\% without hydrostatic adjustment (member range 31--100\%) and in uncorrected FCN3 (33--100\%), and 94.2\% (13--99\%) in the 12 members that fail without the wind limiter; in FCN3 + HADRIAN and the members of the other configurations that remain stable, $D$ exceeds $\tau$ on at most 6.8\% of the steps.

\begin{figure}[!htb]
  \centering
  \includegraphics[width=\textwidth]{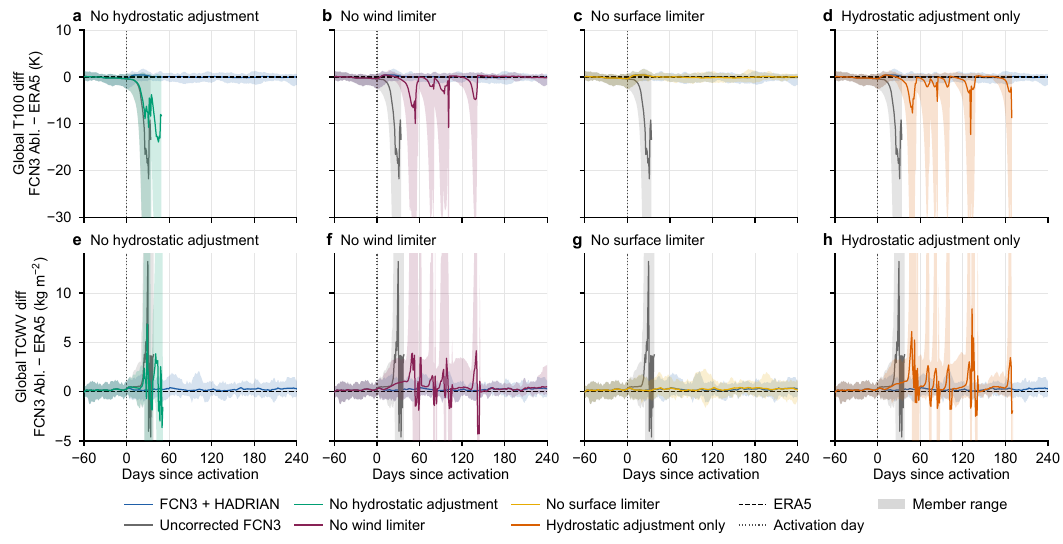}
  \caption{\textbf{Global-mean 100-hPa temperature and TCWV after activation in the ablation configurations.} Global-mean (a--d)~100-hPa temperature and (e--h)~TCWV minus ERA5 against days since activation, for the configurations without (a,e)~hydrostatic adjustment, (b,f)~the wind limiter and (c,g)~the surface limiter, and with (d,h)~hydrostatic adjustment only, each shown with FCN3 + HADRIAN and uncorrected FCN3; 20 members from the four 2021 initialization dates. Lines show the mean and shading the range over the members whose global mean lies within 50--400~K (100-hPa temperature) or $0$--$200\,\mathrm{kg\,m^{-2}}$ (TCWV), drawn where this holds for at least half of the members. FCN3 Abl.\ in the axis labels denotes FCN3 in the configuration of each column. Before day 0, all configurations that keep the activation criterion are identical to uncorrected FCN3.}
  \label{fig:app_abl_routes}
\end{figure}

Both configurations fail in the same way as uncorrected FCN3 (\cref{fig:app_abl_routes}a,b,e,f): within weeks of activation, global-mean 100-hPa temperature falls and TCWV rises, whereas global-mean T2m stays close to ERA5 until it rises abruptly before failure (\cref{fig:app_abl_alwayson}b, members initialized on 15 November 2021). Twenty days after activation, the departures of global-mean 100-hPa temperature and TCWV from ERA5 are $-3.1$~K and $+0.4\,\mathrm{kg\,m^{-2}}$ without hydrostatic adjustment and $-4.1$~K and $+0.6\,\mathrm{kg\,m^{-2}}$ for uncorrected FCN3, against $+0.4$~K and $+0.1\,\mathrm{kg\,m^{-2}}$ for FCN3 + HADRIAN, whose 100-hPa temperature stays within 0.5~K of ERA5 after activation (means over the 20 members). Thirty days after activation, TCWV without the wind limiter is $0.9\,\mathrm{kg\,m^{-2}}$ above ERA5, against $0.2\,\mathrm{kg\,m^{-2}}$ for FCN3 + HADRIAN.

\paragraph{Removal of the surface limiter.}
Without the surface limiter, global-mean 100-hPa temperature and TCWV stay as close to ERA5 as in FCN3 + HADRIAN (\cref{fig:app_abl_routes}c,g). During the 240 days after activation, the member mean of the 100-hPa temperature departure lies between $-0.35$ and $+0.42$~K, against $-0.35$ and $+0.43$~K for FCN3 + HADRIAN, and thirty days after activation TCWV is $0.1\,\mathrm{kg\,m^{-2}}$ above ERA5, against $0.2\,\mathrm{kg\,m^{-2}}$ for FCN3 + HADRIAN. The main change is in near-surface temperature: Northern Hemisphere land T2m over days 1--730 is warmer than in FCN3 + HADRIAN in 15 of the 20 members (\cref{tab:ablation}).

\begin{figure}[!htb]
  \centering
  \includegraphics[width=\textwidth]{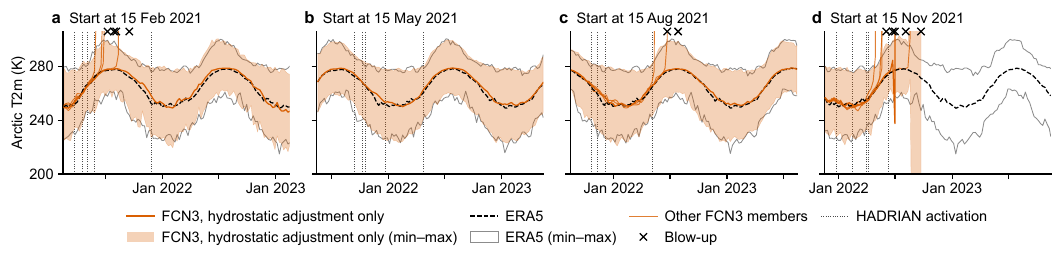}
  \caption{\textbf{Arctic T2m of FCN3 with hydrostatic adjustment only for the four 2021 initialization dates.} Five members each, initialized on (a)~15 February, (b)~15 May, (c)~15 August and (d)~15 November 2021. Curves and bands average the members that complete two years; members that fail are shown individually, and where all five fail, curves and bands follow the members as for uncorrected FCN3 in \cref{fig:crossmodel_arctic}. ERA5 as in \cref{fig:crossmodel_arctic}. Dotted vertical lines mark activation of each member; crosses mark blow-up.}
  \label{fig:app_abl_honly_arctic}
\end{figure}

\paragraph{Hydrostatic adjustment only.}
With hydrostatic adjustment only, 9 of the 20 members complete two years: one of those initialized on 15 February, all five initialized on 15 May, three of the five initialized on 15 August and none of those initialized on 15 November (\cref{fig:ablation}; \cref{fig:app_abl_honly_arctic}). Seven of these nine members also remain stable without the wind limiter (\cref{tab:ablation}). The other two fail without the wind limiter, on days 462 and 700.5, and one member that remains stable without the wind limiter fails with hydrostatic adjustment only, on day 309. Removing the surface limiter as well therefore changes the outcome of three members. After activation, $D$ of the raw predictions rises with hydrostatic adjustment only as it does without the wind limiter (\cref{fig:app_abl_residual}b,d). Global-mean 100-hPa temperature and TCWV after activation in this configuration are shown in \cref{fig:app_abl_routes}d,h.

\paragraph{Always-on correction.}
Applied from the first step, the correction changes the day-1 RMSE of Z500, T850, T2m and MSLP by at most 0.4\% on average over the 20 members (\cref{fig:app_abl_alwayson}a).

\paragraph{Vertical extent of hydrostatic adjustment.}
With always-on correction, adjusting all 13 levels (50--1000~hPa) raises the short-range errors of Z500 and MSLP more than adjusting 50--300~hPa, as in FCN3 + HADRIAN (\cref{fig:app_abl_levels}). The day-1 Z500 RMSE is 1.038 times that of uncorrected FCN3 with all 13 levels (member range 1.016--1.053) and 1.004 times with 50--300~hPa; over days 1--15, the RMSE ratios of Z500, MSLP, T850 and T2m stay within 0.99--1.04 with all 13 levels and within 0.98--1.01 with 50--300~hPa.

% Appendix D: complete seasonal-mean field comparison.
\clearpage
\section{Seasonal-Mean Field Fidelity}\label{app:seasonal_fidelity}

\begin{figure}[p]
\centering
\includegraphics[width=\textwidth]{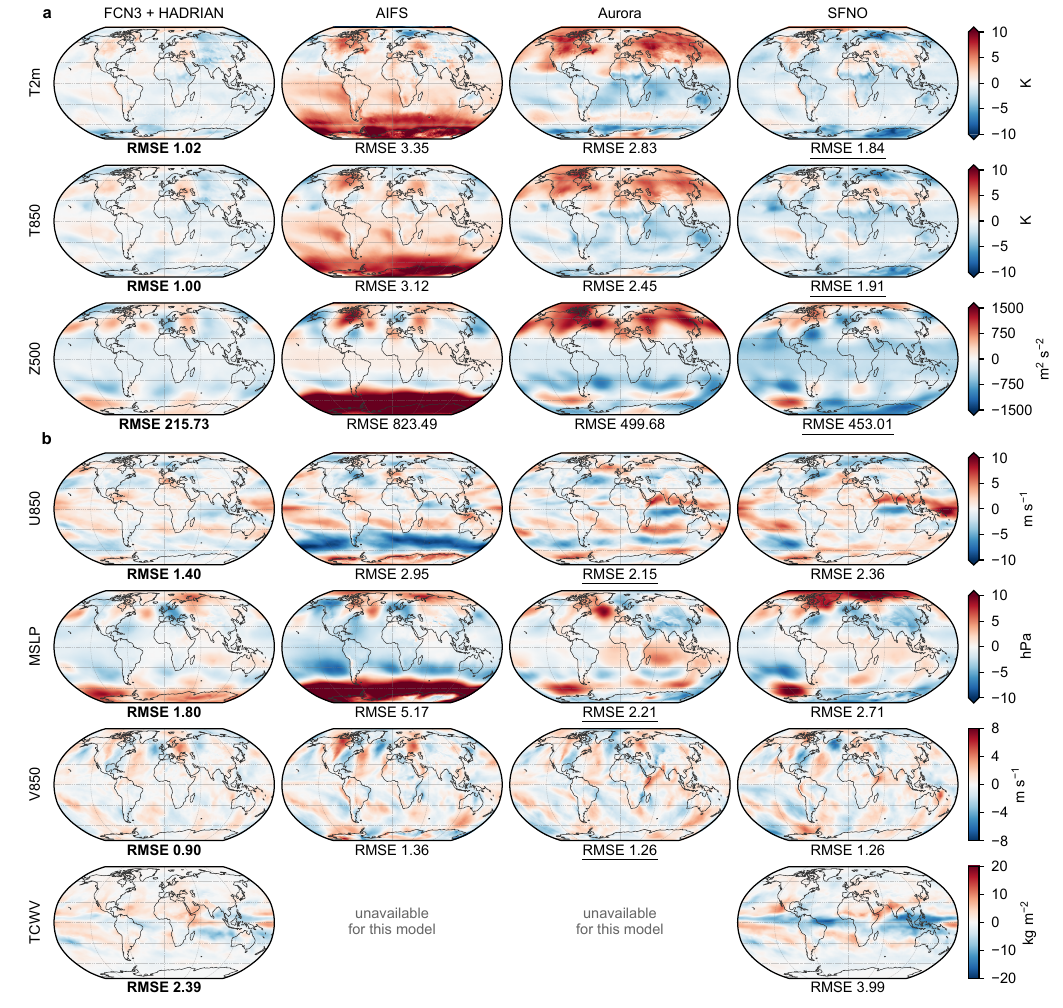}
\caption{\textbf{Seasonal-mean errors in MAM 2022 for the 15 November 2021 initialization.} Differences from ERA5 of the March--May 2022 mean of (a)~T2m, T850 and Z500 and (b)~U850, MSLP, V850 (850-hPa meridional wind) and TCWV for the FCN3 + HADRIAN five-member mean, AIFS, Aurora and SFNO. Labels give the RMSE (bold, lowest; underlined, second lowest; TCWV, lowest only).}
\label{fig:seasonal_i04}
\end{figure}

\begin{figure}[p]
\centering
\includegraphics[width=\textwidth]{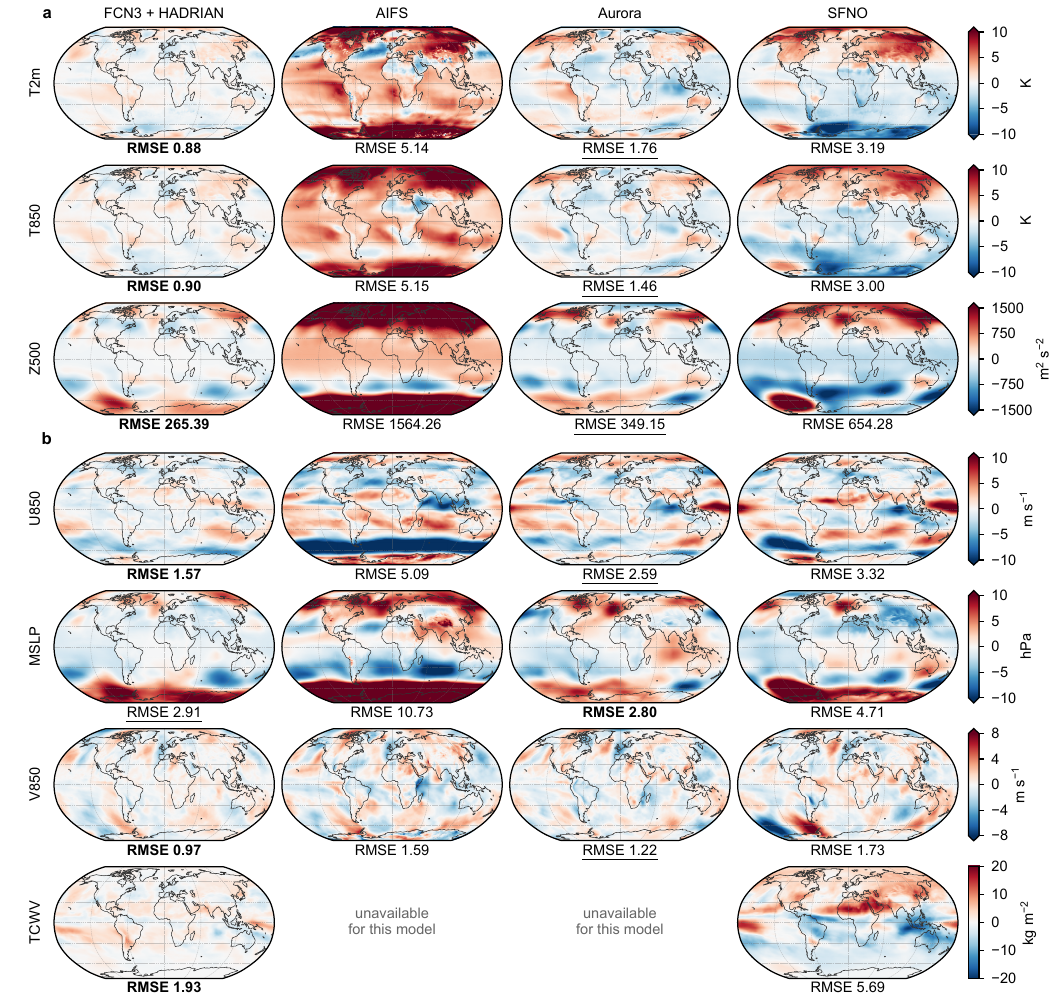}
\caption{\textbf{Seasonal-mean errors over days 273--365 for the 15 November 2021 initialization.} As \cref{fig:seasonal_i04}, for the mean over 15 August--14 November 2022.}
\label{fig:seasonal_late_i04}
\end{figure}

Seasonal-mean fields of the four 2021 initializations are verified against ERA5 in two three-month windows of the first year: the calendar season beginning 106--109 days after initialization (JJA 2021, SON 2021, DJF 2021/22 and MAM 2022 for the 15 February, 15 May, 15 August and 15 November initializations) and days 273--365 after initialization. The five-member FCN3 + HADRIAN ensemble is compared with the AIFS Single 1.1, pretrained Aurora and SFNO trajectories of \cref{sec:spectra}, one per initialization. Seasonal means average the six-hourly fields of each window, and scores are area-weighted over the whole globe, including grid points where the 850-hPa surface lies below ground. \Cref{fig:seasonal_i04,fig:seasonal_late_i04} show the errors for the 15 November 2021 initialization, also used in \cref{fig:crossmodel_arctic}. \Cref{tab:seasonal_rmse,tab:seasonal_late_diagnostics} compare the models over all four initializations: \cref{tab:seasonal_rmse} gives the RMSE, the CRPS and the ensemble dispersion, and \cref{tab:seasonal_late_diagnostics} the global-mean temperature biases and storm-track variance ratios over days 273--365.

\begin{table}[!htbp]
\centering
\caption{\textbf{Scores of seasonal-mean fields, averaged over the four initializations.} The RMSE of FCN3 + HADRIAN uses the five-member ensemble-mean field; the indented row gives the mean RMSE of its individual members and is not ranked. CRPS is computed at each grid cell from the five members and then area-averaged; for a single trajectory it equals the mean absolute error. The spread--skill ratio divides the RMS ensemble spread, the square root of the area-weighted ensemble variance (normalized by $n-1$) averaged over the four initializations, by the RMS of the ensemble-mean RMSEs of the four initializations; coverage is the area fraction of ERA5 values within the five-member range. For a statistically consistent five-member ensemble, the spread--skill ratio is about 0.91 and the coverage 67\%. Column units apply to RMSE and CRPS. \textbf{Bold} and \underline{underlining} mark the lowest and second-lowest RMSE or CRPS, ranked before rounding; for TCWV, only the lowest is marked. --, not predicted by the model.}
\label{tab:seasonal_rmse}
\footnotesize
\setlength{\tabcolsep}{2.6pt}
\renewcommand{\arraystretch}{1.04}
\begin{tabular*}{\textwidth}{@{\extracolsep{\fill}}llrrrrrrr@{}}
\toprule
Score & Model & T2m & T850 & Z500 & U850 & MSLP & V850 & TCWV \\
 & & K & K & m$^2$ s$^{-2}$ & m s$^{-1}$ & hPa & m s$^{-1}$ & kg m$^{-2}$ \\
\midrule
\multicolumn{9}{l}{\textit{(a) Calendar season}} \\
RMSE & FCN3 + HADRIAN & \textbf{1.09} & \textbf{1.09} & \textbf{297.95} & \textbf{1.74} & \textbf{2.78} & \textbf{1.00} & \textbf{2.41} \\
 & \quad Individual members & 1.28 & 1.30 & 359.55 & 2.11 & 3.22 & 1.28 & 2.78 \\
 & AIFS & \underline{1.99} & \underline{1.85} & 493.92 & \underline{2.25} & 3.52 & \underline{1.26} & -- \\
 & Aurora & 2.38 & 2.05 & \underline{485.54} & 2.67 & \underline{3.27} & 1.63 & -- \\
 & SFNO & 2.64 & 2.53 & 507.62 & 2.66 & 3.58 & 1.47 & 4.64 \\
CRPS & FCN3 + HADRIAN & \textbf{0.61} & \textbf{0.63} & \textbf{156.67} & \textbf{0.95} & \textbf{1.39} & \textbf{0.57} & \textbf{1.32} \\
 & AIFS & \underline{1.39} & \underline{1.39} & \underline{289.62} & \underline{1.64} & 2.37 & \underline{0.93} & -- \\
 & Aurora & 1.83 & 1.57 & 346.66 & 1.90 & \underline{2.26} & 1.18 & -- \\
 & SFNO & 1.91 & 1.99 & 414.82 & 1.93 & 2.37 & 1.09 & 3.48 \\
Spread--skill ratio & FCN3 + HADRIAN & 0.68 & 0.72 & 0.75 & 0.77 & 0.67 & 0.87 & 0.65 \\
Coverage (\%) & FCN3 + HADRIAN & 47.40 & 55.33 & 46.21 & 59.80 & 50.41 & 62.25 & 54.22 \\
\addlinespace[3pt]
\multicolumn{9}{l}{\textit{(b) Days 273--365}} \\
RMSE & FCN3 + HADRIAN & \textbf{0.99} & \textbf{1.05} & \textbf{285.74} & \textbf{1.79} & \textbf{2.70} & \textbf{1.02} & \textbf{2.39} \\
 & \quad Individual members & 1.18 & 1.26 & 346.29 & 2.12 & 3.12 & 1.29 & 2.77 \\
 & AIFS & 3.80 & 3.32 & 840.80 & 3.92 & 6.36 & 1.62 & -- \\
 & Aurora & \underline{2.38} & \underline{1.99} & \underline{466.05} & 2.58 & \underline{3.23} & 1.48 & -- \\
 & SFNO & 2.52 & 2.34 & 500.19 & \underline{2.54} & 3.65 & \underline{1.45} & 4.10 \\
CRPS & FCN3 + HADRIAN & \textbf{0.57} & \textbf{0.63} & \textbf{146.27} & \textbf{1.02} & \textbf{1.44} & \textbf{0.58} & \textbf{1.30} \\
 & AIFS & 2.82 & 2.61 & 591.12 & 2.83 & 4.10 & 1.22 & -- \\
 & Aurora & 1.77 & \underline{1.48} & \underline{306.51} & 1.89 & \underline{2.28} & 1.09 & -- \\
 & SFNO & \underline{1.75} & 1.79 & 401.04 & \underline{1.88} & 2.34 & \underline{1.06} & 3.07 \\
Spread--skill ratio & FCN3 + HADRIAN & 0.72 & 0.74 & 0.78 & 0.74 & 0.66 & 0.86 & 0.66 \\
Coverage (\%) & FCN3 + HADRIAN & 46.16 & 51.04 & 52.97 & 56.36 & 45.06 & 60.35 & 54.44 \\
\bottomrule
\end{tabular*}
\end{table}

\begin{table}[!htbp]
\centering
\caption{\textbf{Global-mean temperature bias and storm-track variance ratio over days 273--365.} Storm-track variance ratios (\cref{sec:method}) are computed from the 368 six-hourly Z500 fields for each member, not for the ensemble-mean field, and averaged over members; NH and SH cover latitudes at or poleward of $30^\circ$, and the extratropical ratio combines both. All values are averaged over the four initializations. \textbf{Bold} and \underline{underlining} mark the smallest and second-smallest departures (bias from zero, ratio from one), ranked before rounding.}
\label{tab:seasonal_late_diagnostics}
\footnotesize
\setlength{\tabcolsep}{2.6pt}
\renewcommand{\arraystretch}{1.04}
\begin{tabular*}{\textwidth}{@{\extracolsep{\fill}}lrrrrr@{}}
\toprule
Model & T2m bias (K) & T850 bias (K) & Extratropical ratio & NH ratio & SH ratio \\
\midrule
FCN3 + HADRIAN & \textbf{-0.096} & \textbf{-0.125} & \textbf{0.930} & \textbf{0.884} & \textbf{0.954} \\
AIFS & 1.920 & 1.941 & 0.537 & 0.386 & 0.597 \\
Aurora & \underline{-0.241} & \underline{-0.441} & \underline{0.821} & 0.649 & \underline{0.916} \\
SFNO & -0.512 & -0.581 & 0.817 & \underline{0.751} & 0.864 \\
\bottomrule
\end{tabular*}
\end{table}

\clearpage

% Appendix E: common-start global and polar temperature comparison.
\clearpage
\section{Hydrostatic Imbalance and Temperature Evolution Across Models}\label{app:crossmodel_temperature}

\subsection{Setup}\label{app:crossmodel_setup}
FCN3, GraphCast, GenCast, Aurora, AIFS and FuXi are run with and without HADRIAN for two years from 00~UTC on 15 November 2021, and SFNO, Pangu and DLESyM \citep{cresswellclay2025deep} without it (\cref{app:crossmodel_reference}). Each adaptation to the five other models follows the design of FCN3 + HADRIAN (\cref{sec:method}), with an activation criterion adapted to the model. The ocean 2-m temperature limit of GraphCast is the only stage that acts from the first step. The activation criterion, the form of hydrostatic adjustment and the model-specific limits of each adaptation are summarized in \cref{tab:crossmodel_hosts}. The contribution of each stage is separated for FCN3 in \cref{app:ablation_details}, and those of hydrostatic adjustment and the model-specific limits are compared across models in \cref{app:crossmodel_ablation}. GenCast is run only from 15 November 2021 because each two-year rollout takes about 24 GPU hours. The periods of the training, validation and test data of the nine models are listed in \cref{tab:crossmodel_training_periods}.

In \cref{fig:crossmodel_arctic,fig:crossmodel_global,fig:crossmodel_antarctic}, T2m is compared with ERA5 at matching valid times. Curves show weekly means of cosine-latitude-weighted regional means, and bands the weekly means of the instantaneous regional minimum and maximum; the polar regions comprise land and ocean poleward of $66.5^\circ$. For FCN3 + HADRIAN, curves and bands average the five members; for uncorrected FCN3, whose members are shown individually, the polar band follows one member and, once that member leaves the plotted range or ends, the longest-running member. In \cref{fig:crossmodel_arctic,fig:crossmodel_residual,fig:crossmodel_global,fig:crossmodel_antarctic}, each of the other five models is shown with one trajectory with and one without HADRIAN (for GenCast, both with the same random seed), and SFNO, Pangu and DLESyM with one trajectory each; the comparison reports no ensemble statistics.

\begin{table}[!htb]
\centering
\caption{\textbf{HADRIAN adaptations to the five other AI weather models.} $\beta$ and $c$ are the relaxation factor and cap of \cref{eq:hydrostatic_adjustment}; $r_k$ is the residual of layer $k$ and $r_k^{\max}$ its reference maximum; $\bar r_k$ is the largest regional RMS of $r_k$; $R$ is the large-scale residual of the 50--150-hPa layers divided by its reference maximum. 
Once the activation criterion is met, the correction stays on. FuXi has no activation criterion: its hydrostatic adjustment acts on each step with $D$ above a threshold slightly below $\tau$. 
The activation criteria of GraphCast and GenCast use the statistic $D$ of \cref{eq:residual_statistic} with a threshold of $172.036\,\mathrm{m^2\,s^{-2}}$, calibrated on ERA5 states of 2018--2019 in an earlier configuration, instead of $\tau=157.357\,\mathrm{m^2\,s^{-2}}$; because it is higher than $\tau$, these adaptations activate no earlier than they would with $\tau$. The model-specific limits are derived from climatologies.}
\label{tab:crossmodel_hosts}
\footnotesize
\renewcommand{\arraystretch}{1.25}
\begin{tabularx}{\textwidth}{@{}l>{\raggedright\arraybackslash}p{0.19\textwidth}>{\raggedright\arraybackslash}X>{\raggedright\arraybackslash}X@{}}
\toprule
Model & Activation & Hydrostatic adjustment & Model-specific limits \\
\midrule
GraphCast & $D>172.036\,\mathrm{m^2\,s^{-2}}$ & $\beta=0.5$; columns with $\lvert r_k\rvert>2r_k^{\max}$ & Ocean T2m within its climatological range, from the first step \\
GenCast & $D>172.036\,\mathrm{m^2\,s^{-2}}$ & $\beta=0.2$; change $\le100\,\mathrm{m^2\,s^{-2}}$ per step & 50- and 100-hPa wind and temperature within their climatological ranges; climatological bounds on moisture, spectral energy and increments \\
Aurora & $R>2$ on four consecutive steps & 50--150-hPa layers only & Stratospheric humidity poleward of $45^\circ$ within its climatological range \\
AIFS & $\bar r_k>2r_k^{\max}$ for any $k$ & $\beta=0.5$; columns with $\lvert r_k\rvert>2r_k^{\max}$ & Global-mean 300--1000-hPa thickness relaxed to climatology \\
FuXi & none & $\beta=0.1$, $c=250\,\mathrm{m^2\,s^{-2}}$ & none \\
\bottomrule
\end{tabularx}
\end{table}

\begin{table}[!htb]
\centering
\caption{\textbf{Periods of the training, validation and test data of the nine AI weather models.} Years as stated in the papers that describe the models~\citep{lam2023learning, price2025probabilisti, chen2023fuxi, bi2023accurate, cresswellclay2025deep}; data and abbreviations as in \cref{tab:training_periods}. The first four rows repeat \cref{tab:training_periods}, and its notes apply to them.}
\label{tab:crossmodel_training_periods}
\footnotesize
\renewcommand{\arraystretch}{1.15}
\begin{tabularx}{\textwidth}{@{}>{\raggedright\arraybackslash}p{0.16\textwidth}>{\raggedright\arraybackslash}X>{\raggedright\arraybackslash}p{0.13\textwidth}>{\raggedright\arraybackslash}p{0.22\textwidth}@{}}
\toprule
Model & Training & Validation & Test \\
\midrule
FCN3 & 1980--2016, fine-tuned on 2012--2016 & 2018--2021 & 2017 \\
SFNO & 1979--2015 & 2016--2017 & 2018 \\
Pretrained Aurora & 1979--2020, with other datasets & IFS HRES 2020 & 2021; IFS HRES 2022 \\
AIFS Single 1.1 & 1979--2022, fine-tuned on IFS operational analyses of 2016--2022 & not stated & 2023, against IFS operational analyses \\
GraphCast & 1979--2017, fine-tuned on IFS HRES 2016--2021 & 2016--2017 & 2018--2021 \\
GenCast & 1979--2018 & 2018 & 2019 \\
FuXi & 1979--2015 & 2016--2017 & 2018 \\
Pangu & 1979--2017 & 2019 & 2018 \\
DLESyM & 1983--2016 & 2016--2017 & not stated \\
\bottomrule
\end{tabularx}
\par\smallskip
\begin{minipage}{\textwidth}\footnotesize
GraphCast is run with its operational checkpoint, which according to its checkpoint description was fine-tuned on IFS HRES data of 2016--2021. In the GraphCast paper, the model validated on 2016--2017 was trained on 1979--2015, and the model tested on 2018--2021 was trained on 1979--2017 without fine-tuning. The GenCast model validated on 2018 was trained on 1979--2017 and was retrained on 1979--2018 before it was tested on 2019. The DLESyM years are those of its atmosphere model; the card of the released checkpoint gives 1980--2015 for training, 2016--2017 for testing and 2018--2019 for evaluation.
\end{minipage}
\end{table}

\subsection{Hydrostatic Imbalance and Its Correction}\label{app:crossmodel_hydrostatic}
\begin{figure}[!htbp]
  \centering
  \includegraphics[width=\textwidth]{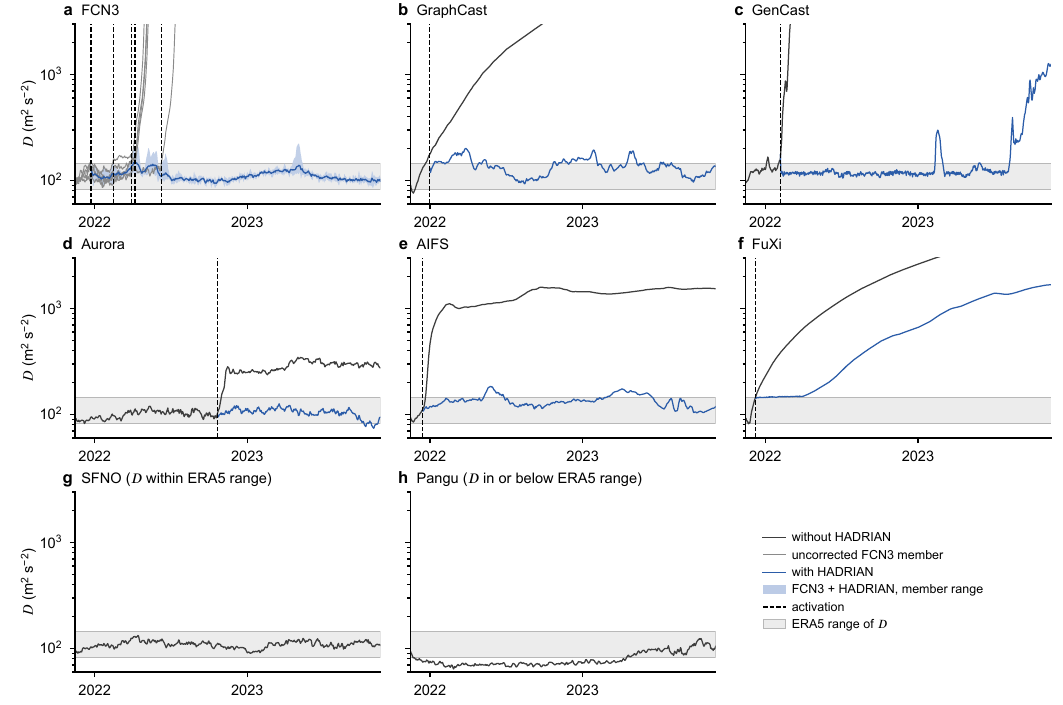}
  \caption{\textbf{Global hydrostatic residual $D$ with and without HADRIAN.} Two-year rollouts from 15 November 2021, as in \cref{fig:crossmodel_arctic,fig:crossmodel_global}. (a--f)~Models run with and without HADRIAN; (g,h)~SFNO and Pangu, run without HADRIAN because their $D$ stays below the ERA5 maximum and hence below the activation threshold $\tau$. Grey curves show uncorrected rollouts (FCN3: five members individually); blue curves show rollouts corrected with HADRIAN, with $D$ of the raw prediction, as in \cref{fig:imbalance_hadrian}c (FCN3: five-member geometric mean and member range). For AIFS, $D$ of the raw prediction is not recorded after activation, and its curve after activation shows $D$ of the corrected state. Dashed vertical lines mark activation, and the grey band shows the 2000--2015 ERA5 range of $D$. Curves are daily geometric means of $D$ recorded during the rollout (\cref{sec:method}). DLESyM is not shown because $D$ cannot be computed from its predicted variables (\cref{app:crossmodel_reference}).}
  \label{fig:crossmodel_residual}
\end{figure}

In FCN3, $D$ of the five uncorrected members grows until the rollouts fail; with HADRIAN, the five-member mean stays at 94--139~m$^2$\,s$^{-2}$ after the last member is activated, near the middle of the ERA5 range, as for all 20 members in \cref{sec:failure} (\cref{fig:crossmodel_residual}a). In uncorrected GraphCast, $D$ crosses $\tau$ on day 41 and grows to $94\tau$, together with a whole-column energy loss; with HADRIAN, activated on day 46, $D$ stays between $0.59\tau$ and $1.3\tau$ and exceeds $\tau$ on 119 of the 684 days after activation (\cref{fig:crossmodel_residual}b). In uncorrected GenCast, $D$ first exceeds $\tau$ on day 52 and exceeds it continuously from day 83 as the imbalance grows in the 50--100-hPa layer, and it grows by more than two orders of magnitude before the rollout fails in its tenth month; with HADRIAN, activated on day 83, $D$ stays within the ERA5 range for most of the first 20 months and then rises above $\tau$ (\cref{fig:crossmodel_residual}c; \cref{app:crossmodel_gencast}).

In uncorrected Aurora, $D$ crosses $\tau$ on day 351 (1 November 2022), when the imbalance arises in the 50--150-hPa layers, and stays above $\tau$, at up to $2.2\tau$, for the rest of the rollout; with HADRIAN, activated on day 341, $D$ stays between $0.47\tau$ and $0.80\tau$, within or below the ERA5 range (\cref{fig:crossmodel_residual}d). From November 2022 to March 2023, the monthly-mean residual of the 50--100-hPa layer is 277--300~m$^2$\,s$^{-2}$ in the corrected state, against 729--776~m$^2$\,s$^{-2}$ without correction and a 2000--2015 ERA5 median of 292~m$^2$\,s$^{-2}$.

In uncorrected AIFS, the imbalance arises in near-polar columns; $D$ crosses $\tau$ on day 37 and reaches $10\tau$. HADRIAN activates on day 28, before $D$ reaches $\tau$, and afterwards $D$ of the corrected state stays between $0.66\tau$ and $1.17\tau$, exceeding $\tau$ on 102 of 702 days (\cref{fig:crossmodel_residual}e). In uncorrected FuXi, $D$ crosses $\tau$ on day 26 and grows to $36\tau$; with HADRIAN, $D$ stays near the upper end of the ERA5 range until day 152 and then grows to almost $11\tau$ (\cref{fig:crossmodel_residual}f; \cref{app:crossmodel_fuxi}).

\FloatBarrier

\subsection{Global-Mean Temperature}\label{app:crossmodel_global}
\begin{figure}[!htbp]
  \centering
  \includegraphics[width=\textwidth,pagebox=cropbox]{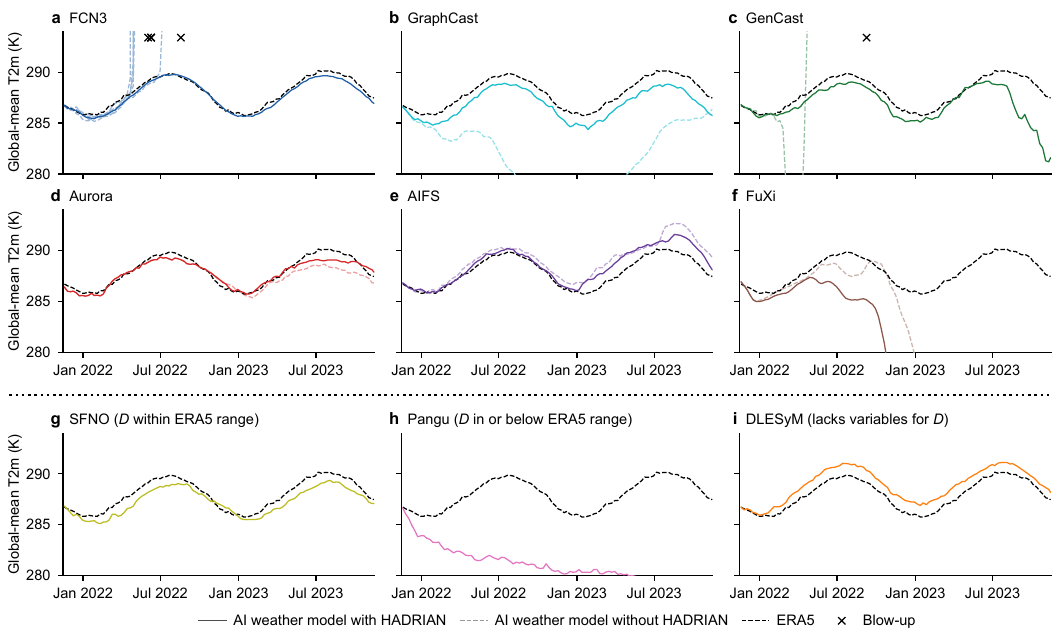}
  \caption{\textbf{Global-mean T2m with and without HADRIAN.} Weekly means for two-year rollouts from 15 November 2021. (a--f)~Six models with HADRIAN (solid) and without HADRIAN (light dashed); FCN3 + HADRIAN is shown as the five-member mean, and uncorrected FCN3 members individually. (g--i)~SFNO, Pangu and DLESyM, run without HADRIAN. Dashed black curves show ERA5; crosses mark blow-up.}
  \label{fig:crossmodel_global}
\end{figure}

With HADRIAN, offsets from ERA5 remain in the global mean (\cref{fig:crossmodel_global}a--f). FCN3 + HADRIAN follows both seasonal cycles with a small cold offset, and GraphCast + HADRIAN remains colder than ERA5, although most of the cold drift of the uncorrected rollout is removed. The second summer maximum of Aurora + HADRIAN stays below ERA5; GenCast and FuXi are discussed in Sections~\ref{app:crossmodel_gencast} and~\ref{app:crossmodel_fuxi}. Without HADRIAN (\cref{fig:crossmodel_global}g--i), SFNO and DLESyM retain the seasonal cycle with cold and warm offsets, respectively, whereas Pangu cools well below ERA5 within the first months.

\FloatBarrier

\subsection{Antarctic Temperature}\label{app:crossmodel_antarctic}
\begin{figure}[!htbp]
  \centering
  \includegraphics[width=\textwidth,pagebox=cropbox]{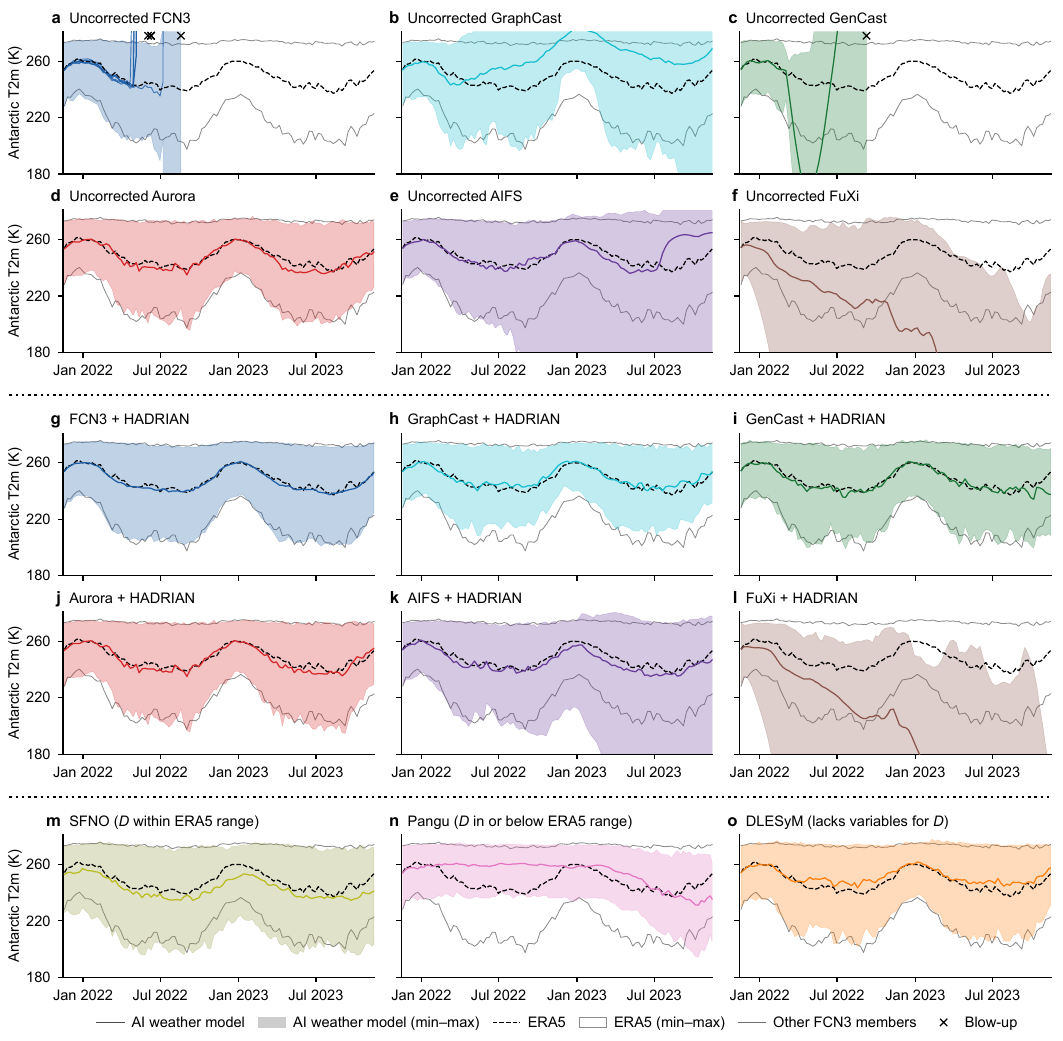}
  \caption{\textbf{Antarctic T2m with and without HADRIAN.} As \cref{fig:crossmodel_arctic}, for land and ocean poleward of $66.5^\circ$S.}
  \label{fig:crossmodel_antarctic}
\end{figure}

The Antarctic comparison (\cref{fig:crossmodel_antarctic}) shows the same improvements as the Arctic one (\cref{fig:crossmodel_arctic}) and differs from it for AIFS and Aurora. FCN3 + HADRIAN follows the ERA5 regional mean and spatial range through both years, HADRIAN largely removes the warm drift of GraphCast, and GenCast + HADRIAN follows the seasonal cycle until its cold departure late in the second year. For AIFS, HADRIAN keeps the Antarctic regional mean close to ERA5 late in the second year, when the Arctic mean falls below ERA5, although the regional minimum remains below that of ERA5. Aurora retains the Antarctic seasonal cycle with a winter cold bias with or without HADRIAN, whereas HADRIAN reduces its Arctic cold bias. Without HADRIAN (\cref{fig:crossmodel_antarctic}m--o), SFNO is generally too cold, DLESyM is too warm in winter, and Pangu has a distorted seasonal cycle.

\FloatBarrier

\subsection{Ablation Across Models}\label{app:crossmodel_ablation}
\begin{figure}[!htbp]
  \centering
  \includegraphics[width=\textwidth]{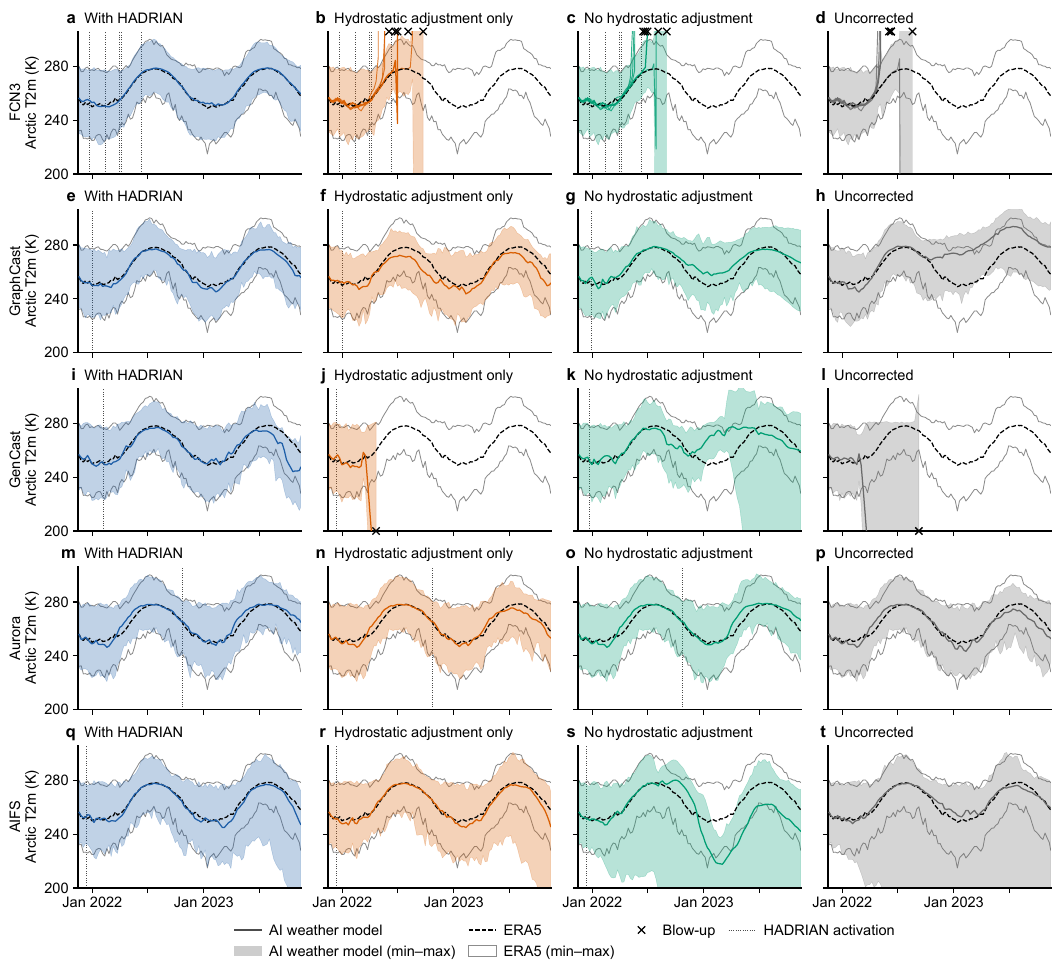}
  \caption{\textbf{Arctic T2m in the ablation across models.} Two-year rollouts from 15 November 2021 with HADRIAN, with hydrostatic adjustment only, without hydrostatic adjustment and uncorrected (columns), for (a--d)~FCN3, (e--h)~GraphCast, (i--l)~GenCast, (m--p)~Aurora and (q--t)~AIFS. Curves, bands and ERA5 as in \cref{fig:crossmodel_arctic}. Dotted vertical lines mark activation; crosses mark blow-up.}
  \label{fig:crossmodel_ablation_arctic}
\end{figure}

\begin{figure}[!htb]
  \centering
  \includegraphics[width=\textwidth]{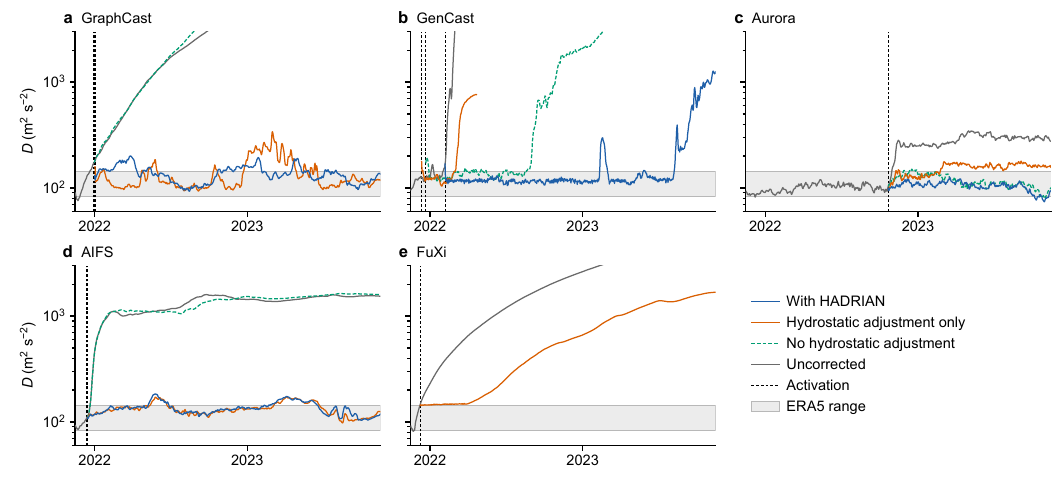}
  \caption{\textbf{Global hydrostatic residual $D$ in the ablation across models.} Two-year rollouts from 15 November 2021 in the configurations of \cref{fig:crossmodel_ablation_arctic}, for (a)~GraphCast, (b)~GenCast, (c)~Aurora and (d)~AIFS, and for (e)~FuXi, whose adaptation consists of hydrostatic adjustment alone, with hydrostatic adjustment only and uncorrected. Curves are daily geometric means of $D$ of the raw prediction, as in \cref{fig:crossmodel_residual}; for AIFS after activation and for Aurora without hydrostatic adjustment, they show $D$ of the corrected state. Dashed vertical lines mark activation, and the grey band shows the ERA5 range of $D$.}
  \label{fig:crossmodel_ablation_residual}
\end{figure}

As a preliminary test of the two parts of the adaptations, the two-year rollouts of FCN3, GraphCast, GenCast, Aurora and AIFS from 15 November 2021 are repeated with hydrostatic adjustment only and without hydrostatic adjustment, the latter keeping the model-specific limits (\cref{fig:crossmodel_ablation_arctic,fig:crossmodel_ablation_residual}; \cref{tab:crossmodel_ablation}). For FCN3, both configurations are those of \cref{app:ablation_details}, where the configuration without hydrostatic adjustment is named ``No hydrostatic adjustment''. FuXi, whose adaptation consists of hydrostatic adjustment alone (\cref{tab:crossmodel_hosts}), is included in \cref{fig:crossmodel_ablation_residual} with and without hydrostatic adjustment. Each configuration is represented by one trajectory, except for FCN3 with five members. The GraphCast, GenCast and AIFS rollouts are not bitwise reproducible and differ from the uncorrected rollouts from the first step.

Without hydrostatic adjustment, $D$ exceeds the ERA5 maximum on 100\% of the days after activation in GraphCast, 99\% in AIFS and 71\% in GenCast, and throughout days 365--730 in all three, as in uncorrected GraphCast and AIFS (\cref{tab:crossmodel_ablation}; \cref{fig:crossmodel_ablation_residual}); for FCN3, see \cref{app:ablation_details}. In Aurora, the humidity limit alone keeps $D$ below the ERA5 maximum on 95\% of the days after activation. Hydrostatic adjustment alone keeps $D$ near the ERA5 range in GraphCast and AIFS but not in Aurora or FuXi; with it alone, the GenCast rollout ends on day 157.5 and all five FCN3 members fail (\cref{fig:crossmodel_ablation_arctic}).

\begin{table}[!htb]
\centering
\caption{\textbf{Hydrostatic residual in the ablation across models.}}
\label{tab:crossmodel_ablation}
\footnotesize
\begin{tabular}{lcccc}
\toprule
Model & With HADRIAN & Hydrostatic adjustment only & No hydrostatic adjustment & Uncorrected \\
\midrule
GraphCast & 137 (35\%) & 116 (27\%) & 6{,}684 (100\%) & 4{,}683 (100\%) \\
GenCast & 119 (19\%) & 133 (40\%)$^{\mathrm{a}}$ & 1{,}865 (71\%) & 35{,}981 (100\%)$^{\mathrm{b}}$ \\
Aurora & 106 (0\%) & 159 (69\%) & 114 (5\%)$^{\mathrm{c}}$ & 287 (98\%) \\
AIFS & 133 (23\%)$^{\mathrm{c}}$ & 131 (21\%)$^{\mathrm{c}}$ & 1{,}451 (99\%)$^{\mathrm{c}}$ & 1{,}434 (99\%) \\
FuXi & --$^{\mathrm{d}}$ & 592 (98\%) & --$^{\mathrm{d}}$ & 2{,}332 (100\%) \\
\bottomrule
\end{tabular}
\par\smallskip
\begin{minipage}{\textwidth}\footnotesize
Median of the daily geometric means of $D$ from activation to the end of each rollout from 15 November 2021, in m$^2$\,s$^{-2}$; in parentheses, the share of these days on which $D$ exceeds the ERA5 maximum of 144.7~m$^2$\,s$^{-2}$. For each uncorrected rollout, the period starts on the activation day of the rollout with HADRIAN of the same model (for FuXi, of the rollout with hydrostatic adjustment only). $D$ of the raw prediction, as in \cref{fig:crossmodel_residual}. $^{\mathrm{a}}$Rollout ends on day 157.5. $^{\mathrm{b}}$Rollout fails in its tenth month, on day 297.5 (\cref{app:crossmodel_hydrostatic}). $^{\mathrm{c}}$$D$ of the corrected state. $^{\mathrm{d}}$The FuXi adaptation consists of hydrostatic adjustment alone (\cref{tab:crossmodel_hosts}), so its rollout with HADRIAN is that with hydrostatic adjustment only; FuXi has no model-specific limits.
\end{minipage}
\end{table}
\FloatBarrier

\subsection{Why HADRIAN Does Not Stabilize FuXi}\label{app:crossmodel_fuxi}
Hydrostatic adjustment is the only stage of the FuXi adaptation (\cref{tab:crossmodel_hosts}). In the two-year rollout from 15 November 2021, it delays the first crossing of $\tau$ by $D$ from day 26 to day 152 but does not delay failure: the weekly global-mean T2m departure from ERA5 on day 365 is $-12.9$~K with HADRIAN and $-1.0$~K without it (\cref{fig:crossmodel_global}f). The per-step limit on the geopotential adjustment is reached in 59\% of columns when the correction is first applied and in 63\% of columns in the second year.

The rollout failure of uncorrected FuXi does not originate in a single balance relation: in the rollout from 15 November 2021, hydrostatic and geostrophic balance break down in succession, interleaved with surface and mass drift. From day 5, 500-hPa geopotential at wavelengths shorter than 250~km has more energy than in ERA5. The 400--500-hPa layer residual leaves its ERA5 range on day 11, 5 days before the minimum T2m of the Southern Hemisphere extratropics first leaves its climatological range (day 16), whereas $D$ crosses $\tau$ only on day 26. The wind field is at first over-smoothed rather than unbalanced, so geostrophic imbalance does not precede hydrostatic imbalance: the 500-hPa ageostrophic wind reaches its minimum, 0.52 and 0.48 times that of ERA5 in the Northern and Southern Hemispheres, on days 12 and 13, and leaves the ERA5 daily range only on day 45 in the Northern Hemisphere and day 60 in the Southern Hemisphere.

\subsection{Late Degradation of GenCast + HADRIAN}\label{app:crossmodel_gencast}
During roughly the first 600 days of the GenCast + HADRIAN rollout from 15 November 2021, the weekly means of the global-mean departures from ERA5 lie between $-1.4$ and $+0.2$~K for T2m, between $-1.3$ and $+0.7$~K for T850 and between $-0.6$ and $+1.4$~K for the 100-hPa temperature. From activation to day 600, the daily geometric mean of the global residual lies within the ERA5 range on 96\% of days and exceeds $\tau$ only on days 454--470, in February 2023, when it reaches 298~m$^2$\,s$^{-2}$ (\cref{fig:crossmodel_residual}c). Uncorrected GenCast from the same initialization fails in its tenth month.

From about day 600, three changes follow within 60 days. Local 100-hPa temperatures first exceed 280~K on day 636, and from day 644 the weekly means of the global-mean 100-hPa temperature are 2.2--25.8~K above ERA5. The daily geometric mean of the global residual stays above $\tau$ from day 633, reaches 394~m$^2$\,s$^{-2}$ on day 638 and first exceeds 1000~m$^2$\,s$^{-2}$ on day 709. The weekly means of the global-mean T2m departure deepen from between $-0.4$ and $-1.4$~K on days 560--602 to between $-1.3$ and $-2.9$~K on days 602--644, and to between $-3.7$ and $-6.4$~K thereafter. Two explanations remain untested: the imbalance generated at each step may grow beyond what the limited per-step correction can remove, or a slow cold drift in GenCast's predicted sea-surface temperature may cool the whole column.

\subsection{Models Run Without HADRIAN}\label{app:crossmodel_reference}
Because HADRIAN leaves states unchanged until it activates, $D$ of an uncorrected rollout shows whether HADRIAN would activate: in SFNO and Pangu, it never reaches $\tau$ (\cref{fig:crossmodel_residual}g,h), and in DLESyM it cannot be computed from the predicted variables. Below, $D$ is compared with its 2000--2015 ERA5 range of 83.2--144.7~m$^2$\,s$^{-2}$ (median 109.9~m$^2$\,s$^{-2}$).

\paragraph{SFNO.}
In two-year uncorrected rollouts from the four 2021 initialization dates, $D$ of SFNO has annual means of 108.2--110.9~m$^2$\,s$^{-2}$, never leaves the 1st--99th percentile range of ERA5 (89.3--136.6~m$^2$\,s$^{-2}$) and peaks at $0.85\tau$; no layer residual grows. SFNO damps small scales at every step~\citep{lehmann2026can}, and its smoothed states keep geopotential and temperature consistent; its 850-hPa zonal-wind power at zonal wavenumbers 40--80 settles at 0.62 times that of ERA5. Its errors are also hydrostatically consistent and therefore do not raise the residual: the winter Arctic 50-hPa temperature is 19--31~K too warm and the polar vortex nearly disappears, yet the Northern Hemisphere 50--100-hPa residual does not rise.

\paragraph{Pangu.}
In the two-year uncorrected rollout from 15 November 2021, $D$ of Pangu stays at or below ERA5 levels (annual means 71.6 and 88.4~m$^2$\,s$^{-2}$; maximum 125.4~m$^2$\,s$^{-2}$). Consistent with medium-range results for other deterministic models \citep{kasteleyn2026physmetrics}, Pangu is at least as hydrostatically balanced as ERA5, and in the rollouts of \citet{lehmann2026can} its small-scale energy is below that of ERA5 for eight of nine variables. Only the 400--500-hPa layer residual exceeds its ERA5 maximum, at 1.4 times the ERA5 median over the first ten days and 1.7 times in the second year; because the ERA5 median for this layer is only 19~m$^2$\,s$^{-2}$, it barely affects the global value. Pangu instead cools throughout the column and loses most of its seasonal cycle: T850 falls to 5.4~K below ERA5 in the first year and 8.4~K below in the second, 500-hPa geopotential height falls by 102 and 152~m with it, and the second-year Arctic T2m seasonal amplitude shrinks to 5.7~K, against 28.4~K in ERA5. This whole-column cooling remains hydrostatically consistent and leaves hydrostatic adjustment no imbalance to remove: if the 1000--500-hPa layer cools by the 8.4~K seen at 850~hPa and the 1000-hPa geopotential height is unchanged, the hypsometric equation gives a 500-hPa geopotential height decrease of about 170~m, comparable with the 152~m observed in the second year.

\paragraph{DLESyM.}
DLESyM predicts geopotential at 1000, 500 and 250~hPa and the 300--700-hPa thickness, but temperature only at 850~hPa and 2~m and moisture only as TCWV. No layer therefore has both bounding geopotentials and an interior virtual-temperature profile, so the residual in \cref{eq:hydrostatic_residual} cannot be evaluated: HADRIAN would have no activation criterion, and hydrostatic adjustment no target thickness. Approximating the 1000--500-hPa layer temperature by T850 would rely on an empirical lapse-rate relation rather than a balance constraint. Its two-year rollout from 15 November 2021 remains finite and passes the available checks, but several of the evaluated variables are missing, so the outcome is not comparable with those of the other models.

% Appendix F: two-year climatology across four models.
\clearpage
\begingroup
% Compact float spacing for the three-table/two-figure appendix section.
\setlength{\intextsep}{6pt plus 2pt minus 1pt}
\setlength{\textfloatsep}{8pt plus 2pt minus 1pt}
\setlength{\floatsep}{8pt plus 2pt minus 1pt}
\section{Two-Year Climatology}
\label{app:climatology}

\Cref{tab:clim_mean,tab:clim_seasons,tab:clim_years} and \cref{fig:clim_mean_maps,fig:clim_storm_maps} compare the climate of all 32 two-year 2021 trajectories of \cref{tab:survival} with ERA5 over 22 complete calendar months per trajectory, excluding the partial months at either end and the first complete calendar month after initialization. These scores use six-hourly states, whereas the biases and RMSE of \cref{tab:survival} use daily 00~UTC states on days 1--730; the values therefore differ, for example the T2m bias of FCN3 + HADRIAN is -0.143~K here and -0.137~K in \cref{tab:survival}.

Biases and storm-track variance ratios are means of per-trajectory values, whereas the RMSE of FCN3 + HADRIAN is computed from the five-member ensemble-mean field, which is smoother than the single trajectory of each comparison model (\cref{sec:method}). The storm-track variance ratios of FCN3 + HADRIAN, which stay close to one (\cref{tab:clim_years,fig:clim_storm_maps}), therefore show that its lower ensemble-mean errors are not accompanied by reduced synoptic-scale variability in individual members.

\begin{table}[!ht]
\caption{\textbf{Two-year mean-state errors.}
Area-weighted bias and RMSE of the 22-month mean field relative to ERA5;
U200 is the 200-hPa zonal wind. Bold and underline mark the smallest
and second-smallest absolute bias or RMSE. --, not predicted by the model;
for TCWV, only the better value is marked.}
\label{tab:clim_mean}
\centering
\fontsize{9}{11}\selectfont
\setlength{\tabcolsep}{2.7pt}\renewcommand{\arraystretch}{1.02}
\begin{tabular*}{\linewidth}{@{\extracolsep{\fill}}llrrrr@{}}
\toprule
Variable (unit) & Metric & FCN3 + HADRIAN & Aurora & SFNO & AIFS \\
\midrule
T2m (K) & Bias & \textbf{-0.143} & \underline{-0.378} & -0.550 & +1.325  \\
 & RMSE & \textbf{0.662} & \underline{1.662} & 2.004 & 2.594 \\
\addlinespace[1.5pt]
T850 (K) & Bias & \textbf{-0.175} & \underline{-0.523} & -0.605 & +1.392  \\
 & RMSE & \textbf{0.607} & \underline{1.317} & 1.780 & 2.135 \\
\addlinespace[1.5pt]
MSLP (hPa) & Bias & -0.100 & +0.724 & \textbf{-0.008} & \underline{+0.097} \\
 & RMSE & \textbf{2.039} & \underline{2.057} & 2.436 & 4.626 \\
\addlinespace[1.5pt]
Z500 (m$^2$\,s$^{-2}$) & Bias & \underline{-53.380} & \textbf{-49.972} & -181.235 & +256.863  \\
 & RMSE & \textbf{183.386} & \underline{290.994} & 368.933 & 523.302  \\
\addlinespace[1.5pt]
U850 (m\,s$^{-1}$) & Bias & -0.087 & \textbf{+0.002} & \underline{-0.026} & -0.386  \\
 & RMSE & \textbf{1.135} & 1.656 & \underline{1.607} & 2.430 \\
\addlinespace[1.5pt]
U200 (m\,s$^{-1}$) & Bias & \underline{-0.241} & -0.418 & -2.491 & \textbf{-0.187}  \\
 & RMSE & \textbf{2.957} & \underline{4.793} & 5.083 & 5.233 \\
\addlinespace[1.5pt]
10-m wind speed (m\,s$^{-1}$) & Bias & \textbf{-0.049} & -0.463 & \underline{-0.266} & -1.535 \\
 & RMSE & \textbf{0.361} & 0.727 & \underline{0.577} & 1.867 \\
\addlinespace[1.5pt]
TCWV (kg\,m$^{-2}$) & Bias & \textbf{+0.069} & -- & -0.475 & -- \\
 & RMSE & \textbf{1.559} & -- & 3.146 & -- \\
\bottomrule
\end{tabular*}\space\space
\end{table}

\begin{table}[!ht]
\caption{\textbf{RMSE of seasonal-mean fields.}
DJF, December--February; MAM, March--May; JJA, June--August; SON, September--November.
Only complete three-month seasons are included, and a season that occurs twice for an
initialization date is averaged over both occurrences, weighted by sample count.
RMSE is computed and ranked as in \cref{tab:clim_mean}.}
\label{tab:clim_seasons}
\centering
\fontsize{9}{11}\selectfont
\setlength{\tabcolsep}{2.7pt}\renewcommand{\arraystretch}{1.02}
\begin{tabular*}{\linewidth}{@{\extracolsep{\fill}}llrrrr@{}}
\toprule
Season & Variable (unit) & FCN3 + HADRIAN & Aurora & SFNO & AIFS \\
\midrule
DJF & T2m (K) & \textbf{1.048} & 2.562 & \underline{2.406} & 3.186  \\
 & Z500 (m$^2$\,s$^{-2}$) & \textbf{329.414} & 546.548 & \underline{417.515} & 671.201  \\
\addlinespace[2pt]
MAM & T2m (K) & \textbf{0.953} & 2.676 & \underline{1.680} & 3.046  \\
 & Z500 (m$^2$\,s$^{-2}$) & \textbf{228.827} & 492.595 & \underline{424.656} & 473.155  \\
\addlinespace[2pt]
JJA & T2m (K) & \textbf{0.917} & \underline{2.268} & 2.422 & 3.680  \\
 & Z500 (m$^2$\,s$^{-2}$) & \textbf{225.581} & \underline{370.857} & 456.908 & 640.263  \\
\addlinespace[2pt]
SON & T2m (K) & \textbf{0.950} & \underline{2.065} & 3.559 & 3.828  \\
 & Z500 (m$^2$\,s$^{-2}$) & \textbf{290.563} & \underline{372.574} & 614.007 & 917.812  \\
\bottomrule
\end{tabular*}\space\space
\end{table}

\begin{figure}[!ht]
\centering
\includegraphics[width=\linewidth]{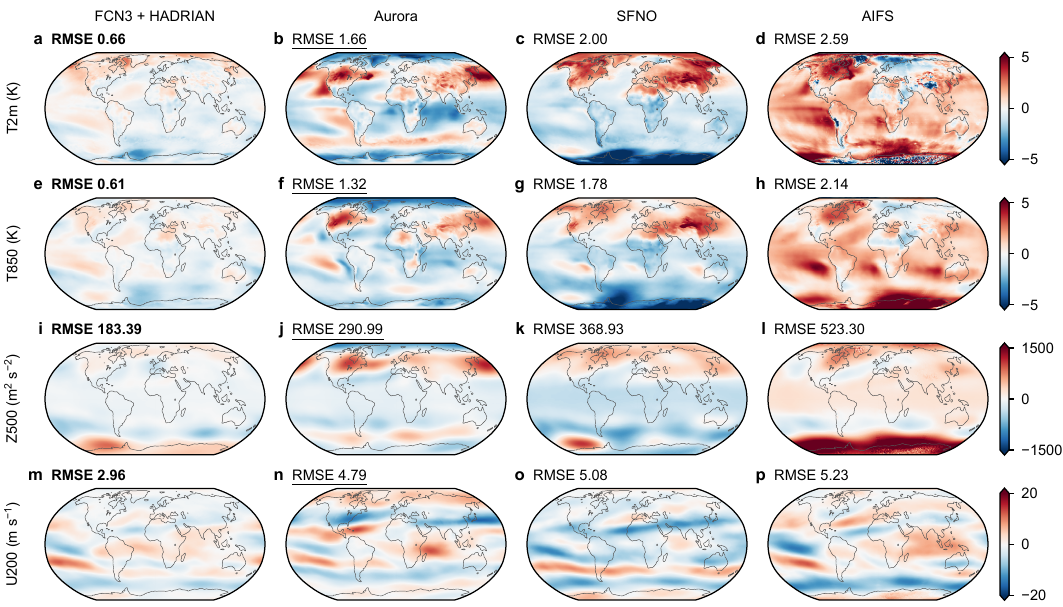}
\caption{\textbf{Two-year mean-state biases.}
22-month mean T2m, T850, Z500 and U200 minus ERA5 for each model, averaged over
members and initialization dates. Panel titles give the RMSE from
\cref{tab:clim_mean} (bold, lowest; underlined, second lowest).}
\label{fig:clim_mean_maps}
\end{figure}

\begin{table}[!ht]
\caption{\textbf{Mean-state errors and synoptic-scale variability in the first and second rollout years.} Both years use the same ten complete calendar months for every initialization date. Storm-track variance ratios are computed from the six-hourly Z500 fields of each year (\cref{sec:method}). Ranking as in \cref{tab:clim_mean}, with ratios ranked by closeness to one.}
\label{tab:clim_years}
\centering
\fontsize{9}{11}\selectfont
\setlength{\tabcolsep}{2.7pt}\renewcommand{\arraystretch}{1.02}
\begin{tabular*}{\linewidth}{@{\extracolsep{\fill}}llrrrr@{}}
\toprule
Metric & Unit & FCN3 + HADRIAN & Aurora & SFNO & AIFS \\
\midrule
\multicolumn{6}{@{}l}{\textbf{First rollout year}} \\
T2m bias & K & \textbf{-0.114} & \underline{-0.193} & -0.538 & +0.596  \\
T2m RMSE & K & \textbf{0.731} & 1.776 & 2.022 & \underline{1.597}  \\
Z500 RMSE & m$^2$\,s$^{-2}$ & \textbf{210.364} & 320.998 & 383.320 & \underline{281.877}  \\
Storm-track variance ratio, NH & 1 & \textbf{0.922} & 0.671 & \underline{0.794} & 0.759  \\
Storm-track variance ratio, SH & 1 & \underline{0.952} & \textbf{0.963} & 0.884 & 0.819  \\
\midrule
\multicolumn{6}{@{}l}{\textbf{Second rollout year}} \\
T2m bias & K & \textbf{-0.171} & \underline{-0.569} & -0.587 & +1.919  \\
T2m RMSE & K & \textbf{0.784} & \underline{1.914} & 2.117 & 4.271  \\
Z500 RMSE & m$^2$\,s$^{-2}$ & \textbf{209.969} & \underline{335.886} & 393.309 & 919.731  \\
Storm-track variance ratio, NH & 1 & \textbf{0.950} & 0.682 & \underline{0.817} & 0.486  \\
Storm-track variance ratio, SH & 1 & \textbf{0.918} & 0.730 & \underline{0.796} & 0.047  \\
\bottomrule
\end{tabular*}\space\space
\end{table}

\begin{figure}[!ht]
\centering
\includegraphics[width=\linewidth]{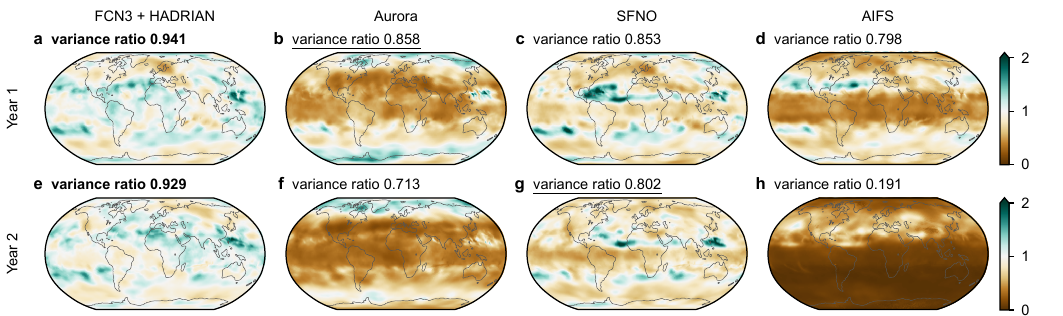}
\caption{\textbf{Synoptic-scale Z500 variability in the first and second rollout years.} Ratio of the variance of 2--6-day band-pass-filtered Z500 at each grid point to that of ERA5 (\cref{sec:method}), averaged over trajectories, for the same months as in \cref{tab:clim_years}. Panel titles give the storm-track variance ratio computed over both hemispheres poleward of $30^\circ$ together (bold, closest to one; underlined, second closest), unlike the separate NH and SH values of \cref{tab:clim_years}.}
\label{fig:clim_storm_maps}
\end{figure}
\clearpage
\endgroup

% Appendix G: extratropical cyclones at lead times of 6--18 months.
\clearpage
\begingroup
% Local spacing as in the other appendices.
\setstretch{1.0}
\setlength{\intextsep}{6pt plus 2pt minus 1pt}
\setlength{\textfloatsep}{8pt plus 2pt minus 1pt}
\setlength{\floatsep}{8pt plus 2pt minus 1pt}
\section{Extratropical Cyclones at Lead Times of 6--18 Months}
\label{app:cyclones}

\begin{figure}[H]
\centering
\includegraphics[width=\linewidth]{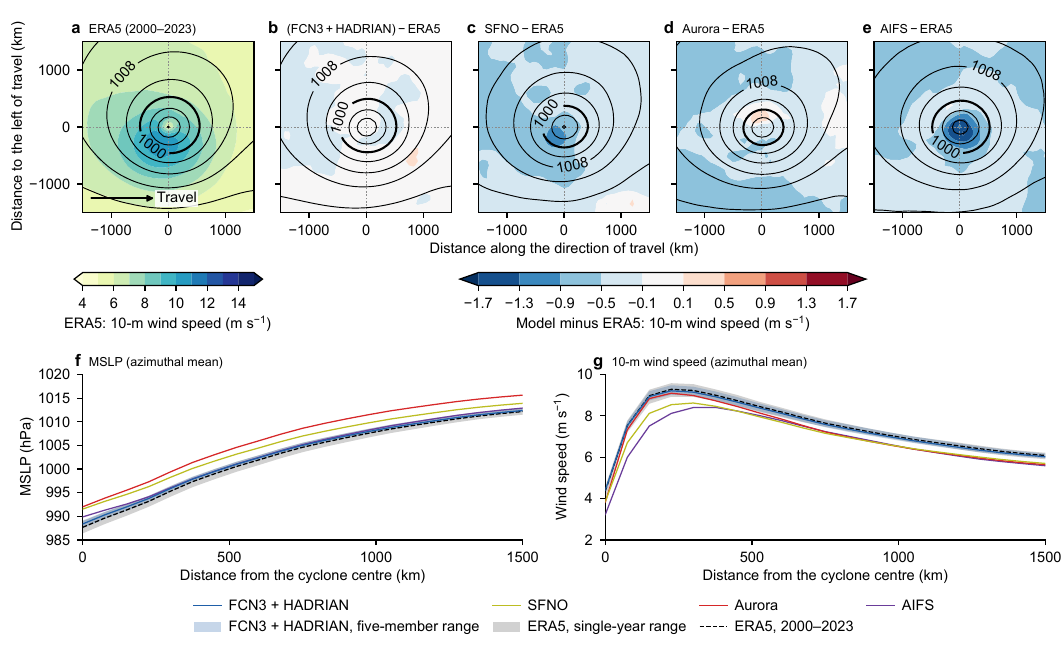}
\caption{\textbf{Cyclone-centred composites at the time of minimum central MSLP.}
Averages over all cyclones first detected over the North Atlantic and Europe at
lead times of 6--18 months, centred on each cyclone and rotated so that its
direction of travel points to the right~\citep{catto2010can}. (a) ERA5 over
2000--2023: 10-m wind speed (shading) and MSLP (contours every 4~hPa). (b--e)
10-m wind speed of each model minus that of ERA5 (shading, white within
$\pm$0.1~m\,s$^{-1}$) and MSLP of the model (contours); 95\% of the differences
between single ERA5 years and their 2000--2023 mean are smaller than
0.33~m\,s$^{-1}$. (f,g) Azimuthal means of MSLP and 10-m wind speed. The blue
band shows the FCN3 + HADRIAN five-member range, the range over its five random
seeds, each with one member per initialization date, and the grey band the range
of the 24 single ERA5 years. The composites contain 30,217 (ERA5), 6,266
(FCN3 + HADRIAN), 1,054 (SFNO), 943 (Aurora) and 757 (AIFS) cyclones.}
\label{fig:cyclone_composites}
\end{figure}

\Cref{fig:cyclone_composites,fig:cyclone_stats} compare extratropical cyclones
over the North Atlantic and Europe at lead times of 6--18 months. FCN3 + HADRIAN
reproduces the structure, number and intensity of ERA5 cyclones within the
year-to-year range of ERA5, whereas SFNO, Aurora and AIFS produce fewer cyclones
with weaker winds. Uncorrected FCN3 is not shown because 18 of its 20 members
blow up within the first year (\cref{sec:failure}).

Cyclones are tracked with the same settings in the rollouts from the four 2021
initialization dates and in ERA5~\citep{hersbach2020era5} for the same windows
in each initialization year from 2000 to 2023, whose range over these 24 years
gives the year-to-year variation. A cyclone is a local MSLP minimum enclosed by a
closed isobar 2~hPa above its centre within
$6^\circ$~\citep{ullrich2021tempestextremes}, linked every six hours to the
nearest centre within 750~km of its extrapolated
position~\citep{hanley2012objective}, first detected within
$30^\circ$--$75^\circ$N, $80^\circ$W--$40^\circ$E, and lasting at least 24~h over
a path of at least 1,000~km~\citep{neu2013imilast,crawford2021sensitivity}.

At the time of minimum central MSLP, the composite 10-m wind speed of
FCN3 + HADRIAN differs from that of ERA5 by less than 0.24~m\,s$^{-1}$
(\cref{fig:cyclone_composites}). Over 6--18 months, FCN3 + HADRIAN produces 0.99
times as many cyclones as ERA5 and 0.99 times its strong-wind area, against
0.60--0.84 and 0.41--0.58 for the comparison models (\cref{fig:cyclone_stats}a,b);
this contrast persists across 12 variations of the tracking settings. Part of it
reflects the probabilistic design of FCN3 rather than the correction, because
deterministic models also underestimate cyclone winds in forecasts of up to
10 days~\citep{charltonperez2024do,dacre2025northern}.

\begin{figure}[p]
\centering
\includegraphics[width=\linewidth]{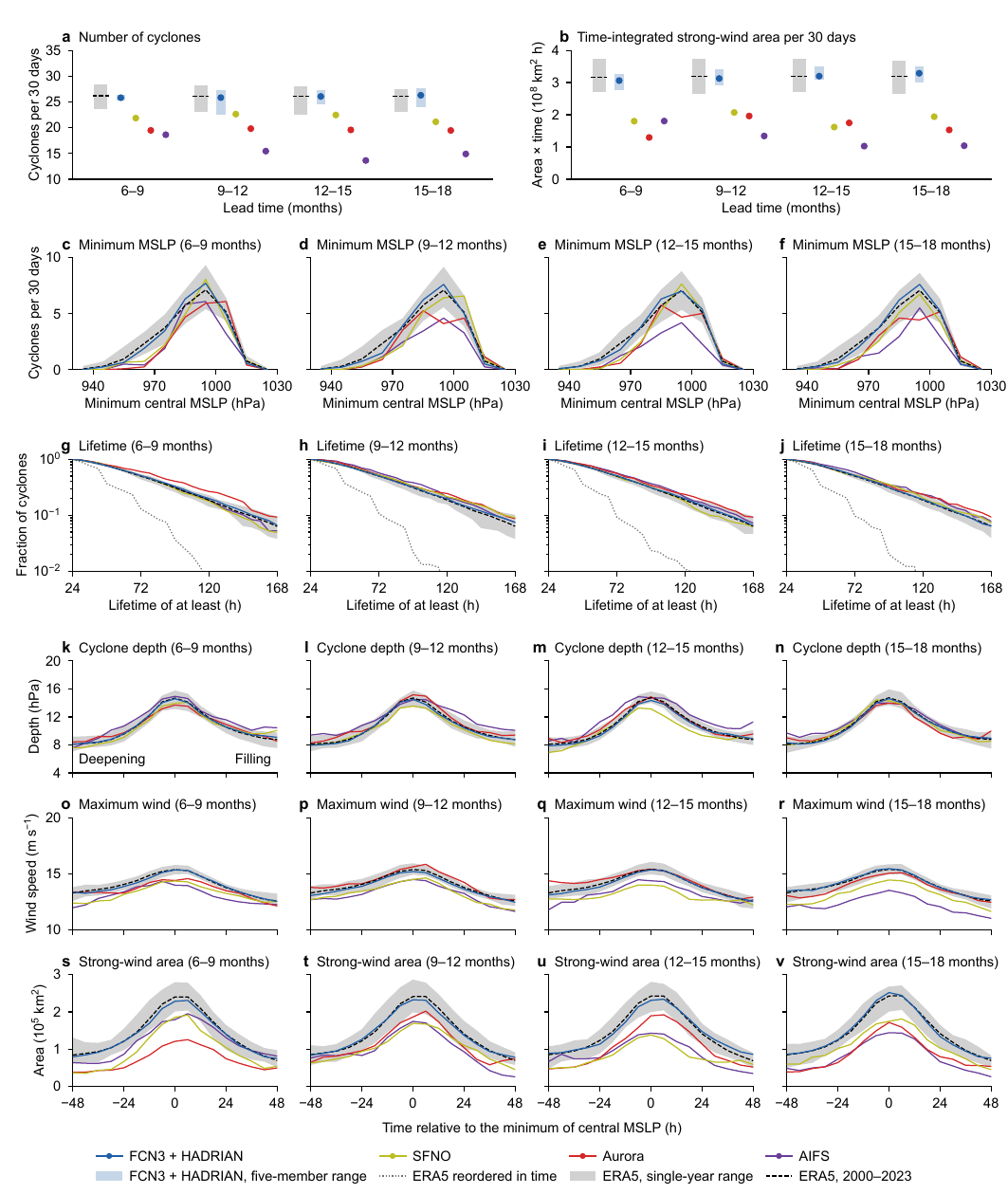}
\caption{\textbf{Cyclone statistics by lead-time interval.} Cyclones of
\cref{fig:cyclone_composites}. (a) Number of cyclones per 30 days.
(b) Strong-wind area per 30 days: the area within 1,000~km of a centre in which
the 10-m wind speed exceeds the local 98th percentile of ERA5 (2000--2019, by
calendar month and UTC hour), summed over all cyclones and six-hourly times and
multiplied by 6~h. (c--f) Distribution of the lifetime-minimum central MSLP.
(g--j) Fraction of cyclones lasting at least the time on the abscissa; dotted
lines show ERA5 with whole days reordered at random within each calendar month.
(k--v) Depth (the pressure of the outermost closed isobar that encloses only this
centre, minus the central pressure), maximum 10-m wind speed within $5^\circ$ and
strong-wind area, averaged over all cyclones, against time from the minimum of
central MSLP. FCN3 + HADRIAN is the mean over its 20 members; in a and b, blue bars
show its five-member range. Dashed lines show ERA5 over 2000--2023, and grey bars
and bands the range of its 24 single years.}
\label{fig:cyclone_stats}
\end{figure}
\endgroup

\end{document}